\documentclass[aps,pra,showpacs,superscriptaddress,tightenlines,twocolumn,lengthcheck]{revtex4-1}
\usepackage{diagbox}
\usepackage{multirow}
\usepackage{amsmath}
\usepackage{amsfonts}
\usepackage{graphicx}
\usepackage{epsfig}
\usepackage{color}
\usepackage{txfonts}
\usepackage[colorlinks,citecolor=blue]{hyperref}
\usepackage{diagbox}
\usepackage{mathrsfs}
\usepackage{xcolor}
\usepackage{amsfonts}
\usepackage{bbm}
\usepackage{dsfont}
\usepackage[T1]{fontenc}

\begin{document}

\title{General dark-state theory for arbitrary multilevel quantum systems}

\author{Xuan Zhao}
\affiliation{Key Laboratory of Low-Dimensional Quantum Structures and Quantum Control of Ministry of Education, Key Laboratory for Matter Microstructure and Function of Hunan Province, Department of Physics and Synergetic Innovation Center for Quantum Effects and Applications, Hunan Normal University, Changsha 410081, China}
\affiliation{Hunan Research Center of the Basic Discipline for Quantum Effects and Quantum Technologies, Hunan Normal University, Changsha 410081, China}

\author{Le-Man Kuang}
\affiliation{Key Laboratory of Low-Dimensional Quantum Structures and Quantum Control of Ministry of Education, Key Laboratory for Matter Microstructure and Function of Hunan Province, Department of Physics and Synergetic Innovation Center for Quantum Effects and Applications, Hunan Normal University, Changsha 410081, China}
\affiliation{Hunan Research Center of the Basic Discipline for Quantum Effects and Quantum Technologies, Hunan Normal University, Changsha 410081, China}

\author{Jie-Qiao Liao}
\email{Contact author: jqliao@hunnu.edu.cn}
\affiliation{Key Laboratory of Low-Dimensional Quantum Structures and Quantum Control of Ministry of Education, Key Laboratory for Matter Microstructure and Function of Hunan Province, Department of Physics and Synergetic Innovation Center for Quantum Effects and Applications, Hunan Normal University, Changsha 410081, China}
\affiliation{Hunan Research Center of the Basic Discipline for Quantum Effects and Quantum Technologies, Hunan Normal University, Changsha 410081, China}
\affiliation{Institute of Interdisciplinary Studies, Hunan Normal University, Changsha 410081, China}

\begin{abstract}

The dark-state effect, caused by destructive quantum interference, is an important physical effect in atomic physics and quantum optics. It not only deepens the understanding of light-atom interactions, but also has wide applications in quantum physics and quantum information.
Therefore, how to efficiently and conveniently determine the number and form of the dark states in multilevel quantum systems with complex transitions is an important and interesting topic in this field. In this work, we present a general theory for determining the dark states in multilevel quantum systems with any coupling configuration using the arrowhead-matrix method. To confirm the dark states in a multilevel system, we first define the upper- and lower-state subspaces, and then diagonalize the Hamiltonians restricted within the two subspaces to obtain the dressed upper and lower states. By further expressing the transitions between the dressed upper and lower states, we can map the multilevel system to a bipartite-graph network, in which the nodes and links are acted by the dressed states and transitions, respectively. Based on the coupling configurations of the network, we can determine the lower dark states with respect to the upper-state subspace.
As examples, we analyze the dark states in three-, four-, and five-level quantum systems, for all possible configurations through the classification of the numbers of upper and lower states. Furthermore, we extend the framework to multilevel quantum systems and discuss the existence of dark states in some typical configurations. 
We also recover the results of the dark-state polaritons in driven three-level systems with the arrowhead-matrix method. Our theory paves the way for manipulating and utilizing the dark states of multilevel quantum systems in atomic physics and quantum optics.
\end{abstract}

\date{\today }
\maketitle

\section{Introduction}

The dark states~\cite{DS1976}, owing to their novel physical properties and wide applications, play a crucial role in modern atomic physics and quantum optics~\cite{ScullyQO1997,CAbook2011}. For example, in a $\Lambda$-type three-level system under two-photon resonance, the dark state is a coherent superposition of the two lower states and hence it is immune to the effect of spontaneous emission due to destructive quantum interference. 
As a result, the dark states provide the physical mechanism underlying many quantum phenomena such as coherent population trapping~\cite{CPT1978,CPT1982,CPT1988,CPT1996}, electromagnetically induced transparency~\cite{EITPRL1991,Arimondo1996,EIT:H1997,DPiEIT2000,EIT2003,EIT2005,EIT2007}, stimulated Raman adiabatic passage~\cite{STIRAP:A1989,STIRAP1998,STIRAP2015,STRIP2017}, quantum state engineering~\cite{DS:jpa2017,DS:jpl2021}, laser cooling~\cite{CEwEIT2000,CuEIT2000,LCiTS2010,LCiTi2012}, and classical and quantum interference~\cite{DS:CIarX}. In particular, the dark-state effect has been demonstrated in various physical platforms, such as cavity-QED systems~\cite{DS_AF2007,DS:prx2022,DS_AF2014,DS_AF2020,DS_AF2021,DS_AF2023,DS_CQEDe2024}, trapped ions~\cite{DS_TiCe2009,DS_TiCe2013,DS_Tie2015,DS_TiCe2020,DS_Tie2024}, and superconducting quantum circuits~\cite{DS_QCe2014,DS_QCe2015,DS_QCe2018,DS_QCe2022,DS_QCe2025}.
	
Recently, there has been a growing interest in studying the dark states in multilevel quantum systems~\cite{DS:prx2022,MSt:PRA2006,CoDS:PRA2019,DS_ML2022,DDSiMS2025}, which naturally offer more complex energy-level structure, and hence they can exhibit richer physical phenomena. 
The dark states in multilevel systems based on the extension of the $\Lambda$-type three-level systems have been widely studied~\cite{ELtMS1998,DSoMM2013,ELtMS2020}, especially with the $\Lambda$-chain configurations constructed by linking multiple $\Lambda$-type three-level structures~\cite{ELtMS:JCP1991,ELtMS:PRA1991,ELtMS1992}. 
Consequently, the dark states in multilevel systems play an important role in many fields such as adiabatic population transfer~\cite{,MAPT1991,EoSTIRAP1997,APT1993,OSTIRAP2005,M_STIRAP2008,ELtMS2020}, quantum computing~\cite{QCaQI2000}, and atom optics~\cite{AO1994,AO1998}.
Since there are more transition paths in multilevel systems, how to efficiently and conveniently determine the dark states in multilevel systems with complex transitions becomes an interesting and important research topic.
Currently, several theoretical works and analytical approaches to identify the dark states in multilevel systems have been proposed~\cite{DC1981,MS:PRA1983,MSN:JMO2014,MSt:PRA2006,MSfNS:PRA2020,SVDiCQED2013,SVDiJC2013,SVDoA2024}, for example, utilizing the symmetries with respect to the decoupling of the system to reduce the multilevel systems~\cite{DC1981}, and adopting the Morris-Shore transformation to reduce the multilevel system to a set of independent nondegenerate two-state systems and a number of uncoupled dark states~\cite{MS:PRA1983,MSN:JMO2014,MSt:PRA2006,MSfNS:PRA2020}. The dark states can also be analyzed based on the singular value decomposition (SVD) of the coupling matrix~\cite{SVDiCQED2013,SVDiJC2013,SVDoA2024}.

In this work, we propose a general theory for studying the dark states in arbitrary multilevel systems based on the arrowhead-matrix method. Our method not only gives the number of the dark states, but also presents their form.
The arrowhead-matrix method, as an efficient and practical method, was originally proposed for analyzing the dark-mode effects in linear bosonic networks~\cite{huang2023dark}.
Concretely, by classifying the types of the modes and transforming the Hamiltonian matrix of the system into an arrowhead matrix, we can utilize the properties of the arrowhead matrix to analyze and determine the dark modes in complex bosonic networks.
Note that the dark-mode effect and relating physical phenomena have recently been studied in multimode optomechanical systems~\cite{DM_GSC2020,DM_OIT2020,DM_OC2022,DM_OE2022,DM_QS2022,DM_QS2023}. 
In a multilevel system, the energy levels and the transitions can be served as the nodes and links, respectively, then a map can be established between the multilevel systems and the linear bosonic networks. This feature motivates us to study the determination of the dark states in multilevel systems with the arrowhead-matrix method. Note that this method has been used to study the dark-state effect in the multimode Jaynes-Cummings model~\cite{DSiFSL}. 

To expound the arrowhead-matrix method in a general manner, we consider the multilevel system with all possible transitions among these energy levels. When some forbidden transition exists, we let the corresponding transition amplitude be zero. 
To perform the arrowhead-matrix method, we first categorize the basis states into two types and define them as the upper and lower states depending on the specific research demands. We point out that the dark states refer to some states in the lower-state subspace decoupled from all the states in the upper-state subspace.
Next, by defining the basis vectors according to the upper and lower states, the Hamiltonian of the multilevel system can be expressed as the block matrix formed by the upper- and lower-state submatrices, as well as the coupling matrix.
Furthermore, we diagonalize both the upper- and lower-state submatrices and express the internal transitions with these dressed upper and lower states. Then we can obtain an arrowhead matrix with diagonalized upper- and lower-state submatrices and a transformed coupling matrix. As a result, we can analyze the dark states with the arrowhead-matrix method~\cite{huang2023dark,DSiFSL}. 
Based on the above method, we concretely study the dark states in the three-, four-, and five-level quantum systems, as well as  some typical coupling configurations in the general multilevel systems. We also recover the results of the dark-state polaritons in driven three-level systems with the arrowhead-matrix method.

The rest of this paper is organized as follows. In Sec.~\ref{sec2}, we introduce the multilevel quantum systems and the arrowhead-matrix method. In Secs.~\ref{sec3}, ~\ref{sec4}, and~\ref{sec5}, we study the dark states in the three-, four-, and five-level systems, respectively. By classifying the system according to the numbers of the upper and lower states, we analyze the dark states in various configurations in detail. 
In Sec.~\ref{sec6}, we extend the analysis to an arbitrary multilevel system, and confirm the parameter conditions under which the dark-state effect appears. We also present the number and form of the dark states in some typical coupling configurations. In Sec.~\ref{DSp}, we rederive the results of the dark-state polaritons in driven three-level systems using the arrowhead-matrix method. Finally, we conclude this work in Sec.~\ref{conclusion}. 
Two Appendices~({\ref{A1} and~\ref{A2}}) are presented to show the detailed derivation of the time-independent Hamiltonian and proof of the assertions for determining the bright and dark states in the system.

\section{The multilevel quantum systems and arrowhead-matrix method}\label{sec2}

In this section, we introduce the multilevel quantum systems and present the arrowhead-matrix method.

\subsection{A general multilevel quantum system}

\begin{figure*}[t]
	\centering\includegraphics[width=0.68\textwidth]{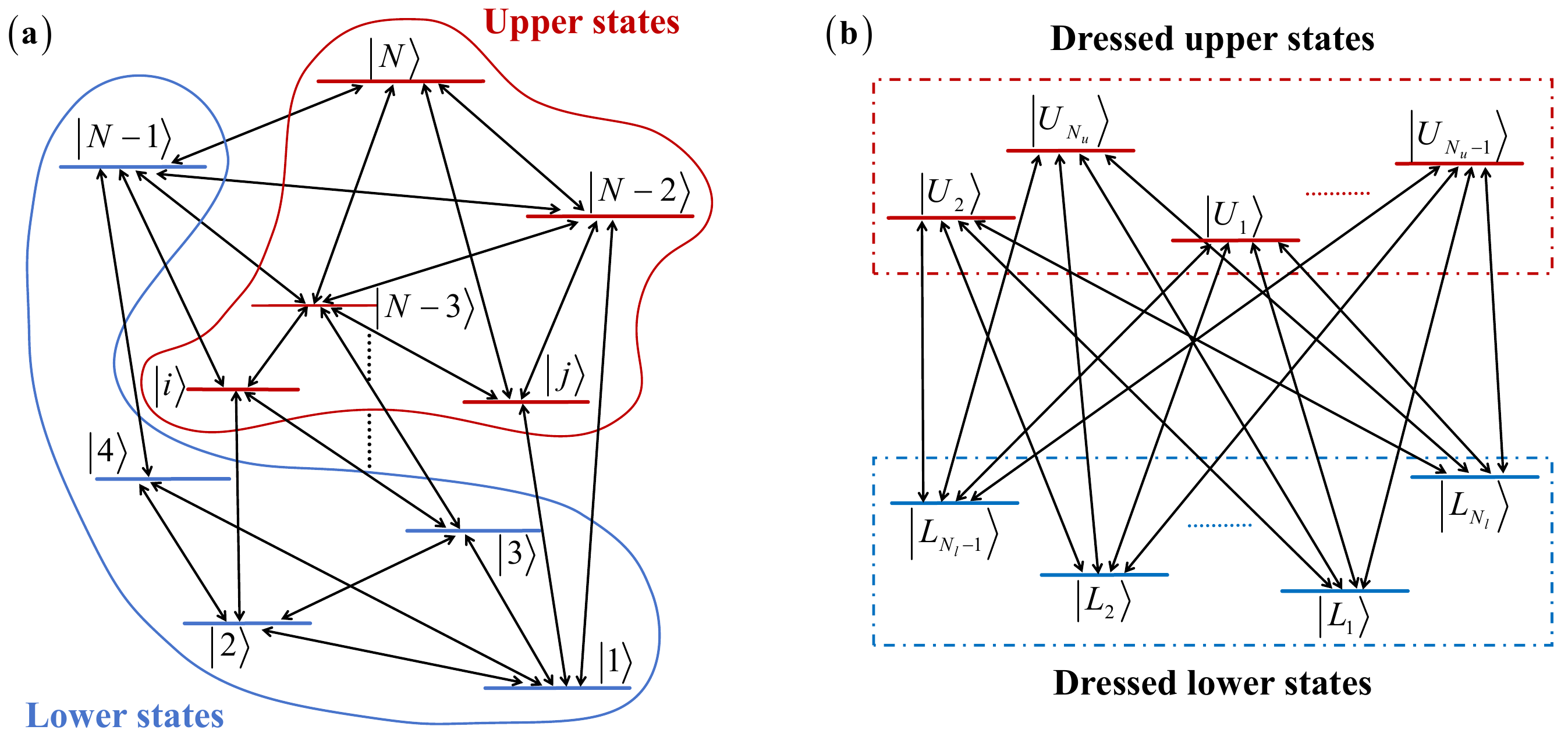}
	\caption{(a) The energy-level diagram of a general multilevel quantum system with all possible transitions among these energy levels (some transitions are omitted for concision), which are divided into two components: $N_{u}$ upper states marked with the red line ($\left\vert u_{1}\right\rangle=\left\vert N\right\rangle$, $\left\vert u_{2}\right\rangle=\left\vert N-2\right\rangle$, $\ldots$, and $\vert u_{N_{u}}\rangle=\left\vert j\right\rangle$) and $N_{l}$ lower states  marked with the blue line ($\left\vert l_{1}\right\rangle=\left\vert N-1\right\rangle$, $\ldots$, $\vert l_{N_{l}-1}\rangle=\left\vert 2\right\rangle$, and $\vert l_{N_{l}}\rangle=\left\vert 1\right\rangle$).
		(b) The bipartite-graph presentation of the $N$-level system with the dressed upper states ($\left\vert U_{1}\right\rangle$, $\left\vert U_{2}\right\rangle$, \ldots, and $\vert U_{N_{u}}\rangle$) and the dressed lower states ($\left\vert L_{1}\right\rangle $, $\left\vert L_{2}\right\rangle $, \ldots, and $\vert L_{N_{l}}\rangle$), where the couplings only exist between the dressed upper states and the dressed lower states.}
	\label{nls}
\end{figure*}

We consider a general multilevel quantum system (choosing an $N$-level system without loss of generality) with possible transitions among all these energy levels~[see Fig.~\ref{nls}(a)]. 
To be general, here we do not consider forbidden transitions among these energy levels. In realistic physical systems, the coupling strength can be taken as zero when the corresponding transition is forbidden. The Hamiltonian of the general $N$-level quantum system can be written as ($\hbar =1$)%
\begin{equation} \label{HN}
H^{[N]}=\sum_{{j}=1}^{N}E_{j}\left\vert
j\right\rangle \left\langle j\right\vert +\sum_{j,j^{\prime }=1,j<j^{\prime }}^{N} (\Omega _{jj^{\prime }}e^{-i\omega _{jj^{\prime
	}}t} \vert
j^{\prime }\rangle \left\langle j\right\vert +\text{H.c.}),
\end{equation}%
where $E_{j}$ is the energy of the $j$th energy level, $\Omega _{jj^{\prime }}$ and $\omega _{jj^{\prime }}$ are, respectively, the transition coefficient and the driving laser frequency associated with the transition $\vert j^{\prime }\rangle \leftrightarrow	
\left\vert j\right\rangle $.
The superscript \textquotedblleft$[N]$\textquotedblright\ in $H^{[N]}$ denotes the $N$-level system.
To analyze the dark states in the quantum systems, we need to confirm the upper and lower states in advance. Here, the upper states are those to be decoupled from, while the remaining states are the lower states. Note that the upper and lower states here are relative concepts, rather than the high- and low-energy levels in realistic physical systems. 

For better analyzing the dark states, we prefer to work in a rotating frame 
with respect to $H_{0}=E_{N}\left\vert
N\right\rangle \left\langle N\right\vert+\sum_{{r}=1}^{N-1}( E_{r}+\Delta _{rN}) \left\vert
r\right\rangle \left\langle r\right\vert $, and then a time-independent Hamiltonian in this rotating frame can be obtained as
\begin{eqnarray}\label{HN_tilde}
	\tilde{H}^{[N]} =\sum_{{r}=1}^{N-1} -\Delta _{rN} \left\vert r\right\rangle
	\left\langle r\right\vert + \sum_{j,j^{\prime }=1,j<j^{\prime
	}}^{N}(\Omega _{jj^{\prime }}\vert j^{\prime }\rangle \left\langle
	j\right\vert +\text{H.c.}) ,
\end{eqnarray}%
where $\Delta _{rN}=E_{N}-E_{r}-\omega _{rN}$ is the detuning of the energy separation $E_{N}-E_{r}$ of the transition $\left\vert r\right\rangle \leftrightarrow \vert N \rangle$ with respect to the driving frequency $\omega_{rN}$ of the field. Note that these detunings should satisfy the relations $ \Delta _{rN}-\Delta _{r^{\prime }N}=\Delta _{rr^{\prime }}$ for $r,r^{\prime }=1,2,\ldots,N-1,$ and $r<r^{\prime
}$, such that the Hamiltonian becomes time-independent in the rotating frame.

Without loss of generality, we assume that the $N$-level quantum system has $N_{u}$ upper states defined as $\{\left\vert u_{1}\right\rangle ,$ $\left\vert u_{2}\right\rangle
,$ $\ldots,$ $\vert u_{N_{u}}\rangle \}$ and $N_{l}$ lower states $%
\{\left\vert l_{1}\right\rangle ,$ $\left\vert l_{2}\right\rangle
,$ $\ldots,$ $\vert l_{N_{l}}\rangle \}$, where $N_{u}$ and $N_{l }$ satisfy the relation $N_{u}+N_{l }=N$. 
Note that the number of the lower states should satisfy $N_{l} > 1$ for maintaining the quantum interference channels when studying the dark-state effect in the lower-state subspace.
Then the Hamiltonian can be rewritten as 
\begin{eqnarray}
	\tilde{H}^{[N]} &=&\sum_{{n_{u}}=1}^{N_{u}}\delta _{n_{u}} \vert u_{n_{u}}\rangle
	\langle u_{n_{u}}\vert + \sum_{n_{u},n_{u}^{\prime }=1,n_{u}<n_{u}^{\prime
	}}^{N_{u}}(\xi _{n_{u}n_{u}^{\prime }}\vert u_{n_{u}}\rangle \langle
	u_{n_{u}^{\prime }}\vert +\text{H.c.}) \notag \\
	&&+\sum_{{n_{l}}=1}^{N_{l}}\omega _{n_{l}} \vert l_{n_{l}}\rangle
	\langle l_{n_{l}}\vert + \sum_{n_{l},n_{l}^{\prime }=1,n_{l}<n_{l}^{\prime
	}}^{N_{l}}(\eta_{n_{l}n_{l}^{\prime }}\vert l_{n_{l}}\rangle \langle
	l_{n_{l}^{\prime }}\vert +\text{H.c.})\notag \\
&&+\sum_{{n_{u}}=1}^{N_{u}}\sum_{{n_{l}}=1}^{N_{l}}(g_{n_{u}n_{l}}\vert u_{n_{u}}\rangle
\langle l_{n_{l}}\vert+\text{H.c.}).
\end{eqnarray}%
We further define the basis vectors for these upper and lower states as
\begin{subequations}
\begin{align}
\left\vert u_{1}\right\rangle =&(
	\underbrace{1_1,0,\ldots,0,}_{N_{u}\text{ upper states }}
	\underbrace{0,0,\ldots,0}_{N_{l}\text{ lower states }}
) ^{T}, \\
\left\vert u_{2}\right\rangle =&\left(
0,1_2,\ldots,0,0,0,\ldots,0\right) ^{T}, \\
&\ldots  \notag \\
\vert u_{n_{u}}\rangle =&\left(
0,0,\ldots,1_{n_{u}},\ldots,0,0,0,\ldots,0\right) ^{T}, \\
&\ldots  \notag \\
\vert u_{N_{u}}\rangle =&\left(
0,0,\ldots,1_{N_{u}},0,0,\ldots,0\right) ^{T}, \\
\left\vert l_{1}\right\rangle =& \left(
0,0,\ldots,0,1_{N_{u}+1},0,\ldots,0\right) ^{T}, \\
\left\vert l_{2}\right\rangle =& \left(
0,0,\ldots,0,0,1_{N_{u}+2},\ldots,0\right) ^{T}, \\
&\ldots  \notag \\
\left\vert l_{n_{l}}\right\rangle =& \left(
0,0,\ldots,0, 0,0,\ldots,1_{N_{u}+n_{l}},\ldots,0\right) ^{T}, \\
&\ldots  \notag \\
\vert l_{N_{l}}\rangle =&\left(
0,0,\ldots,0,0,0,\ldots,1_{N}\right) ^{T},
\end{align}%
\end{subequations}
where the subscript of the element \textquotedblleft1\textquotedblright\ is introduced to denote its position in the vector, and the superscript \textquotedblleft$T$\textquotedblright\ denotes the matrix transpose. 
In this representation, the time-independent Hamiltonian $\tilde{H}^{[N]}$ can be expressed as
\begin{eqnarray}
	\tilde{H}^{[N]}&=&\left( 
	\begin{array}{c|c}
		\mathbf{H}_{u} & \mathbf{c} \\ \hline
		\mathbf{c}^{\dag } & \mathbf{H}_{l}%
	\end{array}%
	\right) \notag \\
	&=&\left(\begin{array}{cccc|cccc}
		\delta_{1} & \xi_{12} & \cdots & \xi_{1N_{u}} & g_{11} & g_{12} & \cdots & g_{1N_{l}}\\
		\xi_{12}^{\ast} & \delta_{2} & \cdots & \xi_{2N_{u}} & g_{21} & g_{22} & \cdots & g_{2N_{l}}\\
		\vdots & \vdots & \ddots & \vdots & \vdots & \vdots & \ddots & \vdots\\
		\xi_{1N_{u}}^{\ast} & \xi_{2N_{u}}^{\ast} & \cdots & \delta_{N_{u}} & g_{N_{u}1} & g_{N_{u}2} & \cdots & g_{N_{u}N_{l}}\\
		\hline g_{11}^{\ast} & g_{21}^{\ast} & \cdots & g_{N_{u}1}^{\ast} & \omega_{1} & \eta_{12} & \cdots & \eta_{1N_{l}}\\
		g_{12}^{\ast} & g_{22}^{\ast} & \cdots & g_{N_{u}2}^{\ast} & \eta_{12}^{\ast} & \omega_{2} & \cdots & \eta_{2N_{l}}\\
		\vdots & \vdots & \ddots & \vdots & \vdots & \vdots & \ddots & \vdots\\
		g_{1N_{l}}^{\ast} & g_{2N_{l}}^{\ast} & \cdots & g_{N_{u}N_{l}}^{\ast} & \eta_{1N_{l}}^{\ast} & \eta_{2N_{l}}^{\ast} & \cdots & \omega_{N_{l}}
	\end{array}\right),
\end{eqnarray}%
where $\mathbf{H}_{u}$, $\mathbf{H}_{l}$, and $\mathbf{c}$ are the submatrices
related to the Hamiltonians in the upper- and lower-state subspaces, and the coupling Hamiltonian between the upper- and lower-state subspaces, respectively. 

According to the dark-mode theorems~\cite{huang2023dark}, we need to diagonalize both the upper- and lower-state submatrices to analyze the dark states. By introducing the unitary matrix $\mathbf{S}_{u}$ ($\mathbf{S}_{l}$), the upper-state (lower-state) submatrix $\mathbf{H}_{u}$ ($\mathbf{H}_{l}$) can be diagonalized as $
\mathbf{H}_{U} =\mathbf{S}_{u} \mathbf{H}_{u}\mathbf{S}_{u}^{\dag }=\text{diag }(\Delta_{1},\Delta_{2},\Delta_{3},\ldots,\Delta_{N_{u}})$ [$
\mathbf{H}_{L} =\mathbf{S}_{l} \mathbf{H}_{l}\mathbf{S}_{l}^{\dag }=\text{diag }(\Omega_{1},\Omega_{2},\Omega_{3},\ldots,\Omega_{N_{l}})  $], and the corresponding coupling matrix $\mathbf{c} $ is transformed into $\mathbf{C} = \mathbf{S}_{u}\mathbf{c}\mathbf{S}_{l}^{\dag }=( \mathbf{C}_{1},\mathbf{C}_{2},\ldots,\mathbf{C}_{N_{l}}) $. 
Then the transformed Hamiltonian can be expressed as 
\begin{eqnarray}\label{HDN_UL}
	\tilde{H}^{[N]}_{D} &=&\sum_{{n_{u}}=1}^{N_{u}}\Delta _{n_{u}} \vert U_{n_{u}}\rangle
	\langle U_{n_{u}}\vert +\sum_{{n_{l}}=1}^{N_{l}}\Omega _{n_{l}} \vert L_{n_{l}}\rangle
	\langle L_{n_{l}}\vert \notag \\
	&&+\sum_{{n_{u}}=1}^{N_{u}}\sum_{{n_{l}}=1}^{N_{l}}(G_{n_{u}n_{l}}\vert U_{n_{u}}\rangle
	\langle L_{n_{l}}\vert+\text{H.c.}).
\end{eqnarray}%
Here, we introduce the diagonalized upper and lower states [namely the dressed upper and lower states in Fig.~\ref{nls}(b)] as $\{\left\vert U_{1}\right\rangle ,$ $\left\vert
U_{2}\right\rangle ,$ $\ldots,$ $\vert U_{N_{u}}\rangle ,$ $\left\vert
L_{1}\right\rangle ,$ $\left\vert L_{2}\right\rangle ,$ $\ldots,$ $\vert
L_{N_{l}}\rangle\} $, which are the eigenstates of the upper- and lower-state Hamiltonians, i.e., $\mathbf{H}_{U}\vert U_{n_{u}}\rangle=\Delta_{n_{u}}\vert U_{n_{u}}\rangle$ for $n_{u}=1,2,\ldots,N_{u}$, and $\mathbf{H}_{L}\vert L_{n_{l}}\rangle=\Omega_{n_{l}}\vert L_{n_{l}}\rangle$ for $n_{l}=1,2,\ldots,N_{l}$.
The subscript \textquotedblleft$D$\textquotedblright\ in $\tilde{H}^{[N]}_{D}$ denotes the diagonalized upper- and lower-state submatrices. Now the $N$-level system can be described by a bipartite graph~[{see Fig.~\ref{nls}(b)}], where the dressed upper and lower states play the role of the nodes, and the transitions between the dressed upper and lower states serve as the links.

Furthermore, we define the basis vectors for these dressed upper and lower states as
\begin{subequations}
		\begin{align}
			\left\vert U_{1}\right\rangle =&(
			\underbrace{1_1,0,\ldots,0,}_{N_{u}}
			\underbrace{0,0,\ldots,0}_{N_{l}}
			) ^{T}, \\
			\left\vert U_{2}\right\rangle =&\left(
			0,1_2,\ldots,0,0,0,\ldots,0\right) ^{T}, \\
			&\ldots  \notag \\
			\vert U_{n_{u}}\rangle =&\left(
			0,0,\ldots,1_{n_{u}},\ldots,0,0,0,\ldots,0\right) ^{T}, \\
			&\ldots  \notag \\
			\vert U_{N_{u}}\rangle =&\left(
			0,0,\ldots,1_{N_{u}},0,0,\ldots,0\right) ^{T}, \\
			\left\vert L_{1}\right\rangle =& \left(
			0,0,\ldots,0,1_{N_{u}+1},0,\ldots,0\right) ^{T}, \\
			\left\vert L_{2}\right\rangle =& \left(
			0,0,\ldots,0,0,1_{N_{u}+2},\ldots,0\right) ^{T}, \\
			&\ldots  \notag \\
			\left\vert L_{n_{l}}\right\rangle =& \left(
			0,0,\ldots,0, 0,0,\ldots,1_{N_{u}+n_{l}},\ldots,0\right) ^{T}, \\
			&\ldots  \notag \\
			\vert L_{N_{l}}\rangle =&\left(
			0,0,\ldots,0,0,0,\ldots,1_{N}\right) ^{T},
		\end{align}%
\end{subequations}
then the Hamiltonian $\tilde{H}^{[N]}_{D}$ can be expressed by the following thick arrowhead matrix ("thick" means the dimension of the arrowhead edge is greater than one):
\begin{eqnarray}\label{HDN}
	\tilde{H}^{[N]}_{D}&=&\left( 
	\begin{array}{c|c}
		\mathbf{H}_{U} & \mathbf{C} \\ \hline
		\mathbf{C}^{\dag } & \mathbf{H}_{L}%
	\end{array}%
	\right)\notag \\
	&=&\left(
	\begin{array}{cccc|cccc}
		\Delta _{1} & 0 & \cdots  & 0 &G_{11}& G_{12} & \cdots  & G_{1N_{l}} \\
		0 & \Delta _{2} & \cdots  & 0 & G_{21}& G_{22} & \cdots  & G_{2N_{l}} \\
		\vdots  & \vdots  & \ddots  & \vdots  & \vdots  & \vdots  & \ddots  & \vdots\\
		0 & 0 & \cdots  & \Delta _{N_{u}} & G_{N_{u}1} &G_{N_{u}2} & \cdots  & G_{N_{u}N_{l}} \\  \hline
		G^{\ast}_{11} & G^{\ast}_{21} & \cdots  & G^{\ast}_{N_{u}1} & \Omega _{1} & 0 & \cdots  & 0 \\
		G^{\ast}_{12} & G^{\ast}_{22} & \cdots  & G^{\ast}_{N_{u}2} & 0 & \Omega _{2} & \cdots  & 0 \\
		\vdots  & \vdots  & \ddots  & \vdots  & \vdots  & \vdots  & \ddots  & \vdots \\
		G^{\ast}_{1N_{l}} & G^{\ast}_{2N_{l}} & \cdots  & G^{\ast}_{N_{u}N_{l}} & 0 & 0 & \cdots & \Omega _{N_{l}}
	\end{array}\right).\notag \\
\end{eqnarray}%
Based on this thick arrowhead matrix, the dark-state effect can be analyzed in detail.

\subsection{The arrowhead-matrix method}\label{assertion}

In this section, we introduce the detailed procedure of the arrowhead-matrix method. To this end, we first present the definition of the dark state.
Typically, the dark state refers to a special quantum state in atomic or molecular systems that cannot absorb or emit light, making it effectively "invisible" to electromagnetic radiation,  and specially, if the system is initially prepared in one of the dark states, the system will stay in this dark state forever.
Here, we generalize the concept of the dark state to a wider sense of decoupling, i.e., \textit{the dark state is a state in a subspace decoupled from the target subspace}. 
Therefore, we can also call the dark state as the decoupled state. 
For describing a dark state, therefore, we need to confirm the target subspace in advance (namely, which state should be specified to be decoupled by the dark states). In this work, we denote the target subspace as the upper-state subspace, then the rest subspace is referred as the lower-state subspace. The dark-state subspace is formed by the set of all the states in the lower-state subspace  decoupled from the target subspace. In addition, we want to point out that the term "dark" only works with respect to the lights that induce the inter-transition between the upper- and lower-state subspaces. For the lights inducing the intratransition within the upper-state (lower-state) subspace, they can still be absorbed or emitted.

For the present $N$-level quantum system, based on the definition of the dark state and Eq.~(\ref{HDN}), we can determine the number and form of the dark states in this system with \textbf{the arrowhead-matrix method}~\cite{huang2023dark,DSiFSL}:

(1) If the $k$th column vector of the coupling matrix $\mathbf{C}$ in Eq.~{(\ref{HDN})} satisfies $\mathbf{C}_{k}=(G_{1k},G_{2k},\ldots,G_{N_{u}k})^{T}=\mathbf{0}$, namely, ${G}_{jk}={0}$ for $j=1,2,\ldots,N_{u}$, then the corresponding basis state $\left\vert L_{k}\right\rangle$ is a dark state with respect to all these dressed upper states.

(2) If all the column vectors of the coupling matrix $\mathbf{C}$ are nonzero $\mathbf{C}_{k=1-N_{l}}\neq \mathbf{0}$ and there are $l$ $(l=2,3,\ldots,N_{l})$ degenerate dressed lower states [i.e., $\Omega_{j=1\text{-}l}=\Omega$, as marked by the red fonts in Eq.~{(\ref{HDN_e})}], then the corresponding Hamiltonian can be expressed as 
\begin{widetext}
	\begin{equation}\label{HDN_e}
	\tilde{H}^{[N]}_{D}=\left(
	\begin{array}{cccc|cccc|c|c|c}
		\Delta _{1} & 0 & \cdots  & 0 &\textcolor{blue}{G_{11}}& \textcolor{blue}{G_{12}} & \textcolor{blue}{\cdots}  & \textcolor{blue}{G_{1l}} & G_{1(l+1)} & \cdots  &   G_{1N_{l}}     \\
		0 & \Delta _{2} & \cdots  & 0 & \textcolor{blue}{G_{21}}& \textcolor{blue}{G_{22}} & \textcolor{blue}{\cdots}  & \textcolor{blue}{G_{2l}} & G_{2(l+1)} & \cdots  &   G_{2N_{l}}\\
		\cdots  & \cdots  & \ddots  & \cdots  & \textcolor{blue}{\cdots}  & \textcolor{blue}{\cdots}  & \textcolor{blue}{\cdots}  & \textcolor{blue}{\cdots}& \cdots  & \cdots  & \cdots \\
		0 & 0 & \cdots  & \Delta _{N_{u}} & \textcolor{blue}{G_{N_{u}1}} &\textcolor{blue}{G_{N_{u}2}} & \textcolor{blue}{\cdots}  & \textcolor{blue}{G_{N_{u}l}} & G_{N_{u}(l+1)} & \cdots  &   G_{N_{u}N_{l}} \\  \hline
		G^{\ast}_{11} & G^{\ast}_{21} & \cdots  & G^{\ast}_{N_{u}1} & \textcolor{red}{\Omega} & 0 & \cdots  & 0 & 0 & \cdots  & 0  \\
		G^{\ast}_{12} & G^{\ast}_{22} & \cdots  & G^{\ast}_{N_{u}2} & 0 & \textcolor{red}{\Omega}  & \cdots  & 0 & 0 & \cdots  & 0 \\
		\cdots  & \cdots  & \cdots  & \cdots  & \cdots  & \cdots  & \ddots  & \cdots   & \cdots  & \cdots  & \cdots \\
		G^{\ast}_{1l} & G^{\ast}_{2l} & \cdots  & G^{\ast}_{N_{u}l} & 0 & 0 & \cdots & \textcolor{red}{\Omega} & 0 & \cdots  & 0 \\ \hline
		G^{\ast}_{1(l+1)} & G^{\ast}_{2(l+1)} & \cdots  & G^{\ast}_{N_{u}(l+1)} & 0 & 0 & \cdots& 0  & \Omega_{l+1} & \cdots  & 0 \\\hline
			\cdots  & \cdots  & \cdots  & \cdots  & \cdots  & \cdots  & \cdots  & \cdots   & \cdots  & \ddots  & \cdots \\\hline
			G^{\ast}_{1 N_{l}} & G^{\ast}_{2 N_{l}} & \cdots  & G^{\ast}_{N_{u} N_{l}} & 0 & 0 & \cdots& 0  & 0 & \cdots  & \Omega_{N_{l}} 
	\end{array}\right).
\end{equation}%
\end{widetext}
Firstly, we can see based on Eq.~{(\ref{HDN_e})} that the dressed lower state without degeneracy (namely the dimension of the degenerate subspace is one) is a bright state, i.e., it will not be decoupled to all these upper states. For example, the dressed lower states $\{\vert U_{l+1}\rangle$, \ldots, $\vert U_{N_{l}}\rangle\}$ are bright states. The dark states only exist in the degenerate dressed-lower-state subspace. In particular, the number and form of these dark states are determined by the coupling submatrix associated with the degenerate dressed-lower-state subspace [i.e., the submatrix marked by blue fonts in Eq.~{(\ref{HDN_e})}].
	
Below, we present the detailed analyses concerning the dark states in the degenerate dressed-lower-state subspace with dimension $l$. 
We denote the dark state in the $l$-dimensional orthogonal degenerate subspace $\{\left\vert L_{1}\right\rangle ,$ $\left\vert L_{2}\right\rangle ,$ $\ldots,$ $\vert L_{l}\rangle \}$ as $\left\vert D\right\rangle=\sum_{i=1}^{l}x_{i}\left\vert L_{i}\right\rangle $. Then the dark state should be decoupled from all these dressed upper states $\{\left\vert U_{1}\right\rangle ,$ $\left\vert
U_{2}\right\rangle ,$ $\ldots,$ $\vert U_{N_{u}}\rangle \}$, namely, $\langle U_{j}\vert \mathbf{C}_{[l]}\left\vert 	D\right\rangle=0$ for $j=1,2,\ldots,N_{u}$, where $\mathbf{C}_{[l]}$ is the coupling submatrix with dimension $N_{u} \times l$ corresponding to the degenerate dressed-lower-state subspace. Here, we should mention that, when we treat the coupling submatrix $	\mathbf{C}_{[l]}$ as a matrix of dimension $N_{u} \times l$, then the dimensions of the dressed upper states $\{ \vert U_{j}\rangle\}$ and lower states $\{ \vert L_{k}\rangle\}$ are reduced to $N_{u} $ and $ l$, respectively. Based on the above analyses, we can obtain
\begin{eqnarray}
	\mathbf{C}_{[l]}\left\vert 	D\right\rangle &=&\sum_{j=1}^{N_{u}}\sum_{k=1}^{l}
	G_{ jk}\vert U_{j}\rangle \langle L_{k}\vert 
	\sum_{i=1}^{l}x_{i}\left\vert L_{i}\right\rangle \notag  \\
	&=&\sum_{j=1}^{N_{u}}\sum_{k=1}^{l}\sum_{i=1}^{l}
	G_{jk}x_{i}\vert U_{j}\rangle \langle L_{k} 
	\vert L_{i}\rangle \notag  \\
	&=&\sum_{j=1}^{N_{u}}\sum_{k=1}^{l}G_{jk}x_{k}\vert U_{j}\rangle.
\end{eqnarray}%
It can be found that only when $\sum_{k=1}^{l}G_{jk}x_{k}=0$, the state $\left\vert D\right\rangle$ is a dark state satisfying $\langle U_{j}\vert \mathbf{C}_{[l]}\left\vert 	D\right\rangle=0$ for $j=1,2,\ldots,N_{u}$. The parameter condition for the appearance of the dark state can be expressed as 
\begin{equation}
	\mathbf{C}_{[l]}\mathbf{x}=\mathbf{0}, 
\end{equation}%
for $\mathbf{x}=(x_{1},x_{2},\ldots,x_{l})^{T}$.
This implies that the dark state spans the null space of the coupling submatrix $\mathbf{C}_{[l]}$.

There exist many methods to obtain the null space of a matrix, for example, by directly solving the defining equation $\mathbf{C}_{[l]}\mathbf{x}=\mathbf{0}$, reducing the matrix to a row-echelon form and then using the linearly dependent relation, and applying the SVD.
Specially, if $\mathbf{C}_{[l]}$ is a square matrix, the existence of the dark states can be determined by checking whether its determinant equals zero or not. Note that many numerical softwares, such as Python, Mathematica, and MATLAB, provide built-in functions for directly
solving the null space of a matrix.

Below, we present the following assertions for determining the dark states and bright states by solving the null space.

(i) \textit{In a degenerate dressed-lower-state subspace, the number of the bright states is equal to the rank of the coupling submatrix associated with the degenerate dressed-lower-state subspace, and the number of the dark states is equal to the dimension of the degenerate dressed-lower-state subspace minus the number of the bright states.} This relation can be obtained according to the Rank-nullity theorem~\cite{Leon2014Applications}.
	
In addition, the form of the bright and dark states can be obtained in terms of the SVD of the coupling submatrix $\mathbf{C}_{[l]}$~\cite{SVDiCQED2013,Leon2014Applications}. 
For the $l$-dimensional degenerate dressed-lower-state subspace with rank $r$ ($r\leq \text{min}\{N_{u},l\}$), the SVD of the corresponding coupling submatrix $\mathbf{C}_{[l]}$ can be expressed as~\cite{SVDiCQED2013} 
\begin{equation}
	\mathbf{C}_{[l]}=	\mathbf{W}	\mathbf{\Sigma }	\mathbf{V}^{\dag},
\end{equation}
where $\mathbf{W}$ is an $N_{u}\times N_{u}$ orthogonal matrix (left singular vectors), $\mathbf{V }$ is an $l\times l$ orthogonal matrix (right singular vectors), and $\mathbf{\Sigma }=\left(\begin{array}{cc}
	\mathbf{\Sigma }_{r}&  \mathbf{0} \\ 
	\mathbf{0} & \mathbf{0} 
\end{array}\right)$ is a rectangular diagonal matrix of dimension $N_{u}\times l$ with $\mathbf{\Sigma }_{r}=\text{diag(}\sigma_{1},\sigma_{2},\ldots,\sigma_{r}\text{)}$ containing the nonzero singular values sorted in descending order.
The coupling submatrix $\mathbf{C}_{[l]}$ can be further decomposed into
\begin{eqnarray}
	\mathbf{C}_{[l]} &=&\left( \sum_{j=1}^{N_{u}}|U_{j}\rangle \langle
	U_{j}|\right) \mathbf{W}\mathbf{\Sigma} \mathbf{V}^{\dag }\left( \sum_{i=1}^{l}\left\vert
	L_{i}\right\rangle \left\langle L_{i}\right\vert \right)   \notag \\
	&=&\sum_{j=1}^{N_{u}}\sum_{i=1}^{l}\langle U_{j}|\mathbf{W}\mathbf{\Sigma} \mathbf{V}^{\dag }\left\vert
	L_{i}\right\rangle |U_{j}\rangle \left\langle L_{i}\right\vert   \notag \\
	&=&\sum_{j,j^{\prime }=1}^{N_{u}}\sum_{i,i^{\prime }=1}^{l}{W}_{jj^{\prime
	}}{\Sigma} _{j^{\prime }i^{\prime }}{V}_{i^{\prime }i}^{\dag }|U_{j}\rangle
	\left\langle L_{i}\right\vert   \notag \\
	&=&\sum_{j=1}^{N_{u}}\sum_{i=1}^{l}\sum_{k=1}^{r}W_{jk}\sigma
	_{k}V_{ki}^{\dag }|U_{j}\rangle \left\langle L_{i}\right\vert   \notag \\
	&=&\sum_{k=1}^{r}\sigma _{k}\left( \sum_{j=1}^{N_{u}}W_{jk}\vert
	U_{j}\rangle \right) \left( \sum_{i=1}^{l}V_{ki}^{\dag }\left\langle
	L_{i}\right\vert \right)   \notag \\
	&=&\sum_{k=1}^{r}\sigma _{k}\vert \tilde{U}_{k}\rangle
	\langle \tilde{L}_{k}\vert ,
\end{eqnarray}%
where $\vert \tilde{U}_{k}\rangle
=\sum_{j=1}^{N_{u}}W_{jk}\vert U_{j}\rangle $ and $\langle 
\tilde{L}_{k}\vert =\sum_{i=1}^{l}V_{ki}^{\dag }\langle
L_{i}\vert$ for $k=1,2,\ldots,r$.
It can be found that only $r$ dressed lower states $\vert \tilde{L}_{k=1-r}\rangle$ corresponding to the nonzero singular values are coupled with the upper states, while other $l-r$ dressed lower states $\vert
\tilde{L}_{k^{\prime}} \rangle =\sum_{i=1}^{l}V_{ik^{\prime}}\vert L_{i} \rangle$ (for $k^{\prime}=r+1,r+2,\ldots,l$) corresponding to the zero singular values are decoupled from the upper states and become dark states.
Since the rank of the coupling submatrix $\mathbf{C}_{[l]}$ is $r$, the number of the bright states is equal to the rank of $\mathbf{C}_{[l]}$, and the number of the dark states is equal to $l-r$.

(ii) In a degenerate dressed-lower-state subspace with dimension $l$, if all the column vectors in $\mathbf{C}_{[l]}$ are linearly dependent (i.e., $\mathbf{C}_{j=2-l}=\lambda _{j}\mathbf{C}_{1}$), then in this degenerate dressed-lower-state subspace, there exists one bright state $\left\vert B_{l-1}\right\rangle$ satisfying 
\begin{eqnarray}\label{Bj}
	\vert B_{j^{\prime}}\rangle =\frac{1}{\mathcal{N}_{j^{\prime}} }( \mathcal{N}_{j^{\prime}-1} \vert
	B_{j^{\prime}-1}\rangle +\lambda _{j^{\prime}+1}^{\ast }\vert L_{j^{\prime}+1}\rangle ), 
\end{eqnarray}
and $l-1$ orthogonal dark states $\left\vert D_{1}\right\rangle $, $\left\vert D_{2}\right\rangle $, $\ldots$, and $\left\vert D_{l-1}\right\rangle $, which can be expressed as
\begin{eqnarray}\label{Dj}
	\vert D_{j^{\prime}}\rangle &=&\frac{1}{\mathcal{N}_{j^{\prime}} }( \lambda _{j^{\prime}+1}\vert
	B_{j^{\prime}-1}\rangle -\mathcal{N}_{j^{\prime}-1} \vert L_{j^{\prime}+1}\rangle ), 
\end{eqnarray}
with the coefficient $\mathcal{N}_{j^{\prime}} =\sqrt{1+\sum_{i^{\prime}=2}^{j^{\prime}+1}\vert \lambda_{i^{\prime}}\vert^{2}}$ for $j^{\prime}=1,2,\ldots,l-1$ and $\vert B_{0}\rangle=\vert L_{1}\rangle$.

(3) If there exist multiple degenerate dressed-lower-state subspaces for all these $N_{l}$ dressed lower states, the Hamiltonian can be expressed as 
\begin{eqnarray}\label{HCLS}
	\tilde{H}^{[N]}_{D}&=&
	\left(\begin{array}{c|c|c|c|c}
		\mathbf{H}_{U} & \mathbf{C}_{[l_{1}]}  & \mathbf{C}_{[l_{2}]}  & \cdots & \mathbf{C}_{[l_{s}]} \\ \hline
		\mathbf{C}_{[l_{1}]}^{\dagger} & \mathbf{H}_{ L}^{ [l_{1}]} & \mathbf{0} & \cdots & \mathbf{0} \\\hline
		\mathbf{C}_{[l_{2}]} ^{\dagger} & \mathbf{0} & \mathbf{H}_{ L}^{[l_{2}]} & \cdots & \mathbf{0} \\\hline
		\cdots & \cdots & \cdots &  \ddots & \cdots \\\hline
		\mathbf{C}_{[l_{s}]} ^{\dagger} & \mathbf{0} & \mathbf{0} & \cdots & \mathbf{H}_{ L}^{[l_{s}]} 
	\end{array}\right),
\end{eqnarray}%
where $\mathbf{0}$ denote the zero matrices, and $l_{k=1-s}$ are the dimensions of the degenerate dressed-lower-state subspaces and satisfy the relation $\sum_{k=1}^{s}l_{k}=N_{l}$ (the dimension could also be $1$, namely, the corresponding dressed lower state is nondegenerate). The submatrix 
 $\mathbf{H}_{ L}^{ [l_{k}]}$ is a diagonal matrix $\mathbf{H}_{ L}^{ [l_{k}]}=\Omega_{k}\mathds{1}_{l_{k}}$, where $\mathds{1}_{k}$ denotes the $k\times k$ identity matrix.
Then the number of the dark states is $N_{l}-R$, where $R=\sum_{k=1}^{s}R_{k}$ is the sum of the ranks of the coupling submatrices $\mathbf{C}_{[l_{k}]}$ corresponding to each degenerate subspace. 
The froms of the bright and dark states in each degenerate dressed-lower-state subspace can be obtained according to the items 2(i) and 2(ii). Note that the dark state only exists within the same degenerate dressed-lower-state subspace, and it will not across different degenerate dressed-lower-state subspaces.

(4) If all the column vectors in the coupling matrix $\mathbf{C}$ satisfy $\mathbf{C}_{k=1-N_{l}}\neq \mathbf{0}$ and there is no degeneracy in these dressed lower states ($\Omega_{k} \neq \Omega_{k^{\prime }}$ for all $k\neq k^{\prime }$), then there is no dark state in the system.

\section{Dark states in the three-level quantum systems} \label{sec3}

\begin{figure}[t]
	\centering\includegraphics[width=0.45\textwidth]{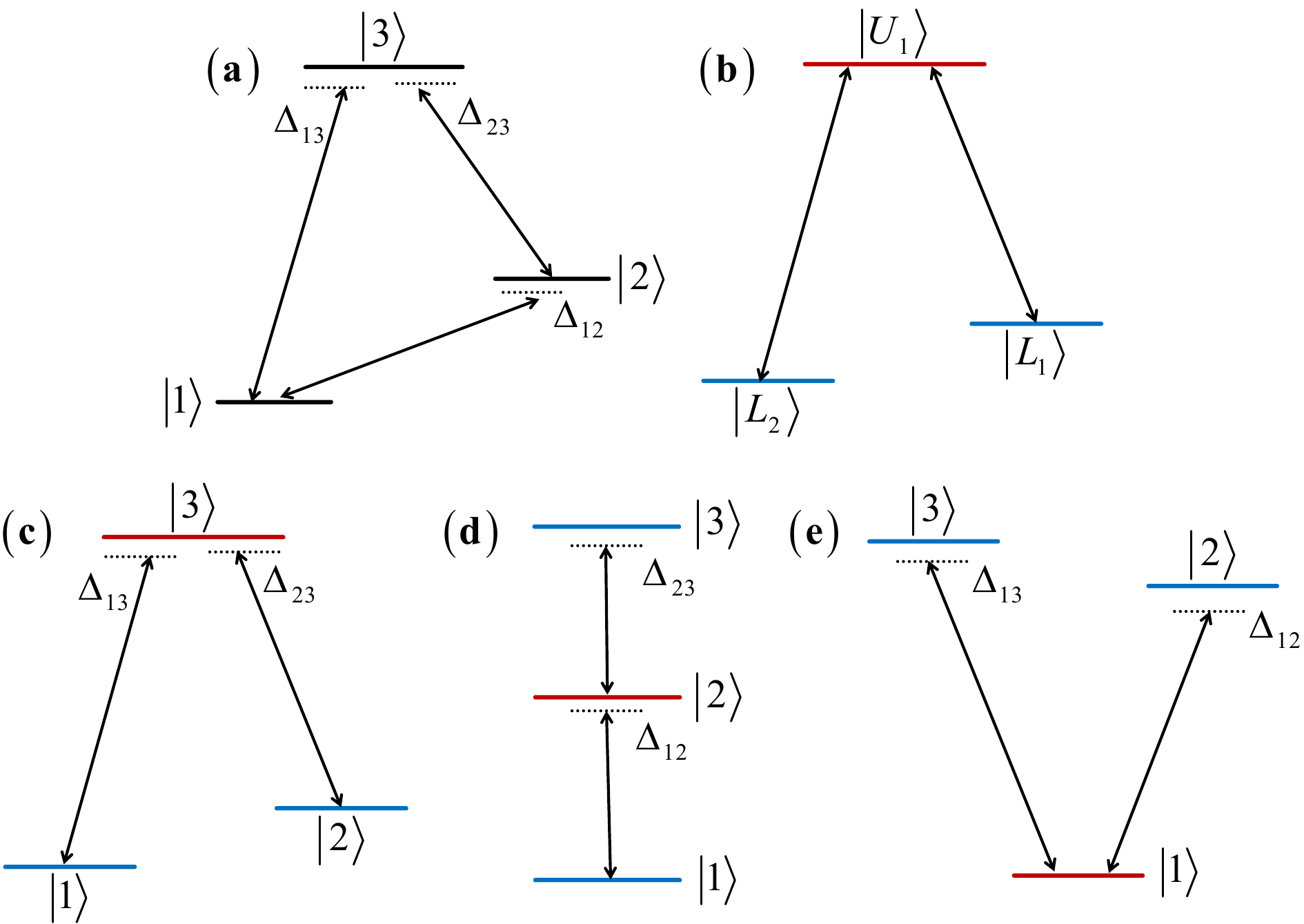}
	\caption{(a) A general $\Delta$-type three-level system with all transitions among three energy levels expressed in the bare-state representation. (b)  The single configuration of the three-level system, according to the numbers of the upper and lower states, expressed in the dressed upper- and lower-state representation.
		Based on the $\Delta$-type three-level system, we can further obtain three specific configurations by cutting one coupling channel (here we only cut one coupling such that these three levels are still connected): (c) $\Lambda$-type three-level system, (d) $\Xi $-type three-level system, and (e) $\mathrm{V}$-type three-level system. The red (blue) levels denote the dressed upper (lower) states of the system. We point out that the selection of the upper and lower states does not depend on the specific high- and low-energy levels, but depends on the specific research topic.}
	\label{tls}
\end{figure}

In this section, we study the dark states in the three-level systems. Without loss of generality, we consider a general $\Delta $-type three-level system with all possible transitions~[see Fig.~\ref{tls}(a)] and analyze the dark states with the arrowhead-matrix method. For the three-level systems, there is only one configuration according to the numbers of the upper and lower states: $N_{u}=1$ and $ N_{l}=2$. Here the upper and lower states can be defined on demand.

The Hamiltonian of the $\Delta $-type three-level system can be described by Eq.~(\ref{HN}) for $N=3$,
\begin{eqnarray}
H^{[3]} &=&E_{1}\left\vert 1\right\rangle \left\langle 1\right\vert
+E_{2}\left\vert 2\right\rangle \left\langle 2\right\vert +E_{3}\left\vert
3\right\rangle \left\langle 3\right\vert+(\Omega _{12} e^{-i\omega _{12}t}\left\vert
2\right\rangle \left\langle 1\right\vert  \notag \\
&&+\Omega _{13}e^{-i\omega _{13}t}\left\vert 3\right\rangle \left\langle 1\right\vert
+\Omega _{23}e^{-i\omega _{23}t}\left\vert 3\right\rangle \left\langle
2\right\vert +\text{H.c.}),
\end{eqnarray}%
and the corresponding time-independent Hamiltonian [Eq.~(\ref{HN_tilde}) for $N=3$] reads
\begin{eqnarray}
\tilde{H}^{[3]} &=&-\Delta _{13}\left\vert 1\right\rangle \left\langle
1\right\vert -\Delta _{23}\left\vert 2\right\rangle \left\langle
2\right\vert +(\Omega _{12}\left\vert 2\right\rangle \left\langle 1\right\vert
\notag \\
&&+\Omega _{23}\left\vert 3\right\rangle \left\langle 2\right\vert +\Omega
_{13}\left\vert 3\right\rangle \left\langle 1\right\vert +\text{H.c.}),
\end{eqnarray}%
where the detunings are introduced by $\Delta_{13}=E_{3}-E_{1}-\omega _{13}$, $\Delta_{23}=E_{3}-E_{2}-\omega _{23}$, and they should satisfy the relation $\Delta_{13}-\Delta _{23}=\Delta _{12}$ such that the Hamiltonian is time-independent.

Here we define the state $\left\vert 3\right\rangle $ as the upper state $\left\vert
u_{1}\right\rangle ,$ and the remaining states $\left\vert 2\right\rangle $ and  $\left\vert 1\right\rangle $ as the lower states $\left\vert l_{1}\right\rangle $ and $%
\left\vert l_{2}\right\rangle $. Therefore the basis vectors can be defined as 
$\left\vert u_{1}\right\rangle =\left\vert 3\right\rangle =(1,0,0)^{T},$ $%
\left\vert l_{1}\right\rangle =\left\vert 2\right\rangle =(0,1,0)^{T},$ and $%
\left\vert l_{2}\right\rangle =\left\vert 1\right\rangle =(0,0,1)^{T}$. Then the Hamiltonian $\tilde{H}^{[3]}$ can be expressed as 
\begin{equation}\label{H3}
\tilde{{H}}^{[3]}=\left( 
\begin{array}{c|c}
\mathbf{H}_{u} & \mathbf{c} \\ \hline
\mathbf{c}^{\dag } & \mathbf{H}_{l}%
\end{array}%
\right) =\left( 
\begin{array}{c|cc}
0 & \Omega _{23} & \Omega _{13} \\ \hline
\Omega _{23}^{\ast } & -\Delta _{23} & \Omega _{12} \\ 
\Omega _{13}^{\ast } & \Omega _{12}^{\ast } & -\Delta _{13}%
\end{array}%
\right) . 
\end{equation}

To analyze the dark states, we transform the Hamiltonian matrix in Eq.~(\ref{H3}) into an arrowhead matrix. To this end, we diagonalize the lower-state submatrix $\mathbf{H}_{l}$ with the unitary matrix 
\begin{equation}\label{Sl}
\mathbf{S}_{l}=\left( 
\begin{array}{cc}
	-1/\sqrt{2} &  e^{i\theta }/\sqrt{2} \\
	 1/\sqrt{2} & e^{i\theta }/\sqrt{2} 
\end{array}%
\right),
\end{equation}
where we consider the case of $\Omega
_{12}=\vert \Omega _{12}\vert e^{i\theta }$ and $\Delta _{13}=\Delta _{23}=\Delta $. 
Based on the eigensystem of $\mathbf{H}_{l}$, here the resonance condition is chosen for satisfying the degenerate condition of the lower states and for simplifying the  calculations.
Then the Hamiltonian becomes%
\begin{eqnarray}\label{H3D}
\tilde{{H}}^{[3]}_{D}&=& \left( 
\begin{array}{c|c}
\mathbf{H}_{U} & \mathbf{C} \\ \hline
\mathbf{C}^{\dag } & \mathbf{H}_{L}%
\end{array}%
\right)  \notag \\
&=&\left( 
\begin{array}{c|cc}
0 & \frac{-\Omega _{23}+e^{-i\theta }\Omega _{13}}{\sqrt{2}} & \frac{\Omega
_{23}+e^{-i\theta }\Omega _{13}}{\sqrt{2}} \\ \hline
\frac{-\Omega _{23}^{\ast }+e^{i\theta }\Omega _{13}^{\ast }}{\sqrt{2}} & 
-\Delta -\left\vert \Omega _{12}\right\vert & 0 \\ 
\frac{\Omega _{23}^{\ast }+e^{i\theta }\Omega _{13}^{\ast }}{\sqrt{2}} & 0 & 
-\Delta +\left\vert \Omega _{12}\right\vert%
\end{array}%
\right) ,
\end{eqnarray}%
where the new basis vectors are given by $\left\vert U_{1}\right\rangle =\left\vert u_{1}\right\rangle =\left\vert
3\right\rangle ,$ $\left\vert L_{1}\right\rangle =(e^{-i\theta }\left\vert
l_{2}\right\rangle -\left\vert l_{1}\right\rangle )/\sqrt{2}, $ and $%
\left\vert L_{2}\right\rangle =(e^{-i\theta }\left\vert l_{2}\right\rangle
+\left\vert l_{1}\right\rangle )/\sqrt{2}$. Based on the arrowhead matrix in Eq.~(\ref{H3D}), we can analyze the dark states in the $\Delta $-type three-level system~[Fig.~\ref{tls}(b)].

(1) Firstly, we focus on the case of zero column vector in the coupling matrix $\mathbf{C}$, where the corresponding dressed lower state is a dark state. In Eq.~(\ref{H3D}), there are two coupling column vectors: $\mathbf{C}_{1}=({-\Omega _{23}+e^{-i\theta }\Omega _{13}})/{\sqrt{2}} $ and $\mathbf{C}_{2}=({\Omega
	_{23}+e^{-i\theta }\Omega _{13}})/{\sqrt{2}}$.
By considering the coupling column vector $\mathbf{C}_{1}=\mathbf{0}$ or $\mathbf{C}_{2}=\mathbf{0}$, we can analyze the dark states in the system as follows.

(i) The case of $\mathbf{C} _{1}=\mathbf{0}$: (a) When $\Omega  _{23}=\Omega  _{13}$ and $\theta =2n\pi $ with $n\in \mathbb{Z}  $, the state $\left\vert L_{1}\right\rangle =(\left\vert
l_{2}\right\rangle -\left\vert l_{1}\right\rangle )/\sqrt{2}$  is decoupled
from the dressed upper state, and it is a dark state. (b) When $\Omega  _{23}=-\Omega  _{13}$ [for example, there is a phase $\phi=(2n+1)\pi$ between $\Omega  _{23}$ and $\Omega  _{13}$] and $\theta =(2n+1)\pi $, the state $\left\vert L_{1}\right\rangle =(\left\vert
l_{2}\right\rangle+\left\vert l_{1}\right\rangle )/\sqrt{2}$ is a dark state.

(ii) The case of $\mathbf{C} _{2}=\mathbf{0}$: (a) When $\Omega  _{23}=\Omega  _{13}$ and $\theta =(2n+1)\pi $, the state $\left\vert L_{2}\right\rangle =(-\left\vert
l_{2}\right\rangle +\left\vert l_{1}\right\rangle )/\sqrt{2}$ is decoupled
from the dressed upper state, and it is a dark state. (b) When $\Omega  _{23}=-\Omega  _{13}$ and $\theta =2n\pi $, the state $\left\vert L_{2}\right\rangle =(\left\vert
l_{2}\right\rangle +\left\vert l_{1}\right\rangle )/\sqrt{2}$  is a dark state.

Therefore, we can see that when $\Omega _{23}=\Omega _{13}$, there is always a dark state 
$(\left\vert l_{2}\right\rangle -\left\vert l_{1}\right\rangle )/\sqrt{2}$ for $\theta =n\pi $; and when $\Omega _{23}=-\Omega _{13}$, there is always a dark state $(\left\vert l_{2}\right\rangle +\left\vert l_{1}\right\rangle )/\sqrt{2}$ for $\theta =n\pi $.
 
(2) Next, we consider the degenerate condition of the dressed lower states and analyze the dark states in the degenerate dressed-lower-state subspace.

(i) In the case of $\vert \Omega _{12}\vert=0$ (the corresponding phase is also taken to be zero, i.e., $\theta=0$), these two dressed lower states are degenerate, and the Hamiltonian can be rewritten as 
\begin{equation}
	\tilde{H}_{D}^{[3]}=-\Delta 
	 (\vert L_{1}\rangle \langle L_{1}\vert
	+\vert L_{2}\rangle \langle L_{2}\vert ) +[\vert
	U_{1}\rangle ( \mathbf{C}_{1}\langle L_{1}\vert+\mathbf{C}_{2}\langle L_{2}\vert )+\text{H.c.}] .
\end{equation}
We can define two orthogonal states
\begin{subequations}\label{BD}
	\begin{align} 
	\left\vert B_1\right\rangle  = &\frac{1}{\mathcal{N}_{1}}( \mathbf{C}_{1}^{\ast }\left\vert
	L_{1}\right\rangle +\mathbf{C}_{2}^{\ast }\left\vert L_{2}\right\rangle )
	,  \\
\left\vert D_1\right\rangle = &\frac{1}{\mathcal{N}_{1}}(\mathbf{C}_{2}\left\vert
	L_{1}\right\rangle -\mathbf{C}_{1}\left\vert L_{2}\right\rangle ) , 
	\end{align}
	\end{subequations}
where the normalization constant is introduced by $\mathcal{N}_{1}=\sqrt{\vert \mathbf{C}_{1}\vert ^{2}+\vert \mathbf{C}_{2}\vert 	^{2}}$, and the states satisfy the relation $\left\vert B_1\right\rangle \left\langle B_1\right\vert
+\left\vert D_1\right\rangle \left\langle D_1\right\vert =\left\vert
L_{1}\right\rangle \left\langle L_{1}\right\vert +\left\vert
L_{2}\right\rangle \left\langle L_{2}\right\vert $.
Hence, the Hamiltonian $\tilde{H}_{D}^{[3]}$ becomes 
	\begin{equation}
		\tilde{H}_{D}^{[3]}=-\Delta  (\vert B_1\rangle \langle B_1\vert +\vert
		D_1\rangle \langle D_1\vert )+(\mathcal{N}_{1}\vert U_{1}\rangle
		\langle B_1\vert +\text{H.c.}),
	\end{equation}%
and it can be found that the state $\vert B_1\rangle$ is a bright state coupled with the dressed upper state $\vert U_{1}\rangle$, and the state $\vert D_1\rangle$ is a dark state decoupled from the dressed upper state. Then, the dark state in this case can be expressed as 
	\begin{eqnarray}\label{D3}
		\left\vert D_{1}^{[3]}\right\rangle  &=&\frac{1}{\mathcal{N}_{1}^{[3]}}\left( \frac{%
			\Omega _{13}+\Omega _{23}}{\sqrt{2}}\left\vert L_{1}\right\rangle -\frac{%
			\Omega _{13}-\Omega _{23}}{\sqrt{2}}\left\vert L_{2}\right\rangle \right)
		\notag \\
		&=&\frac{\Omega _{23}\left\vert l_{2}\right\rangle -\Omega _{13}\left\vert
			l_{1}\right\rangle }{\mathcal{N}_{1}^{[3]}},
	\end{eqnarray}
	with $\mathcal{N}_{1}^{[3]}=\sqrt{\vert \Omega
		_{13}\vert ^{2}+\vert \Omega _{23}\vert ^{2}}$.
We point out that the present case is right the $\Lambda $-type three-level system under the two-photon resonance~[Fig.~\ref{tls}(c)]. 

(ii) In the case of $\vert \Omega _{12}\vert\neq 0$, there is no degeneracy for the dressed lower states, and when all the coupling column vectors $\mathbf{C}_{1}$ and $\mathbf{C}_{2}$ are nonzero, there is no dark state.

Here we emphasize again that the upper and lower states are
not the high- and low-energy levels in a realistic system. As the example in this section, we define the state $\vert 3\rangle $ as the upper state and other two states as the lower states. Similarly, we can also define the state $\vert 2\rangle $ or $\vert 1\rangle $ as the upper state with the remaining states as the lower states, and the dark state can also be analyzed with the same method.
We find that when the state $\vert 2\rangle $ is defined as the upper state, the conditions for the existence of the dark state are $\Delta _{12}=\Delta _{23}=\Delta $ and $\Omega _{13}=0$, which are the same as the $\Xi $-type three-level system~[Fig.~\ref{tls}(d)]. When the state $\vert 1\rangle $ is defined as the upper state, the conditions for the existence of the dark state are $\Delta _{12}=\Delta _{13}=\Delta $ and $\Omega _{23}=0$, which are the same as the $\mathrm{V}$-type three-level system~[Fig.~\ref{tls}(e)]. In this paper, we define these three cases as the same configuration, because they all consist of one dressed upper state and two dressed lower states, and there exists one dark state composed of the two dressed lower states when they are degenerate.

\section{Dark states in the four-level quantum systems}\label{sec4} 
 
\begin{figure}[t]
	\centering\includegraphics[width=0.45\textwidth]{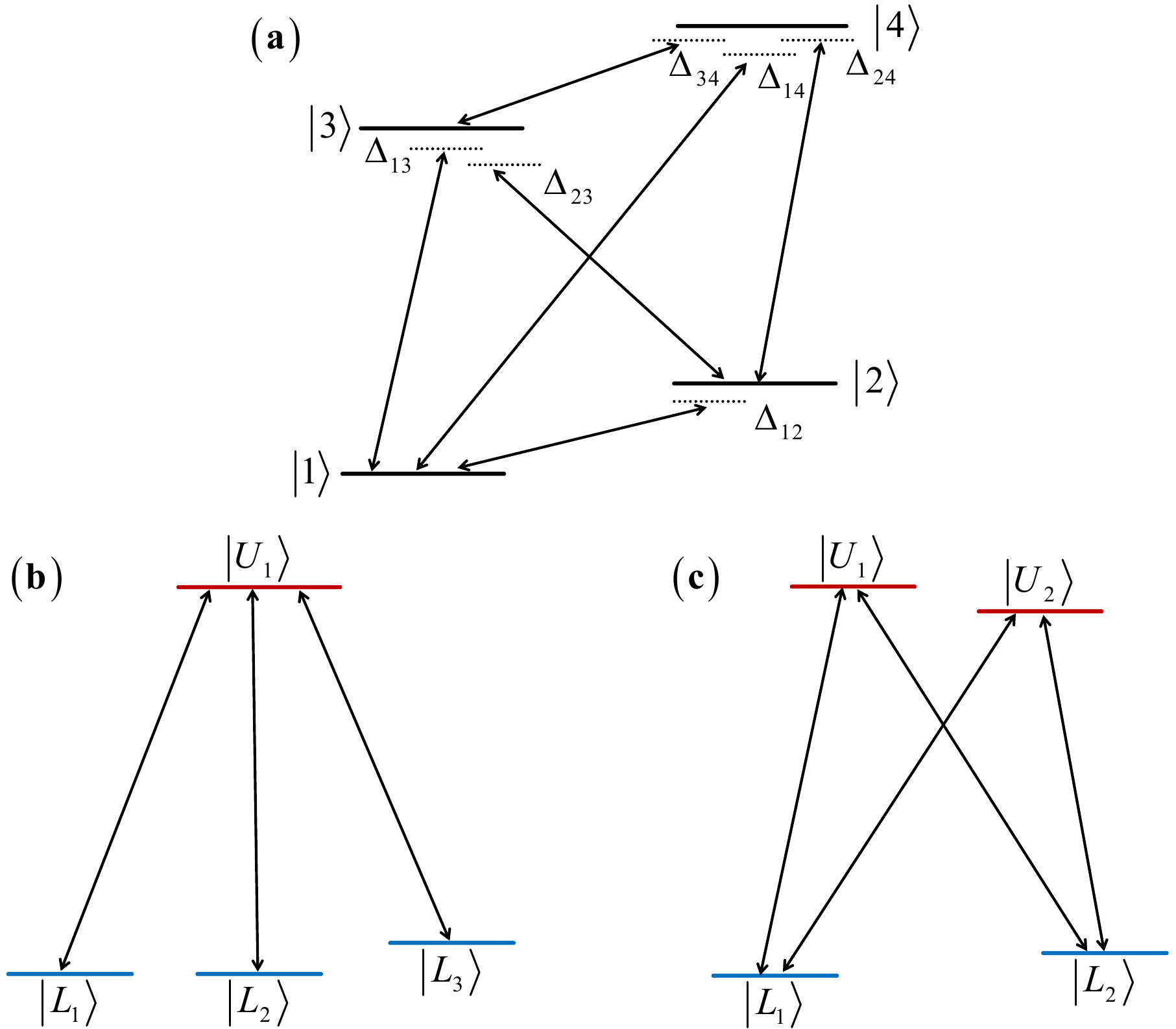}
	\caption{(a) Schematic of a general four-level quantum system with all possible transitions among four energy levels expressed in the bare-state representation. According to the numbers of the upper and lower states, it can be divided into two configurations expressed in the dressed upper- and lower-state representation. (b) Configuration 1: one upper state and three lower states with real symmetric couplings $\Omega _{12}=\Omega
		_{13}=\Omega _{23}=\Omega $ and under the resonance condition $\Delta _{14}=\Delta _{24}=\Delta _{34}=\Delta $. (c) Configuration 2: two upper states and two lower states with $\Omega _{34}=0$ and under the resonance condition $\Delta _{14}=\Delta _{24}=\Delta $.
		The red (blue) levels denote the dressed upper (lower) states of the system. Note that the upper and lower states can be chosen on demand in different configurations, and we only present one representative case as an example.}
	\label{4ls}
\end{figure}

In this section, we analyze the dark states in the four-level quantum systems using the arrowhead-matrix method. According to the numbers of the upper and
lower states, there are two basic configurations of the four-level systems: (a) $N_{u}=1$ and $ N_{l}=3$ and (b) $N_{u}=2$ and $ N_{l}=2$. We first present a general four-level system  with all possible transitions~[see Fig.~\ref{4ls}(a)], and then discuss the two configurations in detail.
Here we consider that the dimension of the lower-state subspace is greater than one for keeping the necessary quantum interference channels.

The Hamiltonian of the general four-level quantum system can be described by Eq.~(\ref{HN}) for $N=4$,
\begin{eqnarray}\label{H4}
H^{[4]} =\sum_{{j}=1}^{4}E_{j}\left\vert
j\right\rangle \left\langle j\right\vert +\sum_{j,j^{\prime }=1,j<j^{\prime }}^{4} (\Omega _{jj^{\prime }}e^{-i\omega _{jj^{\prime
	}}t}\vert
j^{\prime }\rangle \left\langle j\right\vert +\text{H.c.}),
\end{eqnarray}%
and the corresponding time-independent Hamiltonian [Eq.~(\ref{HN_tilde}) for $N=4$] reads
\begin{eqnarray}
\tilde{H}^{[4]}=\sum_{r=1}^{3}-\Delta _{r4}\left\vert r\right\rangle
\left\langle r\right\vert +\sum_{j,j^{\prime }=1,j<j^{\prime
}}^{4}( \Omega _{jj^{\prime }}\vert j^{\prime }\rangle \left\langle
j\right\vert +\text{H.c.}),
\end{eqnarray}
where the detunings are given by $\Delta_{r4}=E_{4}-E_{r}-\omega _{r4}$, and for obtaining the time-independent Hamiltonian, they should satisfy the conditions $\Delta _{r4}-\Delta _{r^{\prime }4} =\Delta _{rr^{\prime }}$ for $r,r^{\prime}=1,2,3,$ and $r<r^{\prime}$.
Below, we analyze the dark and bright states in these two configurations.

\subsection{Configuration 1: $\boldsymbol{N_{u}=1$ and $ N_{l}=3}$}\label{Conf[4,1]}

For the configuration with one upper state (i.e., $\left\vert u_{1}\right\rangle
=\left\vert 4\right\rangle$) and three lower states (i.e., $\left\vert l_{1}\right\rangle =\left\vert 3\right\rangle$, $\left\vert l_{2}\right\rangle =\left\vert 2\right\rangle $, and $\left\vert l_{3}\right\rangle =\left\vert
1\right\rangle$), we define the basis states and vectors as follows: $\left\vert u_{1}\right\rangle
=\left\vert 4\right\rangle =( 1,0,0,0) ^{T},$ $\left\vert
l_{1}\right\rangle =\left\vert 3\right\rangle =( 0,1,0,0) ^{T},$ $%
\left\vert l_{2}\right\rangle =\left\vert 2\right\rangle =(
0,0,1,0) ^{T},$ and $\left\vert l_{3}\right\rangle =\left\vert
1\right\rangle =( 0,0,0,1) ^{T}$. Then the Hamiltonian can be
expressed as 
\begin{equation} \label{H41}
\tilde{{H}}^{[4,1]}=\left( 
\begin{array}{c|c}
\mathbf{H}_{u} & \mathbf{c} \\ \hline
\mathbf{c}^{\dag } & \mathbf{H}_{l}%
\end{array}%
\right) =\left( 
\begin{array}{c|ccc}
0 & \Omega _{34} & \Omega _{24} & \Omega _{14} \\ \hline
\Omega _{34}^{\ast } & -\Delta _{34} & \Omega _{23} & \Omega _{13} \\ 
\Omega _{24}^{\ast } & \Omega _{23}^{\ast } & -\Delta _{24} & \Omega _{12}
\\ 
\Omega _{14}^{\ast } & \Omega _{13}^{\ast } & \Omega _{12}^{\ast } & -\Delta
_{14}%
\end{array}%
\right) ,  
\end{equation}
where the superscript \textquotedblleft$[4,1]$\textquotedblright\ denotes the configuration 1 in the four-level systems.

Next we diagonalize the lower-state submatrix $\mathbf{H}_{l}$. 
For simplicity, we consider the case of real symmetric couplings $\Omega _{12}=\Omega
_{13}=\Omega _{23}=\Omega $ among these lower states. In addition, we consider the two-photon resonance ($\Delta _{14}=\Delta _{24}=\Delta _{34}=\Delta $) for these transitions between the upper state $\vert 4 \rangle$ and the lower states $\{\vert 1 \rangle,\vert 2 \rangle,\vert 3 \rangle\}$, then the single-photon resonance ($\Delta _{12}=\Delta _{13}=\Delta _{23}=0$) exists within the lower-state subspace. With the
unitary matrix 
\begin{equation}\label{Sl3}
	\mathbf{S}_{l}=\left( 
	\begin{array}{ccc}
		-{1}/{\sqrt{2}} & 0 & 1/\sqrt{2}\\
		-1/\sqrt{6} &  2/\sqrt{6}& -1/\sqrt{6} \\ 
	1/\sqrt{3} & 1/\sqrt{3} & 1/\sqrt{3}%
	\end{array}%
	\right) ,  
\end{equation}
the Hamiltonian can be transformed into an arrowhead matrix
\begin{eqnarray}\label{HD41}
\tilde{{H}}^{[4,1]}_{D}&=& \left( 
\begin{array}{c|c}
	\mathbf{H}_{U} & \mathbf{C} \\ \hline
	\mathbf{C}^{\dag } & \mathbf{H}_{L}%
\end{array}%
\right)  \notag \\
&=&{\left( 
\begin{array}{c|ccc}
0 & \frac{\Omega _{14}-\Omega _{34}}{\sqrt{2}} & \frac{2\Omega _{24}-
\Omega _{34}-\Omega _{14} }{\sqrt{6}} & \frac{\Omega _{34}+\Omega
_{14}+\Omega _{24}}{\sqrt{3}} \\ \hline
\frac{\Omega _{14}^{\ast }-\Omega _{34}^{\ast }}{\sqrt{2}} & -\Delta -\Omega
& 0 & 0 \\ 
\frac{2\Omega _{24}^{\ast }- \Omega _{34}^{\ast }-\Omega _{14}^{\ast
} }{\sqrt{6}} & 0 & -\Delta -\Omega & 0 \\ 
\frac{\Omega _{34}^{\ast }+\Omega _{14}^{\ast }+\Omega _{24}^{\ast }}{\sqrt{3%
}} & 0 & 0 & -\Delta +2\Omega%
\end{array}%
\right)}, \notag \\ 
\end{eqnarray}%
with these new basis vectors $\left\vert U_{1}\right\rangle =\left\vert u_{1}\right\rangle =\left\vert
4\right\rangle , $ $\left\vert L_{1}\right\rangle =(\left\vert
l_{3}\right\rangle -\left\vert l_{1}\right\rangle )/\sqrt{2},$ $\left\vert
L_{2}\right\rangle =(2\left\vert l_{2}\right\rangle -\left\vert
l_{1}\right\rangle -\left\vert l_{3}\right\rangle )/\sqrt{6}$, and $%
\left\vert L_{3}\right\rangle =(\left\vert l_{1}\right\rangle +\left\vert
l_{2}\right\rangle +\left\vert l_{3}\right\rangle )/\sqrt{3}.$ Based on the
arrowhead-matrix method, we can analyze the dark states for configuration 1~[see Fig.~\ref{4ls}(b)].

(1) We first consider the zero column vector in the coupling matrix $\mathbf{C}=(\mathbf{C}_{1},\mathbf{C}_{2},\mathbf{C}_{3})$ with $\mathbf{C}_{1}={(\Omega _{14}-\Omega _{34})}/{\sqrt{2}}$, $\mathbf{C}_{2}=  {(2\Omega _{24}-	\Omega _{34}-\Omega _{14}) }/{\sqrt{6}} $, and $\mathbf{C}_{3}= {(\Omega _{34}+\Omega 	_{14}+\Omega _{24})}/{\sqrt{3}}$. 

(i) The case of $\mathbf{C}_{1}=\mathbf{0}$: When $\Omega _{14}=\Omega _{34}$, the state $\left\vert
L_{1}\right\rangle =(\left\vert l_{3}\right\rangle -\left\vert
l_{1}\right\rangle )/\sqrt{2}$ is decoupled from the dressed upper state and it is a dark state.

(ii) The case of $\mathbf{C}_{2}=\mathbf{0}$: When $2\Omega _{24}=\Omega _{34}+\Omega _{14} ,$ the state $\left\vert L_{2}\right\rangle =( 2\left\vert
l_{2}\right\rangle -\left\vert l_{1}\right\rangle -\left\vert
l_{3}\right\rangle ) /\sqrt{6}$ becomes a dark state.

(iii) The case of $\mathbf{C}_{3}=\mathbf{0}$: 
When $\Omega _{34}+\Omega _{14}+\Omega _{24}=0,$ the state $%
\left\vert L_{3}\right\rangle =( \left\vert l_{1}\right\rangle
+\left\vert l_{2}\right\rangle +\left\vert l_{3}\right\rangle ) /\sqrt{%
3}$ becomes a dark state. 

(2) Next, we consider the case of a degenerate dressed-lower-state subspace.

(i) There is a two-dimensional degenerate dressed-lower-state subspace $\{\vert L_{1}\rangle, \vert L_{2}\rangle\}$ for any $\Omega \neq 0$. Based on Eqs.~(\ref{BD}), we can obtain the dark state in this degenerate-state subspace
\begin{eqnarray}\label{D1[41]}
	\left\vert D_{1}^{[4,1]}\right\rangle  &=&\frac{1}{\mathcal{N}_{1}^{[4,1]}}(%
	\mathbf{C}_{2}\left\vert L_{1}\right\rangle -\mathbf{C}_{1}\left\vert
	L_{2}\right\rangle )  \notag \\
	&=&\frac{1}{\sqrt{3}\mathcal{N}_{1}^{[4,1]}}[(\Omega _{14}-\Omega
	_{24})\left\vert l_{1}\right\rangle -(\Omega _{14}-\Omega _{34})\left\vert
	l_{2}\right\rangle   \notag \\
	&&+(\Omega _{24}-\Omega _{34})\left\vert l_{3}\right\rangle ],
\end{eqnarray}
where the coefficient is introduced by $\mathcal{N}_{1}^{[4,1]}=\sqrt{\vert \mathbf{C}_{1}\vert ^{2}+\vert \mathbf{C}_{2}\vert
	^{2}}$. 

(ii) Furthermore, when $\Omega =0$, these three dressed lower states $\left\vert
L_{1}\right\rangle$, $\left\vert L_{2}\right\rangle$, and $\left\vert
L_{3}\right\rangle$ are degenerate. According to the definitions in Eqs.~(\ref{BD}), we introduce
\begin{subequations}\label{B2D2}
	\begin{align} 
		\left\vert B_{2}\right\rangle =&\frac{1}{\mathcal{N}_{2} }( \mathcal{N}_{1} \left\vert
		B_1\right\rangle +\mathbf{C}_{3}^{\ast }\left\vert L_{3}\right\rangle )\notag \\
		=&\frac{1}{\mathcal{N}_{2} }( \mathbf{C}_{1}^{\ast }\left\vert
		L_{1}\right\rangle +\mathbf{C}_{2}^{\ast }\left\vert L_{2}\right\rangle +\mathbf{C}_{3}^{\ast }\left\vert L_{3}\right\rangle )
		,\\
		\left\vert D_{2}\right\rangle =&\frac{1}{\mathcal{N}_{2} }( \mathbf{C}_{3}\left\vert
		B_{1}\right\rangle -\mathcal{N}_{1} \left\vert L_{3}\right\rangle )\notag \\
		=&\frac{%
			1}{\mathcal{N}_{1} \mathcal{N}_{2} }[ \mathbf{C}_{3}(\mathbf{C}_{1}^{\ast }\left\vert L_{1}\right\rangle
		+\mathbf{C}_{2}^{\ast }\left\vert L_{2}\right\rangle ) -\mathcal{N}_{1} ^{2}\left\vert
		L_{3}\right\rangle] ,
		\end{align} 
\end{subequations}
where the coefficient is introduced by $\mathcal{N}_{2} =\sqrt{\mathcal{N}_{1} ^2+\vert \mathbf{C}_{3}\vert
	^{2}}=\sqrt{\vert \mathbf{C}_{1}\vert ^{2}+\vert \mathbf{C}_{2}\vert
	^{2}+\vert \mathbf{C}_{3}\vert
	^{2}}$, and these states satisfy the relation 
$\vert B_{2}\rangle \langle B_{2}\vert
+\vert
D_{1}\rangle \langle D_{1}\vert  +\vert 	D_{2}\rangle \langle D_{2}\vert 
	 =\vert 	L_{1}\rangle \langle L_{1}\vert +\vert 	L_{2}\rangle \langle L_{2}\vert +\vert 	L_{3}\rangle \langle L_{3}\vert$. 
 It can be found that only the state $\vert B_{2}\rangle$ is coupled with the dressed upper state and becomes a bright state, and the states $\vert D_{1}\rangle$ and $\left\vert D_{2}\right\rangle$ are decoupled from the dressed upper state and become the dark states. Therefore, there are two dark states in this case: the state $\left\vert D_{1}^{[4,1]}\right\rangle $ given in Eq.~(\ref{D1[41]}) and the state $\left\vert D_{2}^{[4,1]}\right\rangle $ can be expressed as 
\begin{equation}\label{D2[41]}
	\left\vert D_{2}^{[4,1]}\right\rangle =\frac{\mathbf{C}_{3}( \mathbf{C}_{1}^{\ast }\left\vert
		L_{1}\right\rangle +\mathbf{C}_{2}^{\ast }\left\vert L_{2}\right\rangle
		) -( \mathcal{N}_{1}^{[4,1]}) ^{2}\left\vert
		L_{3}\right\rangle}{\mathcal{N}_{1}^{[4,1]}\mathcal{N}_{2}^{[4,1]}},
\end{equation}%
with $\mathcal{N}_{2}^{[4,1]}=\sqrt{\vert \mathbf{C}_{1}\vert ^{2}+\vert \mathbf{C}_{2}\vert
	^{2}+\vert \mathbf{C}_{3}\vert	^{2}}$. This dark state can be further represented by the bare states based on the relations $\left\vert L_{1}\right\rangle =(\left\vert
	l_{3}\right\rangle -\left\vert l_{1}\right\rangle )/\sqrt{2},$ $\left\vert
	L_{2}\right\rangle =(2\left\vert l_{2}\right\rangle -\left\vert
	l_{1}\right\rangle -\left\vert l_{3}\right\rangle )/\sqrt{6}$, and $%
	\left\vert L_{3}\right\rangle =(\left\vert l_{1}\right\rangle +\left\vert
	l_{2}\right\rangle +\left\vert l_{3}\right\rangle )/\sqrt{3}.$

\subsection{Configuration 2: $\boldsymbol{N_{u}=2$ and $N_{l}=2}$}\label{Conf42}

For the configuration with two upper states (i.e., $\left\vert u_{1}\right\rangle
=\left\vert 4\right\rangle$ and $\left\vert u_{2}\right\rangle
=\left\vert 3\right\rangle$) and two lower states (i.e., $\left\vert l_{1}\right\rangle =\left\vert 2\right\rangle$ and $\left\vert l_{2}\right\rangle
=\left\vert 1\right\rangle$), we define the basis states and vectors as follows: $\left\vert u_{1}\right\rangle
=\left\vert 4\right\rangle =( 1,0,0,0)^{T},$ $\left\vert
u_{2}\right\rangle =\left\vert 3\right\rangle =( 0,1,0,0) ^{T},$ $%
\left\vert l_{1}\right\rangle =\left\vert 2\right\rangle =(
0,0,1,0) ^{T},$ and $\left\vert l_{2}\right\rangle =\left\vert
1\right\rangle =( 0,0,0,1) ^{T}$. Then the Hamiltonian can be
expressed as
\begin{equation}
\tilde{{H}}^{[4,2]}=\left( 
\begin{array}{c|c}
\mathbf{H}_{u} & \mathbf{c} \\ \hline
\mathbf{c}^{\dag } & \mathbf{H}_{l}%
\end{array}%
\right) =\left( 
\begin{array}{cc|cc}
0 & \Omega _{34} & \Omega _{24} & \Omega _{14} \\ 
\Omega _{34}^{\ast } & -\Delta _{34} & \Omega _{23} & \Omega _{13} \\ \hline
\Omega _{24}^{\ast } & \Omega _{23}^{\ast } & -\Delta _{24} & \Omega _{12}
\\ 
\Omega _{14}^{\ast } & \Omega _{13}^{\ast } & \Omega _{12}^{\ast } & -\Delta
_{14}%
\end{array}%
\right) .  \label{H42}
\end{equation}%
The form of the lower-state submatrix $\mathbf{H}_{l}$ is similar to Eq.~(\ref{H3}). 
Similarly, we consider that $\Delta _{14}=\Delta _{24}=\Delta $ and $\Omega
_{12}=\left\vert \Omega _{12}\right\vert e^{i\theta }$. With the
unitary matrix in Eq.~{(\ref{Sl})} and the assumption $\Omega _{34}=0$, the Hamiltonian with dressed upper and lower states becomes%
\begin{eqnarray}
\tilde{{H}}^{[4,2]}_{D}&=& \left( 
\begin{array}{c|c}
	\mathbf{H}_{U} & \mathbf{C} \\ \hline
	\mathbf{C}^{\dag } & \mathbf{H}_{L}%
\end{array}%
\right)  \notag \\
&=&\left( 
\begin{array}{cc|cc}
0 & 0 & \frac{-\Omega _{24}+e^{-i\theta }\Omega _{14}}{\sqrt{2}}
& \frac{\Omega _{24}+e^{-i\theta }\Omega _{14}}{\sqrt{2}} \\ 
0 & -\Delta _{34} & \frac{-\Omega _{23}+e^{-i\theta
}\Omega _{13}}{\sqrt{2}} & \frac{\Omega _{23}+e^{-i\theta }\Omega _{13}}{ 
\sqrt{2}} \\ \hline
\frac{-\Omega _{24}^{\ast }+e^{i\theta }\Omega _{14}^{\ast }}{\sqrt{2}} & 
\frac{-\Omega _{23}^{\ast }+e^{i\theta }\Omega _{13}^{\ast }}{\sqrt{2}} & 
-\Delta -\left\vert \Omega _{12}\right\vert & 0 \\ 
\frac{\Omega _{24}^{\ast }+e^{i\theta }\Omega _{14}^{\ast }}{\sqrt{2}} & 
\frac{\Omega _{23}^{\ast }+e^{i\theta }\Omega _{13}^{\ast }}{\sqrt{2}} & 0 & 
-\Delta +\left\vert \Omega _{12}\right\vert%
\end{array}
\right) ,\notag \\ 
\end{eqnarray}
with these new basis vectors $\left\vert U_{1}\right\rangle =\left\vert u_{1}\right\rangle =\left\vert
4\right\rangle $, $\left\vert U_{2}\right\rangle =\left\vert u_{2}\right\rangle =\left\vert 3\right\rangle,$
$\left\vert L_{1}\right\rangle =(e^{-i\theta }\left\vert l_{2}\right\rangle
-\left\vert l_{1}\right\rangle )/\sqrt{2}, $ and $\left\vert
L_{2}\right\rangle =(e^{-i\theta }\left\vert l_{2}\right\rangle +\left\vert
l_{1}\right\rangle )/\sqrt{2}$. Using the thick arrowhead matrix, we can analyze the dark states for configuration 2~[see Fig.~\ref{4ls}(c)].

(1) We determine the zero column vector in the coupling matrix $\mathbf{C}$.

(i) The case of $\mathbf{C}_{1}=\mathbf{0}$: (a) When $\Omega _{23}=\Omega _{13}$, $\Omega _{24}=\Omega _{14}$, and $\theta =2n\pi $ for $n\in \mathbb{Z} $, the state $\vert L_{1}\rangle
=(\vert l_{2}\rangle -\vert l_{1}\rangle )/\sqrt{2}$
is decoupled from all the dressed upper states and becomes a dark state. (b) When $\Omega _{23}=-\Omega _{13}$, $\Omega _{24}=-\Omega _{14}$, and $%
\theta =(2n+1)\pi $, the state $\vert L_{1}\rangle =(\vert
l_{2}\rangle+\vert l_{1}\rangle )/\sqrt{2}$ becomes a dark state.

(ii) The case of $\mathbf{C}_{2}=\mathbf{0}$: (a) When $\Omega _{23}=\Omega _{13}$, $\Omega _{24}=\Omega _{14}$, and $\theta =(2n+1)\pi $, the state $\vert L_{2}\rangle =(-\vert
l_{2}\rangle +\vert l_{1}\rangle )/\sqrt{2} $ is decoupled
from all the dressed upper states and becomes a dark state. (b) When $\Omega _{23}=-\Omega _{13}$, $\Omega _{24}=-\Omega _{14}$, and $\theta =2n\pi $, the state $\vert L_{2}\rangle =(\vert l_{2}\rangle +\vert l_{1}\rangle )/\sqrt{2} $ becomes a dark state. 

Therefore, when $\Omega _{23}=\Omega _{13}$ and $\Omega _{24}=\Omega _{14}$,  there is always a dark state $(\left\vert
l_{2}\right\rangle -\left\vert l_{1}\right\rangle )/\sqrt{2}$ for $%
\theta =n\pi $; when $\Omega _{23}=-\Omega _{13}$ and $\Omega _{24}=-\Omega
_{14}$, there is always a dark state $(\left\vert
l_{2}\right\rangle +\left\vert l_{1}\right\rangle )/\sqrt{2}$ for $%
\theta =n\pi $.

(2) Next, we consider the case of a degenerate dressed-lower-state subspace. 

In the case of $\vert \Omega _{12}\vert=0$ (the corresponding phase $\theta=0$), these two dressed lower states $\left\vert L_{1}\right\rangle$ and $\left\vert L_{2}\right\rangle$ are degenerate. The dark state exists when the two coupling column vectors $\mathbf{C}_{2}$ and $\mathbf{C}_{1}$ are linearly dependent~\cite{huang2023dark,DSiFSL}. For example, when $\Omega _{24}=2\Omega _{14} $ and $\Omega _{23}=2\Omega _{13}$, the coupling column vectors satisfy $\mathbf{C}_{2}=-3\mathbf{C}_{1}$. Based on Eqs.~(\ref{Bj}) and (\ref{Dj}), we can obtain the bright state $\left\vert B_{1}\right\rangle$ and dark state $\left\vert D_{1}\right\rangle$ as
\begin{subequations}
	\begin{align} 
	\left\vert B_{1}\right\rangle  =&\frac{1}{\sqrt{10}}( \left\vert
	L_{1}\right\rangle -3\left\vert L_{2}\right\rangle ),  \\
		\left\vert D_{1}\right\rangle  =&\frac{1}{\sqrt{10}}(  3\left\vert L_{1}\right\rangle +\left\vert
	L_{2}\right\rangle )  .
	\end{align}	
\end{subequations}
Therefore, the dark state in this case can be expressed as
\begin{equation}\label{D42}
	\left\vert D^{[4,2]}_{1}\right\rangle =\frac{1}{\sqrt{10}}(  3\left\vert L_{1}\right\rangle +\left\vert
	L_{2}\right\rangle )
	=\frac{1}{\sqrt{5}}( 2\left\vert l_{2}\right\rangle -\left\vert
	l_{1}\right\rangle ).
\end{equation}%
Here, the superposition coefficients are determined by the coupling strengths between the upper- and lower-state subspaces.

\section{Dark states in the five-level quantum systems}\label{sec5}

\begin{figure}[t!]
	\centering\includegraphics[width=0.48\textwidth]{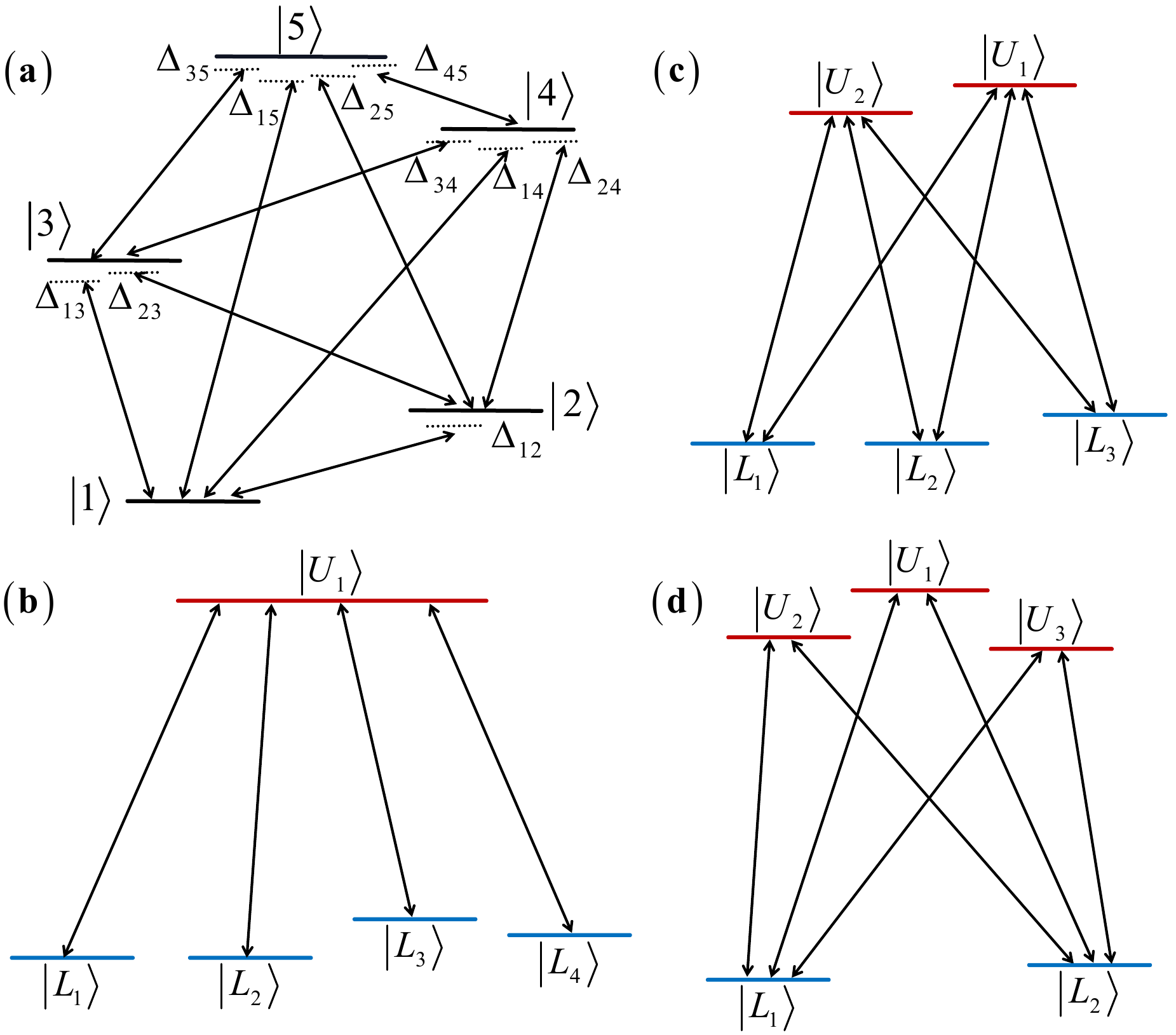}
	\caption{(a) Schematic of a general five-level quantum system expressed in the bare-state representation. According to the numbers of the upper and lower states, it can be divided into three configurations expressed in the dressed upper- and lower-state representation. (b) Configuration 1: one upper state and four lower states with real symmetric couplings $\Omega _{34}=\Omega _{24}=\Omega _{12}=\Omega _{13}=\Omega _{1}$ and $\Omega _{23}=\Omega_{14}=\Omega _{2}$ and under the resonance condition $\Delta _{45}=\Delta
		_{35}=\Delta _{25}=\Delta _{15}=\Delta $. (c) Configuration 2: two upper states and three lower states with real symmetric couplings $%
		\Omega _{23}=\Omega _{12}=\Omega
		_{13}=\Omega $ and the resonance condition $\Delta _{35}=\Delta _{25}=\Delta _{15}=\Delta $. (d) Configuration 3: three upper states and two lower states with real symmetric couplings $\Omega _{45}=\Omega _{35}=\Omega _{34}=\Omega $ and the resonance conditions $\Delta_{25}=\Delta _{15}=\Delta$ and $\Delta _{45}=\Delta _{35}=0$.
		The red (blue) levels denote the dressed upper (lower) states. Similarly, the upper and lower states can be chosen in different configurations on demand, and we only present one case as an example.}\label{5ls}
\end{figure}

In this section, we study the dark states in the five-level quantum systems. According to the numbers of the upper and lower states, the five-level systems can be divided into three configurations: (a) $N_{u}=1$ and $ N_{l}=4$, (b) $N_{u}=2$ and $ N_{l}=3$, and (c) $N_{u}=3$ and $ N_{l}=2$. The Hamiltonian of a general five-level system~[see Fig.~\ref{5ls}(a)] can be described by Eq.~(\ref{HN}) for $N=5$, 
\begin{equation}
H^{[5]}=\sum_{{j}=1}^{5}E_{j}\left\vert
j\right\rangle \left\langle j\right\vert +\sum_{j,j^{\prime }=1,j<j^{\prime }}^{5}(\Omega _{jj^{\prime }}e^{-i\omega _{jj^{\prime
	}}t}\vert
j^{\prime }\rangle \left\langle j\right\vert +\text{H.c.}),
\end{equation}
and the corresponding time-independent Hamiltonian [Eq.~(\ref{HN_tilde}) for $N=5$] reads
\begin{eqnarray}
\tilde{H}^{[5]}=\sum_{r=1}^{4}-\Delta _{r5}\left\vert r\right\rangle
\left\langle r\right\vert +\sum_{j,j^{\prime }=1,j<j^{\prime
}}^{5} (\Omega _{jj^{\prime }}\vert j^{\prime }\rangle \left\langle
j\right\vert +\text{H.c.}),
\end{eqnarray}
where the detunings are defined by $\Delta_{r5}=E_{5}-E_{r}-\omega _{r5}$, and to obtain the time-independent Hamiltonian, the detunings conditions $\Delta _{r5}-\Delta _{r^{\prime }5} =\Delta _{rr^{\prime }}$ for $r,r^{\prime}=1,2,3,4$ with $r<r^{\prime}$ should be satisfied.
Below, we analyze the dark and bright states for these three configurations.

\subsection{Configuration 1: $\boldsymbol{N_{u}=1$ and $ N_{l}=4}$}

For the configuration with one upper state (i.e., $\left\vert u_{1}\right\rangle
=\left\vert 5\right\rangle$) and four lower states (i.e., $\left\vert
l_{1}\right\rangle =\left\vert 4\right\rangle, $ $\left\vert l_{2}\right\rangle =\left\vert 3\right\rangle,$ $\left\vert l_{3}\right\rangle =\left\vert
2\right\rangle,$ and $\left\vert
l_{4}\right\rangle =\left\vert 1\right\rangle $), we define the basis states and vectors as follows: $\left\vert u_{1}\right\rangle
=\left\vert 5\right\rangle =( 1,0,0,0,0) ^{T},$ $\left\vert
l_{1}\right\rangle =\left\vert 4\right\rangle =( 0,1,0,0,0) ^{T},$
$\left\vert l_{2}\right\rangle =\left\vert 3\right\rangle =(
0,0,1,0,0) ^{T},$ $\left\vert l_{3}\right\rangle =\left\vert
2\right\rangle =( 0,0,0,1,0) ^{T},$ and $\left\vert
l_{4}\right\rangle =\left\vert 1\right\rangle =( 0,0,0,0,1) ^{T}$.
Then the Hamiltonian can be expressed as 
\begin{equation}
\tilde{\mathbf{H}}^{[5,1]}=\left( 
\begin{array}{c|c}
\mathbf{H}_{u} & \mathbf{c} \\ \hline
\mathbf{c}^{\dag } & \mathbf{H}_{l}%
\end{array}%
\right) =\left( 
\begin{array}{c|cccc}
0 & \Omega _{45} & \Omega _{35} & \Omega _{25} & \Omega _{15} \\ \hline
\Omega _{45}^{\ast } & -\Delta _{45} & \Omega _{34} & \Omega _{24} & \Omega
_{14} \\ 
\Omega _{35}^{\ast } & \Omega _{34}^{\ast } & -\Delta _{35} & \Omega _{23} & 
\Omega _{13} \\ 
\Omega _{25}^{\ast } & \Omega _{24}^{\ast } & \Omega _{23}^{\ast } & -\Delta
_{25} & \Omega _{12} \\ 
\Omega _{15}^{\ast } & \Omega _{14}^{\ast } & \Omega _{13}^{\ast } & \Omega
_{12}^{\ast } & -\Delta _{15}%
\end{array}%
\right) .
\end{equation}

To analyze the dark-state effect, we need to diagonalize the lower-state submatrix $\mathbf{H}_{l}$. 
For simplicity, below we consider a reduced symmetry case. Concretely, we consider the real symmetric couplings $\Omega _{34}=\Omega _{24}=\Omega _{12}=\Omega _{13}=\Omega _{1}$ and $\Omega _{23}=\Omega_{14}=\Omega _{2}$ among these lower states. Here we choose two independent coupling strengths to include more general cases with diagonal lower-state subspace. For the detunings, we consider the resonance condition $\Delta _{45}=\Delta
_{35}=\Delta _{25}=\Delta _{15}=\Delta $ (the detunings of the upper state with respect to the four lower states are identical), then other detunings satisfy $\Delta _{ij}=0$ for $i,j=1,2,3,4$ and $i \neq j$, namely, the single-photon transitions within the lower-state subspace are resonant.
Therefore, with the unitary matrix%
\begin{equation}
	\mathbf{S}_{l}=\frac{1}{2}\left( 
	\begin{array}{cccc}
		-\sqrt{2} & 0 & 0 & \sqrt{2} \\ 
		0 & -\sqrt{2} & \sqrt{2} & 0 \\ 
		1 & -1 & -1 & 1 \\ 
		1 & 1 & 1 & 1%
	\end{array}%
	\right),
\end{equation}%
the transformed Hamiltonian can be obtained as 
\begin{equation}
\tilde{{H}}_D^{[5,1]}=\left( 
\begin{array}{c|c}
	\mathbf{H}_{U} & \mathbf{C} \\ \hline
	\mathbf{C}^{\dag } & \mathbf{H}_{L}%
\end{array}%
\right) ,
\end{equation}%
where the submatrices are given by
\begin{subequations}
	\begin{align}
		\mathbf{H}_{U} =&0, \\
		\mathbf{H}_{L}=&\text{diag}( -\Delta -\Omega _{2},-\Delta -\Omega _{2},-\Delta -2\Omega _{1}+\Omega _{2},\notag \\
		&-\Delta +2\Omega _{1}+\Omega _{2}) , \\
       \mathbf{C} =&(\mathbf{C}_{1},\mathbf{C}_{2},\mathbf{C}_{3},\mathbf{C}_{4})\notag \\
       =&[({\Omega _{15}-\Omega _{45}})/{\sqrt{2}} , ({\Omega _{25}-\Omega _{35}})/{\sqrt{2}} , (\Omega _{45}-\Omega _{35} \notag \\
	    &-\Omega _{25}+\Omega _{15}	)/{2} , ({\Omega _{45}+\Omega _{35}+\Omega _{25}+\Omega _{15}})/{2}].
	\end{align}%
\end{subequations}%
Here the new basis vectors are obtained as $\left\vert U_{1}\right\rangle =\left\vert u_{1}\right\rangle =\left\vert
5\right\rangle ,$ $\left\vert L_{1}\right\rangle =( \left\vert
l_{4}\right\rangle -\left\vert l_{1}\right\rangle ) /\sqrt{2},$ $\left\vert
L_{2}\right\rangle =( \left\vert l_{3}\right\rangle -\left\vert
l_{2}\right\rangle ) /\sqrt{2},$ $\left\vert L_{3}\right\rangle =(
\left\vert l_{1}\right\rangle -\left\vert l_{2}\right\rangle -\left\vert
l_{3}\right\rangle +\left\vert l_{4}\right\rangle ) /2,$ and $\left\vert
L_{4}\right\rangle =( \left\vert l_{1}\right\rangle +\left\vert
l_{2}\right\rangle +\left\vert l_{3}\right\rangle +\left\vert
l_{4}\right\rangle ) /2$. Based on the arrowhead matrix, we can
analyze the dark states in the system~[see Fig.~\ref{5ls}(b)].

(1) We consider the case where the coupling column vector is zero. 

(i) The case of $\mathbf{C}_{1}=\mathbf{0}$: When $\Omega _{15}=\Omega _{45},$ the state $\left\vert
L_{1}\right\rangle =(\left\vert l_{4}\right\rangle -\left\vert
l_{1}\right\rangle )/\sqrt{2}$ is decoupled from the dressed upper state and becomes a dark state.

(ii) The case of $\mathbf{C}_{2}=\mathbf{0}$: When $\Omega _{25}=\Omega _{35},$ the state $\left\vert
L_{2}\right\rangle =(\left\vert l_{3}\right\rangle -\left\vert
l_{2}\right\rangle )/\sqrt{2}$ becomes a dark state.

(iii) The case of $\mathbf{C}_{3}=\mathbf{0}$: When $\Omega _{45}+\Omega _{15}=\Omega _{35}+\Omega _{25},$ the state $\left\vert L_{3}\right\rangle= (\left\vert l_{1}\right\rangle
-\left\vert l_{2}\right\rangle -\left\vert l_{3}\right\rangle +\left\vert
l_{4}\right\rangle )/2 $ becomes a dark state.

(iv) The case of $\mathbf{C}_{4}=\mathbf{0}$: When $\Omega _{45}+\Omega _{35}+\Omega _{25}+\Omega _{15}=0,$
the state $\left\vert L_{4}\right\rangle =(\left\vert l_{1}\right\rangle
+\left\vert l_{2}\right\rangle +\left\vert l_{3}\right\rangle +\left\vert
l_{4}\right\rangle )/2$ becomes a dark state.

(2) We consider the case with the degenerate dressed lower states.

(i) There is always a two-dimensional degenerate dressed-lower-state subspace $\{\vert
L_{1}\rangle, \vert L_{2}\rangle\}$, hence the dark state can be obtained based on the definitions in Eqs.~(\ref{BD}) as
\begin{eqnarray}\label{D151}
		\left\vert D^{[5,1]}_{1}\right\rangle &= &\frac{1}{\mathcal{N}^{[5,1]}_{1}}(\mathbf{C}_{2}\left\vert L_{1}\right\rangle -\mathbf{C}_{1}\left\vert L_{2}\right\rangle )\notag \\ 
		&= & \frac{1}{2\mathcal{N}^{[5,1]}_{1}}[( \Omega _{35}-\Omega
		_{25}) (\left\vert l_{1}\right\rangle -\left\vert l_{4}\right\rangle )\notag \\ 
		&&-(\Omega
		_{45}-\Omega _{15})(\left\vert l_{2}\right\rangle -\left\vert l_{3}\right\rangle )],
\end{eqnarray}
where the coefficient $\mathcal{N}^{[5,1]}_{1}=\sqrt{ \vert \mathbf{C}_{1}\vert ^{2}+\vert \mathbf{C}_{2}\vert^{2}}$ is introduced.

(ii) Furthermore, in the case of $\Omega _{1}=0$, there are two two-dimensional degenerate dressed-lower-state subspaces $\{\vert L_{1}\rangle, \vert L_{2}\rangle\}$ and $\{\vert L_{3}\rangle, \vert L_{4}\rangle\}$. In addition to the dark state $\left\vert D^{[5,1]}_{1}\right\rangle$ in the subspace $\{\vert L_{1}\rangle, \vert L_{2}\rangle\}$, there exists another dark state in the degenerate subspace $\{\vert L_{3}\rangle, \vert L_{4}\rangle\}$,
\begin{eqnarray}
\left\vert D^{[5,1]}_{2}\right\rangle &=& \frac{1}{\mathcal{N}^{[5,1]}_{2}}(\mathbf{C}_{4}\left\vert L_{3}\right\rangle -\mathbf{C}_{3}\left\vert L_{4}\right\rangle ) \notag \\
 &=& \frac{1 }{2\mathcal{N}^{[5,1]}_{2}}[(\Omega _{35}+\Omega _{25})(\left\vert l_{1}\right\rangle +\left\vert
l_{4}\right\rangle )\notag \\
&&-(\Omega _{15}+\Omega _{45})(\left\vert l_{2}\right\rangle
+\left\vert l_{3}\right\rangle)],
\end{eqnarray}
with the coefficient $\mathcal{N}^{[5,1]}_{2}=\sqrt{ \vert \mathbf{C}_{3}\vert^{2}+\vert \mathbf{C}_{4}\vert^{2}}$.

(iii) In the case of $\Omega _{1}=\Omega _{2},$ there is a three-dimensional
degenerate subspace $\{\vert L_{1}\rangle, \vert L_{2}\rangle,\vert L_{3}\rangle\}$, similar to the case considered in Eqs.~(\ref{B2D2}) and~(\ref{D2[41]}). As a result, there are two dark states. One is $\left\vert D^{[5,1]}_{1}\right\rangle$ given in Eq.~(\ref{D151}) and the other is given by
\begin{equation}
\left\vert D_{3}^{[5,1]}\right\rangle =\frac{\mathbf{C}_{3}( \mathbf{C}_{1}^{\ast }\left\vert
	L_{1}\right\rangle +\mathbf{C}_{2}^{\ast }\left\vert L_{2}\right\rangle
	) -(\mathcal{N}_{1}^{[5,1]})^2 \left\vert L_{3}\right\rangle}{\mathcal{N}_{1}^{[5,1]} \mathcal{N}_{3}^{[5,1]}},
\end{equation}
with $\mathcal{N}_{3}^{[5,1]}=\sqrt{\vert \mathbf{C}_{1}\vert ^{2}+\vert \mathbf{C}_{2}\vert^{2}+\vert \mathbf{C}_{3}\vert^{2}}$. 
This dark state can be further expressed with the bare states using the relations $\left\vert L_{1}\right\rangle =( \left\vert
	l_{4}\right\rangle -\left\vert l_{1}\right\rangle ) /\sqrt{2},$ $\left\vert
	L_{2}\right\rangle =( \left\vert l_{3}\right\rangle -\left\vert
	l_{2}\right\rangle ) /\sqrt{2},$ and $\left\vert L_{3}\right\rangle =(
	\left\vert l_{1}\right\rangle -\left\vert l_{2}\right\rangle -\left\vert
	l_{3}\right\rangle +\left\vert l_{4}\right\rangle ) /2$.
	
(iv) In the case of $\Omega _{1}=\Omega _{2}=0$, there exists a four-dimensional degenerate dressed-lower-state subspace $\{\vert L_{1}\rangle,$ $ \vert L_{2}\rangle, \vert L_{3}\rangle,\vert L_{4}\rangle\}$. Based on Eqs.~(\ref{BD}) and~(\ref{B2D2}), we further introduce 
\begin{subequations}\label{B3D3}
	\begin{align} 
		\vert B_{3}\rangle =&\frac{1}{\mathcal{N}_{3} }( \mathcal{N}_{2} \left\vert
		B_2\right\rangle +\mathbf{C}_{4}^{\ast }\left\vert L_{4}\right\rangle )\notag \\
		=&\frac{1}{\mathcal{N}_{3} }( \mathbf{C}_{1}^{\ast }\left\vert
		L_{1}\right\rangle +\mathbf{C}_{2}^{\ast }\left\vert L_{2}\right\rangle  + \mathbf{C}_{3}^{\ast }\left\vert L_{3}\right\rangle+\mathbf{C}_{4}^{\ast }\left\vert L_{4}\right\rangle)
		,\\
		\vert D_{3}\rangle =&\frac{1}{\mathcal{N}_{3} }( \mathbf{C}_{4}\left\vert
		B_{2}\right\rangle -\mathcal{N}_{2} \left\vert L_{4}\right\rangle )\notag \\
		=&\frac{%
			1}{\mathcal{N}_{3} \mathcal{N}_{2} }[ \mathbf{C}_{4}(\mathbf{C}_{1}^{\ast }\left\vert L_{1}\right\rangle
		+\mathbf{C}_{2}^{\ast }\left\vert L_{2}\right\rangle+\mathbf{C}_{3}^{\ast }\left\vert L_{3}\right\rangle ) -\mathcal{N}_{2} ^{2}\left\vert
		L_{4}\right\rangle ],
	\end{align} 
\end{subequations}
with $\mathcal{N}_{3} =\sqrt{\mathcal{N}_{2} ^2+\vert \mathbf{C}_{4}\vert
	^{2}}=\sqrt{\vert \mathbf{C}_{1}\vert ^{2}+\vert \mathbf{C}_{2}\vert
	^{2}+\vert \mathbf{C}_{3}\vert
	^{2}+\vert \mathbf{C}_{4}\vert
	^{2}}$. These states satisfy the relation $\vert B_{3}\rangle
\langle B_{3}\vert+\vert D_1\rangle \langle D_1\vert +\vert D_{2}\rangle \langle
 D_{2}\vert  +\vert D_{3} \rangle \langle D_{3}\vert =\vert L_{1}\rangle \langle
L_{1}\vert +\vert L_{2}\rangle \langle L_{2}\vert +\vert L_{3}\rangle \langle
L_{3}\vert +\vert L_{4}\rangle \langle L_{4}\vert $, and only the state $\vert B_{3}\rangle$ is coupled with the dressed upper state. Therefore, the states $\vert D_{1}\rangle $, $\vert  D_{2}\rangle $, and $\vert D_{3}\rangle$ are dark states, which can be written as $\left\vert D^{[5,1]}_{1}\right\rangle$, $\left\vert D^{[5,1]}_{3}\right\rangle$, and
\begin{equation}
	\left\vert D_{4}^{[5,1]}\right\rangle =\frac{\mathbf{C}_{4}( \mathbf{C}_{1}^{\ast }\left\vert
		L_{1}\right\rangle +\mathbf{C}_{2}^{\ast }\left\vert L_{2}\right\rangle +	\mathbf{C}_{3}^{\ast }\left\vert L_{3}\right\rangle ) -(\mathcal{N}
		_{3}^{[5,1]})^2\left\vert L_{4}\right\rangle}{\mathcal{N}_{4}^{[5,1]}\mathcal{N}
	_{3}^{[5,1]}},
\end{equation}%
with $\mathcal{N}_{4}^{[5,1]}=\sqrt{\vert \mathbf{C}_{1}\vert ^{2}+\vert \mathbf{C}_{2}\vert
	^{2}+\vert \mathbf{C}_{3}\vert
	^{2}+\vert \mathbf{C}_{4}\vert
	^{2}}$. 
Similarly, the dark state can also be further expressed by the bare states.

\subsection{Configuration 2: $\boldsymbol{N_{u}=2$ and $ N_{l}=3}$}

For the configuration with two upper states (i.e., $\left\vert u_{1}\right\rangle
=\left\vert 5\right\rangle$ and $\left\vert
u_{2}\right\rangle =\left\vert 4\right\rangle $) and three lower states (i.e., $%
\left\vert l_{1}\right\rangle =\left\vert 3\right\rangle ,$
$\left\vert l_{2}\right\rangle =\left\vert 2\right\rangle , $ and $\left\vert l_{3}\right\rangle =\left\vert 1\right\rangle $), we define the basis states and vectors as follows: $\left\vert u_{1}\right\rangle
=\left\vert 5\right\rangle =( 1,0,0,0,0) ^{T},$ $\left\vert
u_{2}\right\rangle =\left\vert 4\right\rangle =( 0,1,0,0,0) ^{T},$ $%
\left\vert l_{1}\right\rangle =\left\vert 3\right\rangle =( 0,0,1,0,0) ^{T},$
$\left\vert l_{2}\right\rangle =\left\vert 2\right\rangle =( 0,0,0,1,0)
^{T}, $ and $\left\vert l_{3}\right\rangle =\left\vert 1\right\rangle =(
0,0,0,0,1) ^{T}$. Then the Hamiltonian can be expressed as 
\begin{equation}
\tilde{{H}}^{[5,2]}=\left( 
\begin{array}{c|c}
\mathbf{H}_{u} & \mathbf{c} \\ \hline
\mathbf{c}^{\dag } & \mathbf{H}_{l}%
\end{array}%
\right)=\left( 
\begin{array}{cc|ccc}
0& \Omega _{45} & \Omega _{35} & \Omega _{25} & \Omega _{15} \\ 
\Omega _{45}^{\ast } & -\Delta _{45} & \Omega _{34} & \Omega _{24} & \Omega
_{14} \\ \hline
\Omega _{35}^{\ast } & \Omega _{34}^{\ast } & -\Delta _{35} & \Omega _{23} & 
\Omega _{13} \\ 
\Omega _{25}^{\ast } & \Omega _{24}^{\ast } & \Omega _{23}^{\ast } & -\Delta
_{25} & \Omega _{12} \\ 
\Omega _{15}^{\ast } & \Omega _{14}^{\ast } & \Omega _{13}^{\ast } & \Omega
_{12}^{\ast } & -\Delta _{15}%
\end{array}%
\right) .  \label{H52}
\end{equation}%
The form of the lower-state submatrix $\mathbf{H}_{l}$ is similar to Eq.~(\ref{H41}%
). 
Similarly, here we consider the real symmetric couplings $%
\Omega _{23}=\Omega _{12}=\Omega
_{13}=\Omega $ and the resonance condition $\Delta _{35}=\Delta _{25}=\Delta _{15}=\Delta $ (other detunings satisfy $\Delta _{ij}=0$ for $i,j=1,2,3$ and $i \neq j$). In particular, we choose $\Omega _{45}=\Omega _{45}^{\ast}=0$ so that $\vert U_{1}\rangle$ and $\vert U_{2}\rangle$ are exactly the two upper bare states, $\vert u_{1} \rangle$ and $\vert u_{2} \rangle$, for simplicity. Based on the unitary matrix in Eq.~(\ref{Sl3}), the Hamiltonian with dressed upper and lower states can be obtained as%
\begin{equation}\label{HD52}
\tilde{{H}}^{[5,2]}_{D}=\left( 
\begin{array}{c|c}
\mathbf{H}_{U} & \mathbf{C} \\ \hline
\mathbf{C}^{\dag } & \mathbf{H}_{L}%
\end{array}%
\right),
\end{equation}%
where the submatrices take the form as
\begin{subequations}\label{HD52_3}
\begin{align}
\mathbf{H}_{U} =&\text{diag}(0,-\Delta_{45} ) 
, \\
\mathbf{H}_{L} =&\text{diag}(-\Delta -\Omega ,-\Delta -\Omega ,-\Delta +2\Omega) 
, \\
\mathbf{C}=&(\mathbf{C}_{1},\mathbf{C}_{2},\mathbf{C}_{3}) \notag \\
=&\left( 
\begin{array}{ccc}
\frac{\Omega _{15}-\Omega _{35}}{\sqrt{2}} & \frac{2\Omega _{25}-\Omega
_{35}-\Omega _{15}}{\sqrt{6}} & \frac{\Omega _{35}+\Omega _{25}+\Omega _{15}%
}{\sqrt{3}} \\ 
\frac{\Omega _{14}-\Omega _{34}}{\sqrt{2}} & \frac{2\Omega _{24}-\Omega
_{34}-\Omega _{14}}{\sqrt{6}} & \frac{\Omega _{34}+\Omega _{24}+\Omega _{14}%
}{\sqrt{3}}%
\end{array}%
\right) .
\end{align}%
\end{subequations}
In Eq.~(\ref{HD52}), these new basis vectors are given by $\vert U_{1}\rangle =\vert u_{1}\rangle =\vert
5\rangle $, $\vert U_{2}\rangle =\vert u_{2}\rangle =\vert
4\rangle  $, $\vert L_{1}\rangle =(\vert
l_{3}\rangle -\vert l_{1}\rangle )/\sqrt{2},$ $\vert
L_{2}\rangle =(2\vert l_{2}\rangle -\vert
l_{1}\rangle -\vert l_{3}\rangle )/\sqrt{6}$, and $%
\vert L_{3}\rangle =(\vert l_{1}\rangle +\vert
l_{2}\rangle +\vert l_{3}\rangle )/\sqrt{3}.$
Now we can analyze the dark states with the arrowhead-matrix method for the configuration 2~[see Fig.~\ref{5ls}(c)].

(1) Consider the case of zero coupling column vector.

(i) The case of $\mathbf{C}_{1}=\mathbf{0}$: When $\Omega _{15}=\Omega _{35}\ $and $\Omega _{14}=\Omega
_{34},$ the state $\left\vert L_{1}\right\rangle =( \left\vert
l_{3}\right\rangle -\left\vert l_{1}\right\rangle ) /\sqrt{2}$ is decoupled
from the dressed upper states and becomes a dark state.

(ii) The case of $\mathbf{C}_{2}=\mathbf{0}$: When $2\Omega _{25}=\Omega _{35}+\Omega _{15}\ $and $2\Omega
_{24}=\Omega _{34}+\Omega _{14},$ the state $\left\vert L_{2}\right\rangle
=( 2\left\vert l_{2}\right\rangle -\left\vert l_{1}\right\rangle -\left\vert
l_{3}\right\rangle ) /\sqrt{6}$ becomes a dark state.

(iii) The case of $\mathbf{C}_{3}=\mathbf{0}$: When $\Omega _{35}+\Omega _{25}+\Omega _{15}=\Omega
_{34}+\Omega _{24}+\Omega _{14}=0,$ the state $\left\vert L_{3}\right\rangle=( \left\vert l_{1}\right\rangle
+\left\vert l_{2}\right\rangle +\left\vert l_{3}\right\rangle ) /\sqrt{%
	3} $ becomes a dark state. 

(2) Consider the case of a degenerate dressed-lower-state subspace. 

(i) There is always a two-dimensional degenerate subspace $\{\vert L_{1}\rangle, \vert L_{2}\rangle\}$, and when the coupling column vectors are linearly dependent, i.e., $\mathbf{C}_{2}=\gamma \mathbf{C}_{1}$, there exists a dark state%
\begin{eqnarray}\label{D521}
\left\vert D^{[5,2]}_{1}\right\rangle &=&\frac{1}{\sqrt{1+\vert \gamma
\vert ^{2}}}( \gamma \left\vert L_{1}\right\rangle -\left\vert
L_{2}\right\rangle ) \notag \\
&=&\frac{ ( 1-\sqrt{3}\gamma ) \left\vert
	l_{1}\right\rangle -2\left\vert l_{2}\right\rangle+( \sqrt{3}\gamma +1)
	\left\vert l_{3}\right\rangle}{\sqrt{6}\sqrt{1+|\gamma |^{2}}}.
\end{eqnarray}

(ii) In the case of $\Omega =0,$ there exists a three-dimensional degenerate subspace $\{\vert L_{1}\rangle, \vert L_{2}\rangle,\vert L_{3}\rangle\}$. 
We know that there is at least one dark state because the dimension of the dressed-lower-state subspace is greater than the dressed-upper-state subspace~\cite{huang2023dark,DSiFSL}. As an example, we consider the case of $\Omega_{25}=\Omega _{34}=0$ and $\Omega _{35}=\Omega _{15}=\Omega _{24}=\Omega _{14}=\Omega$, then the coupling matrix in Eq.~(\ref{HD52_3}) becomes
\begin{eqnarray}
	\mathbf{C} =\frac{\Omega}{\sqrt{6}}\left( 
	\begin{array}{ccc}
		0 & -2 & 2\sqrt{2} \\ 
		\sqrt{3} & 1 & 2\sqrt{2}%
	\end{array}%
	\right).
\end{eqnarray}%
Based on the above coupling matrix, we can obtain its SVD as $\mathbf{C}=\mathbf{W} \mathbf{\Sigma } \mathbf{V }^{\dag}$ with the matrices
\begin{subequations}
	\begin{align}
		\mathbf{W}=& \left( 
		\begin{array}{cc}
			 {1}/{\sqrt{2}}&  {-1}/{\sqrt{2}} \\ 
			 {1}/{\sqrt{2}} &  {1}/{\sqrt{2}}
		\end{array}%
		\right), \\
		\mathbf{\Sigma }=&\left( 
		\begin{array}{ccc}
			\sqrt{3}& 0 & 0\\ 
			0& 1 & 0
		\end{array}%
		\right),\\
		\mathbf{V }=&\left( 
		\begin{array}{ccc}
			 {1}/{(2\sqrt{3})}&  {1}/{2} &  {-\sqrt{2}}/{\sqrt{3}}\\ 
			 {-1}/{6}&  {\sqrt{3}}/{2} &  {\sqrt{2}}/{3}\\
			 {2\sqrt{2}}/{3}& 0 &  {1}/{{3}}
		\end{array}%
		\right).
	\end{align}%
\end{subequations}
The right singular vectors in $\mathbf{V }$ corresponding to the nonzero singular values are the orthogonal bright states
\begin{subequations}
		\begin{align}
	\left\vert B_{1} \right\rangle=& \frac{1}{2\sqrt{3}}\left\vert L_{1}\right\rangle -\frac{1}{{6}}\left\vert L_{2}\right\rangle+\frac{2\sqrt{2}}{{3}}%
		\left\vert L_{3}\right\rangle, \label{H52_B1} \\
	\left\vert B_{2} \right\rangle =&\frac{1}{2}\left\vert L_{1}\right\rangle +\frac{\sqrt{3}}{2}	\left\vert L_{2}\right\rangle,\label{H52_B2}
\end{align}%
\end{subequations}
and the remaining right singular vector in $\mathbf{V }$ is the orthogonal dark state
\begin{eqnarray}\label{D1-52}
	\left\vert D\right\rangle =\frac{-\sqrt{2}}{\sqrt{3}}	\left\vert L_{1}\right\rangle+\frac{\sqrt{2}}{{3}}\left\vert L_{2}\right\rangle +\frac{1}{{3}}\left\vert
		L_{3}\right\rangle.
\end{eqnarray}
Therefore, the dark state in this case can be obtained as 
\begin{eqnarray}\label{D522}
	\left\vert D^{[5,2]}_{2}\right\rangle 
	=\frac{1}{\sqrt{3}}%
	(\left\vert l_{1}\right\rangle +\left\vert l_{2}\right\rangle-\left\vert l_{3}\right\rangle
	 ).
\end{eqnarray}

In particular, if these three column vectors in the coupling matrix in Eq.~(\ref{HD52_3}) are linearly dependent with each other, there exists one bright state and two dark states. For example, when $\Omega _{15}=2\Omega _{35}=2\Omega _{25}$ and $\Omega _{14}=2\Omega _{34}=2\Omega _{24}$, the coupling matrix becomes 
\begin{equation}\label{C123-ex}
	\mathbf{C}=\left( 
	\begin{array}{ccc}
		\frac{\Omega _{35}}{\sqrt{2}} & \frac{-\Omega
			_{35}}{\sqrt{6}} & \frac{4\Omega _{35}		}{\sqrt{3}} \\ 
		\frac{\Omega _{34}}{\sqrt{2}} & \frac{-\Omega
			_{34}}{\sqrt{6}} & \frac{4\Omega _{34}		}{\sqrt{3}}%
	\end{array}%
	\right) ,
\end{equation}
and the coupling column vectors satisfy $\mathbf{C}_2= (-1/\sqrt{3})\mathbf{C}_1$ and $\mathbf{C}_3=(4\sqrt{2}/\sqrt{3})\mathbf{C}_1$.
Based on Eqs.~(\ref{Bj}) and (\ref{Dj}), we can find that there is one bright state $\left\vert B\right\rangle$ given in Eq.~(\ref{H52_B1}) and two dark states $\left\vert D_{1}\right\rangle$ and $\left\vert D_{2}\right\rangle$ with the same forms as Eqs.~(\ref{H52_B2}) and~(\ref{D1-52}).
Therefore, these two dark states can be expressed as $\left\vert D^{[5,2]}_{2}\right\rangle $ in Eq.~(\ref{D522}) and 
\begin{eqnarray}
	\left\vert D^{[5,2]}_{3}\right\rangle  =\frac{1}{\sqrt{2}}( -\left\vert l_{1}\right\rangle +\left\vert
	l_{2}\right\rangle ).
\end{eqnarray}%
Here, we can see that the dark state could only involve partial basis vectors in the lower-state subspace.

\subsection{Configuration 3: $\boldsymbol{N_{u}=3$ and $ N_{l}=2}$}

For the configuration with three upper states (i.e., $\left\vert u_{1}\right\rangle
=\left\vert 5\right\rangle$, $\left\vert
u_{2}\right\rangle =\left\vert 4\right\rangle $, and $\left\vert u_{3}\right\rangle =\left\vert 3\right\rangle$) and two lower states (i.e., $%
\left\vert l_{1}\right\rangle =\left\vert 2\right\rangle $ and $\left\vert l_{2}\right\rangle =\left\vert 1\right\rangle $), we define the basis states and vectors as follows: $\left\vert u_{1}\right\rangle
=\left\vert 5\right\rangle =( 1,0,0,0,0) ^{T},$ $\left\vert
u_{2}\right\rangle =\left\vert 4\right\rangle =( 0,1,0,0,0) ^{T},$
$\left\vert u_{3}\right\rangle =\left\vert 3\right\rangle =(
0,0,1,0,0) ^{T},$ $\left\vert l_{1}\right\rangle =\left\vert
2\right\rangle =( 0,0,0,1,0) ^{T},$ and $\left\vert
l_{2}\right\rangle =\left\vert 1\right\rangle =( 0,0,0,0,1) ^{T} $%
. Then the Hamiltonian can be expressed as 
\begin{equation}
\tilde{{H}}^{[5,3]}=\left( 
\begin{array}{c|c}
\mathbf{H}_{u} & \mathbf{c} \\ \hline
\mathbf{c}^{\dag } & \mathbf{H}_{l}%
\end{array}%
\right) =\left( 
\begin{array}{ccc|cc}
0 & \Omega _{45} & \Omega _{35} & \Omega _{25} & \Omega _{15} \\ 
\Omega _{45}^{\ast } & -\Delta _{45} & \Omega _{34} & \Omega _{24} & \Omega
_{14} \\ 
\Omega _{35}^{\ast } & \Omega _{34}^{\ast } & -\Delta _{35} & \Omega _{23} & 
\Omega _{13} \\ \hline
\Omega _{25}^{\ast } & \Omega _{24}^{\ast } & \Omega _{23}^{\ast } & -\Delta
_{25} & \Omega _{12} \\ 
\Omega _{15}^{\ast } & \Omega _{14}^{\ast } & \Omega _{13}^{\ast } & \Omega
_{12}^{\ast } & -\Delta _{15}%
\end{array}%
\right) .  \label{H53}
\end{equation}%
In this case, the form of the lower-state submatrix $\mathbf{H}_{l}$ is similar to Eqs.~(\ref{H3}) and~(\ref{H42}). Similarly, we consider the case of $\Delta
_{25}=\Delta _{15}=\Delta$ and $\Omega
_{12}=\left\vert \Omega _{12}\right\vert e^{i\theta }$. 
The upper-state submatrix $\mathbf{H}_{u}$ is similar to Eq.~(\ref{H41}), and we similarly consider that $\Omega _{45}=\Omega _{45}^{\ast }=\Omega _{35}=\Omega
_{35}^{\ast }=\Omega _{34}=\Omega _{34}^{\ast }=\Omega $ and $\Delta _{45}=\Delta _{35}=0$. Therefore, we introduce the unitary matrices in Eqs.~{(\ref{Sl3})} and~{(\ref{Sl})} to diagonalize the upper- and lower-state submatrices, respectively. Then the submatrices $\mathbf{H}_{U} $, $\mathbf{H}_{L} $, and $\mathbf{C}$ in the Hamiltonian $\tilde{{H}}^{[5,3]}_{D}$ can be obtained as
\begin{subequations}
\begin{align}
\mathbf{H}_{U}=&\text{diag(}-\Omega,-\Omega,2\Omega \text{)}, \\
\mathbf{H}_{L}=&\text{diag(}-\Delta -\vert \Omega _{12}\vert ,
-\Delta +\vert \Omega _{12}\vert \text{)},\\
\mathbf{C}=&(\mathbf{C}_{1} , \mathbf{C}_{2} ),
\end{align}%
\end{subequations}
with the coupling column vectors
\begin{subequations}
	\begin{align}
\mathbf{C}_{1}=&\left( 
\begin{array}{c}
	\frac{( -\Omega _{23}+e^{-i\theta }\Omega _{13}) -( -\Omega
		_{25}+e^{-i\theta }\Omega _{15}) }{2}  \\ 
	\frac{ 2( -\Omega
		_{24}+e^{-i\theta }\Omega _{14})-( -\Omega _{25}+e^{-i\theta }\Omega _{15}) -( -\Omega _{23}+e^{-i\theta
		}\Omega _{13}) }{2\sqrt{3}}  \\ 
	\frac{( -\Omega _{25}+e^{-i\theta }\Omega _{15}) +( -\Omega
		_{24}+e^{-i\theta }\Omega _{14}) +( -\Omega _{23}+e^{-i\theta
		}\Omega _{13}) }{\sqrt{6}} %
\end{array}%
\right) ,\\
\mathbf{C}_{2}=&\left( 
\begin{array}{c}
	\frac{( \Omega
		_{23}+e^{-i\theta }\Omega _{13}) -( \Omega _{25}+e^{-i\theta
		}\Omega _{15}) }{2} \\ 
	  \frac{ 2( \Omega
	  	_{24}+e^{-i\theta }\Omega _{14})-( \Omega _{25}+e^{-i\theta }\Omega _{15}) -( \Omega _{23}+e^{-i\theta
	  	}\Omega _{13}) }{2\sqrt{3}}\\
	 \frac{( \Omega _{25}+e^{-i\theta
		}\Omega _{15}) +( \Omega _{24}+e^{-i\theta }\Omega _{14})
		+( \Omega _{23}+e^{-i\theta }\Omega _{13}) }{\sqrt{6}}%
\end{array}%
\right) .
\end{align}%
\end{subequations}
The corresponding new basis vectors are given by $\left\vert U_{1}\right\rangle =(\left\vert
u_{3}\right\rangle -\left\vert u_{1}\right\rangle )/\sqrt{2},$ $\left\vert
U_{2}\right\rangle =(2\left\vert u_{2}\right\rangle -\left\vert
u_{1}\right\rangle -\left\vert u_{3}\right\rangle )/\sqrt{6},$ $\left\vert U_{3}\right\rangle =(\left\vert
u_{1}\right\rangle +\left\vert u_{2}\right\rangle +\left\vert
u_{3}\right\rangle )/\sqrt{3}$, $\left\vert
L_{1}\right\rangle =(e^{-i\theta }\left\vert l_{2}\right\rangle -\left\vert
l_{1}\right\rangle )/\sqrt{2}, $ and $\left\vert L_{2}\right\rangle
=(e^{-i\theta }\left\vert l_{2}\right\rangle +\left\vert l_{1}\right\rangle
)/\sqrt{2}$.

With the arrowhead-matrix method, we can analyze the dark states for the configuration 3~[see Fig.~\ref{5ls}(d)].

(1) Consider the case of zero coupling column vector.

(i) The case of $\mathbf{C}_{1}=\mathbf{0}$: (a) When $\Omega _{23}=\Omega _{13}$, $\Omega _{24}=\Omega _{14}$, $ \Omega _{25}=\Omega _{15}$, and $\theta =2n\pi $ for $n\in\mathbb{Z} $, the state $%
\left\vert L_{1}\right\rangle =(\left\vert l_{2}\right\rangle -\left\vert
l_{1}\right\rangle )/\sqrt{2}$ is decoupled from all the dressed upper states and becomes a dark state. (b) When $\Omega _{23}=-\Omega _{13}$, $\Omega _{24}=-\Omega _{14}$%
, $\Omega _{25}=-\Omega _{15}$, and $\theta =(2n+1)\pi $, the state $\left\vert
L_{1}\right\rangle =(\left\vert l_{2}\right\rangle+\left\vert
l_{1}\right\rangle )/\sqrt{2}$ becomes a dark state. 

(ii) The case of $\mathbf{C}_{2}=\mathbf{0}$: (a) When $\Omega _{23}=\Omega _{13}$, $\Omega _{24}=\Omega _{14}$, $%
\Omega _{25}=\Omega _{15}$, and $\theta =(2n+1)\pi $, the state $\left\vert
L_{2}\right\rangle =(-\left\vert l_{2}\right\rangle +\left\vert
l_{1}\right\rangle )/\sqrt{2} $ is decoupled from all the dressed upper states and becomes a dark state. (b) When $\Omega _{23}=-\Omega _{13}$, $\Omega _{24}=-\Omega _{14}$%
, $\Omega _{25}=-\Omega _{15}$, and $\theta =2n\pi $, the
state $\left\vert L_{2}\right\rangle =(\left\vert l_{2}\right\rangle
+\left\vert l_{1}\right\rangle )/\sqrt{2} $ becomes a dark state.

Therefore, when $\Omega _{23}=\Omega _{13}$, $\Omega _{24}=\Omega _{14}$, and $%
\Omega _{25}=\Omega _{15}$, there is always a dark state $(\left\vert l_{2}\right\rangle -\left\vert
l_{1}\right\rangle )/\sqrt{2}$ for $\theta =n\pi $; when $\Omega _{23}=-\Omega _{13}$, $\Omega _{24}=-\Omega _{14}$%
, and $\Omega _{25}=-\Omega _{15}$, there is always a dark state $(\left\vert l_{2}\right\rangle
+\left\vert l_{1}\right\rangle )/\sqrt{2}$ for $\theta =n\pi $.

(2) Consider the case of a degenerate dressed-lower-state subspace. 

In the case of $\left\vert \Omega _{12}\right\vert =0$ (consider the corresponding phase is zero $\theta=0$), the two dressed lower states are degenerate. The dark state only exists when the two coupling vectors $\mathbf{C}_{2}$ and $\mathbf{C}_{1}$ are linearly dependent. We consider the case of $\Omega _{25}=2\Omega _{15}$, $\Omega _{24}=2\Omega _{14}$, and $\Omega
_{23}=2\Omega _{13}$, therefore, the coupling matrix becomes
\begin{equation}
	\mathbf{C} =\left( 
	\begin{array}{cc}
		\frac{-\Omega _{13}+\Omega _{15}}{2} & \frac{3( \Omega _{13}-\Omega
			_{15}) }{2} \\ 
		\frac{-2\Omega _{14}+\Omega _{15}+\Omega _{13}}{2\sqrt{3}} & \frac{3( 2\Omega _{14}-\Omega _{15}-\Omega _{13}) }{2\sqrt{3}} \\ 
		\frac{-\Omega _{15}-\Omega _{14}-\Omega _{13}}{\sqrt{6}} & \frac{3(
			\Omega _{15}+\Omega _{14}+\Omega _{13}) }{\sqrt{6}}%
	\end{array}%
	\right),
\end{equation}%
where the coupling column vectors satisfy $\mathbf{C}_{2}=-3\mathbf{C}_{1}$. Similarly, based on Eqs.~(\ref{Bj}) and (\ref{Dj}), we can obtain the bright state $\left\vert B_{1}\right\rangle$ and dark state $\left\vert D_{1}\right\rangle$ as:
\begin{subequations}
\begin{align}
	\left\vert B_{1}\right\rangle  =&\frac{1}{\sqrt{10}}( \left\vert
	L_{1}\right\rangle -3\left\vert L_{2}\right\rangle ), \\
		\left\vert D_{1}\right\rangle  =&\frac{1}{\sqrt{10}}(  3\left\vert L_{1}\right\rangle +\left\vert
	L_{2}\right\rangle )  .
\end{align}%
\end{subequations}
Therefore, the dark state in this case can be written as 
\begin{eqnarray}
	\left\vert D_{1}^{[5,3]}\right\rangle  =\frac{1}{\sqrt{5}}(2\left\vert l_{2}\right\rangle -\left\vert
	l_{1}\right\rangle ),
\end{eqnarray}%
which is a superposition of the two lower states.

\begin{figure*}[t!]
	\centering\includegraphics[width=0.8\textwidth]{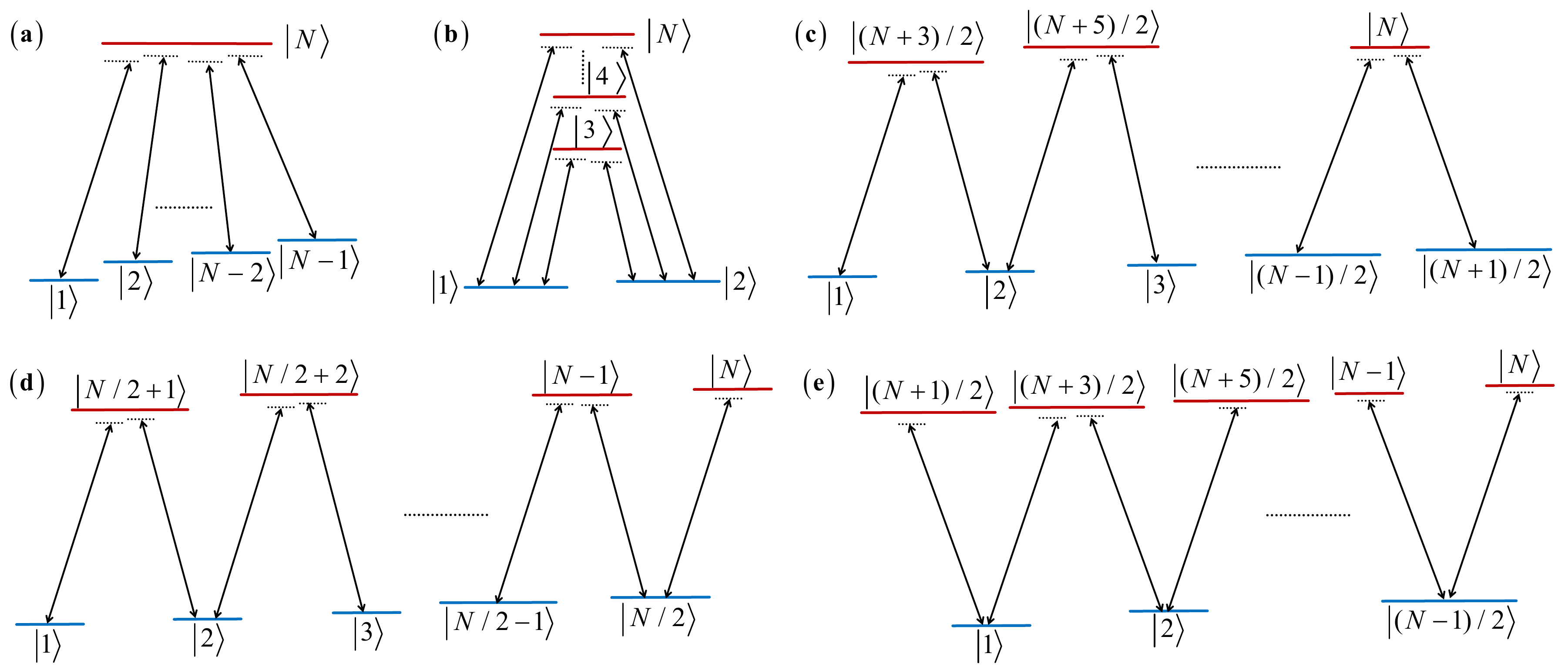}
	\caption{Schematic of five typical coupling configurations of the $N$-level quantum systems. (a) Configuration 1: multipod quantum system with one upper state and $N-1$ lower states. (b) Configuration 2: shared-lower-state multiple-$\Lambda$ system with $N-2$ upper states and two lower states. (c) Configuration 3: $\Lambda$-chain system with a zigzag coupling satisfying $N_{l}=N_{u}+1$. (d) Configuration 4: shared-edge $\mathrm{N}$-chain system with a zigzag coupling satisfying $N_{l}=N_{u}$. (e) Configuration 5: $\mathrm{V}$-chain system with a zigzag coupling satisfying $N_{l}=N_{u}-1$. The red (blue) lines denote the upper (lower) states, and the detunings are omitted for concision.}\label{N}
\end{figure*}

\section{Dark states in the $\boldsymbol{N}$-level quantum systems}\label{sec6}

In this section, we study the dark states in the $N$-level quantum systems. For the $N$-level systems, it is difficult to list all the coupling configurations. As a result, here we only present five typical configurations by classifying the upper and lower states: (a) $N_{u}=1$ and $ N_{l}=N-1$, (b) $N_{u}=N-2$ and $ N_{l}=2$, (c) $N_{l}=N_{u}+1 $, (d) $N_{l}=N_{u} $, and (e) $N_{l}=N_{u}-1 $. Here we consider the cases where there is no intracoupling within the upper- and lower-state subspaces.

\subsection{Configuration 1: $\boldsymbol{N_{u}=1$ and $ N_{l}=N-1}$}

We consider the multipod quantum system with one upper state (i.e., $\left\vert
u_{1}\right\rangle =\vert N\rangle $) and $N-1$ lower
states (i.e., $\vert l_{1}\rangle =\vert N-1\rangle ,$ $\vert l_{2}\rangle =\vert
N-2\rangle ,$ $\ldots,$ and $\vert
l_{N-1}\rangle =\vert 1\rangle$), and without the intracoupling among these lower states~[see Fig.~\ref{N}(a)]. The Hamiltonian of this $N$-level system can be written as
\begin{equation}
H^{[N,1]} =E_{N}\vert N\rangle \langle N\vert +\sum_{j=1}^{N-1}E_{j}\vert j\rangle \langle
j\vert  +\sum_{j=1}^{N-1}(\Omega _{jN}e^{-i\omega _{jN}t}\vert N\rangle \langle j\vert
+\text{H.c.}).
\end{equation}%
In a rotating frame with respect to $H_{0}= E_{N}\vert
N\rangle \langle N\vert
+\sum_{{j}=1}^{N-1}( E_{j}+\Delta _{jN}) \vert
j\rangle \langle j\vert $, the corresponding time-independent Hamiltonian reads
\begin{equation}
	\tilde{H}^{[N,1]}=\sum_{j=1}^{N-1}-\Delta _{jN}\left\vert j\right\rangle
	\left\langle j\right\vert +\sum_{j=1}^{N-1}(\Omega _{jN}\left\vert N\right\rangle
	\left\langle j\right\vert +\text{H.c.}) ,
\end{equation}%
where the detunings are introduced as $\Delta_{jN}=E_{N}-E_{j}-\omega _{jN}$ for $j=1,2,\ldots,N-1$.

It can be seen from Fig.~\ref{N}(a) that the physical model diagram can be described by a bipartite graph. To unify describe the system with the dressed upper and lower states, we define the basis vectors $\vert
U_{1}\rangle =\vert
u_{1}\rangle =\vert N\rangle =( 1,0,0,\ldots,0)
^{T},$ $\vert L_{1}\rangle =\vert
l_{1}\rangle =\vert N-1\rangle =(
0,1,0,\ldots,0) ^{T},$ $\vert L_{2}\rangle =\vert
l_{2}\rangle =\vert
N-2\rangle =( 0,0,1,\ldots,0) ^{T},$ $\ldots,$ and $\vert
L_{N-1}\rangle =\vert
l_{N-1}\rangle =\vert 1\rangle =( 0,0,0,\ldots,1)
^{T}$. Here, $\vert u_{j}\rangle$ and $\vert l_{j^{\prime}}\rangle$ denote the bare upper and lower states. In particular, these dressed states in the upper- and lower-state subspaces are actually these bare states, because there is no intracoupling within the two subspaces. Then the Hamiltonian with the dressed upper and lower states can be expressed as
\begin{equation}\label{HDN1}
\tilde{{H}}_{D}^{[N,1]}=\left( 
\begin{array}{c|cccc}
0 & \Omega _{(N-1)N} & \Omega _{(N-2)N} & \ldots & \Omega _{1N} \\ \hline
\Omega _{(N-1)N}^{\ast } & -\Delta _{(N-1)N} & 0 & \ldots & 0 \\ 
\Omega _{(N-2)N}^{\ast } & 0 & -\Delta _{(N-2)N} & \ldots & 0 \\ 
\ldots & \ldots & \ldots & \ldots & \ldots \\ 
\Omega _{1N}^{\ast } & 0 & 0 & \ldots & -\Delta _{1N}%
\end{array}%
\right) .
\end{equation}%
The dark states in the system can be analyzed based on the arrowhead matrix in Eq.~(\ref{HDN1}). 
Note that here we do not consider the case with the zero coupling strengths, because the corresponding configuration will be changed. 

Consider the case with degenerate dressed-lower-state subspace: $\Delta _{(N-1)N}=\Delta _{(N-2)N}=\ldots=\Delta _{1N}$, then these $N-1$ dressed lower states are degenerate and the number of the dark states is $N-2$~\cite{huang2023dark}.
Based on Eqs.~(\ref{BD}),~(\ref{B2D2}), and~(\ref{B3D3}), we can introduce these states
\begin{subequations}\label{BjDj}
	\begin{align} 
\vert B_{j^{\prime}}\rangle =&\frac{1}{\mathcal{N}_{ j^{\prime}} }( \mathcal{N}_{ j^{\prime}-1} \vert B_{j^{\prime}-1}\rangle +\mathbf{C}_{j^{\prime}+1}^{\ast }\vert L_{j^{\prime}+1}\rangle ), \\
\vert D_{j^{\prime}}\rangle =&\frac{1}{\mathcal{N}_{ j^{\prime}} }( \mathbf{C}_{j^{\prime}+1}\vert	B_{j^{\prime}-1}\rangle -\mathcal{N}_{ j^{\prime}-1} \vert L_{j^{\prime}+1}\rangle ), 
	\end{align} 
\end{subequations}
where $j^{\prime}=1,2,\ldots,N-2$ and $\mathcal{N}_{ j^{\prime}}=\sqrt{\sum_{i=1}^{j^{\prime}+1}\vert  \mathbf{C}_{i}\vert^{2}}$. 
These states satisfy the relation $\vert B_{N-2}\rangle
\langle B_{N-2}\vert+\sum_{i^{\prime}=1}^{N-2}\vert D_{i^{\prime}}\rangle \langle
D_{i^{\prime}}\vert  =\sum_{i=1}^{N-1}\vert L_{i}\rangle \langle L_{i}\vert $, where only the state $\vert B_{N-2}\rangle$ is coupled with the dressed upper state, and other $N-2$ states $\vert D_{j^{\prime}}\rangle $ are dark states. 

We can also consider the cases of partial degeneracy of these dressed lower states. For example, when $\Delta _{(N-1)N}=\Delta _{(N-2)N}=\ldots=\Delta _{(N-r)N}$ for $2\leq r\leq N-1$, namely, these $r$ dressed lower states $\{\vert L_{1}\rangle ,\vert L_{2}\rangle , \ldots, \vert L_{r}\rangle\}$ are degenerate, then there are $r-1$ dark states: $\vert D_{1}\rangle$, $\vert D_{2}\rangle$, \ldots, and $\vert D_{r-1}\rangle$.

\subsection{Configuration 2: $\boldsymbol{N_{u}=N-2$ and $ N_{l}=2}$}

We consider the shared-lower-state multiple-$\Lambda$ system with $N-2$ upper states (i.e., $\vert u_{1}\rangle =\vert N\rangle ,$ $\vert u_{2}\rangle =\vert N-1\rangle,$ \ldots, $\vert u_{N-1}\rangle =\vert 4\rangle $, and $\vert u_{N-2}\rangle =\vert
3\rangle $) and two lower states (i.e., $\vert l_{1}\rangle =\vert 2\rangle $ and $\vert
l_{2}\rangle =\vert 1\rangle $)~[see Fig.~\ref{N}(b)], the Hamiltonian can be written as 
\begin{eqnarray}
H^{[N,2]} &=&\sum_{j=3}^{N}E_{j}\left\vert j\right\rangle \left\langle
j\right\vert 
+E_{2}\left\vert 2\right\rangle \left\langle 2\right\vert
+E_{1}\left\vert 1\right\rangle \left\langle 1\right\vert  \notag \\
&&+\sum_{j=3}^{N}(\Omega _{1j}e^{-i\omega _{1j}t}\left\vert j\right\rangle \left\langle 1\right\vert
+\Omega _{2j}e^{-i\omega _{2j}t}\left\vert j\right\rangle \left\langle
2\right\vert +\text{H.c.}) .\notag \\
\end{eqnarray}%
In a rotating frame with respect to $H_{0}=\sum_{j=3}^{N}E_{j}\vert
j\rangle \langle j\vert+ (E_{2}+\Delta _{2})\vert
2\rangle \langle 2\vert+(E_{1}+\Delta _{1})\vert
1\rangle \langle 1\vert  ,$ the time-independent Hamiltonian can be obtained as
\begin{eqnarray}
\tilde{H}^{[N,2]}&=&-\Delta _{2}\left\vert 2\right\rangle \left\langle 2\right\vert
-\Delta _{1}\left\vert 1\right\rangle \left\langle
1\right\vert \notag \\
&&
+\sum_{j=3}^{N}(\Omega _{1j}\left\vert j\right\rangle \left\langle
1\right\vert +\Omega _{2j}\left\vert j\right\rangle \left\langle
2\right\vert +\text{H.c.}),
\end{eqnarray}%
where the detunings should satisfy $\Delta _{2}=\Delta _{2j}=E_{j}-E_{2}-\omega _{2j}$ and $\Delta _{1}=\Delta _{1j}=E_{j}-E_{1}-\omega _{1j}$ for $j=3,4,\ldots,N$. Similarly, here we can directly define the following basis vectors $\vert U_{1}\rangle =\vert u_{1}\rangle =\vert N\rangle =(
1,0,\ldots,0,0,0) ^{T},$ $\vert U_{2}\rangle =\vert u_{2}\rangle =\vert N-1\rangle =(
0,1,\ldots,0,0,0) ^{T},$ $\ldots,$ $\vert U_{N-2}\rangle =\vert u_{N-2}\rangle =\vert
3\rangle =( 0,0,\ldots,1,0,0) ^{T},$ $\vert L_{1}\rangle
=\vert l_{1}\rangle= \vert 2\rangle =( 0,0,\ldots,0,1,0) ^{T},$ and $\vert
L_{2}\rangle =\vert
l_{2}\rangle =\vert 1\rangle =( 0,0,\ldots,0,0,1)
^{T}$. Then the Hamiltonian with dressed upper and lower states can be expressed as
\begin{equation}
\tilde{{H}}_{D}^{[N,2]}=\left( 
\begin{array}{cccc|cc}
0 & 0 & \ldots & 0 & \Omega _{2N} & \Omega _{1N} \\ 
0 & 0 & \ldots & 0 & \Omega _{2(N-1)} & \Omega _{1(N-1)} \\ 
\ldots & \ldots & \ldots & \ldots & \ldots & \ldots \\ 
0 & 0 & \ldots & 0 & \Omega _{23} & \Omega _{13} \\ \hline
\Omega _{2N}^{\ast } & \Omega _{2(N-1)}^{\ast } & \ldots & \Omega _{23}^{\ast }
&- \Delta _{2} & 0 \\ 
\Omega _{1N}^{\ast } & \Omega _{1(N-1)}^{\ast } & \ldots & \Omega _{13}^{\ast }
& 0 & -\Delta _{1}%
\end{array}%
\right).
\end{equation}%
According to the arrowhead-matrix method, the dark state only exists when the dressed lower states are
degenerate $\Delta _{1}=\Delta _{2}$, namely, $\Delta _{1j}=\Delta _{2j}$. When the two
coupling vectors $\mathbf{C}_{2}$ and $\mathbf{C}_{1}$ are linearly
dependent, i.e., $\mathbf{C}_{2}=\gamma \mathbf{C}_{1}$, there exists a dark state
\begin{equation}
\left\vert D^{[N,2]}_{1}\right\rangle =\frac{1}{\sqrt{1+\vert \gamma \vert  ^{2}}}(
\gamma\left\vert L_{1}\right\rangle - \left\vert L_{2}\right\rangle )=\frac{1}{\sqrt{1+\vert \gamma \vert  ^{2}}}(
\gamma\left\vert l_{1}\right\rangle - \left\vert l_{2}\right\rangle ).
\end{equation}
We mention that this state expressed by the dressed lower states has the same form as that in Eq.~(\ref{D521}) for the five-level system with two degenerate dressed lower states.

\subsection{Configuration 3: $\boldsymbol{N_{l}=N_{u}+1}$}\label{expanded M-type}

We consider the $\Lambda$-chain system with a zigzag coupling configuration and satisfying $N_{l}=N_{u}+1$ (i.e., $\vert u_{1}\rangle =\vert N\rangle,$ $\vert u_{2}\rangle =\vert N-1\rangle,$ $\ldots,$ $\vert u_{(N-1)/2}\rangle =\vert (N+3)/2\rangle $, $\vert l_{1}\rangle =\vert (N+1)/2 \rangle $, $\vert l_{2}\rangle =\vert (N-1)/2 \rangle $, $\ldots$, and $\vert
l_{(N+1)/2}\rangle =\vert 1\rangle $)~[see Fig.~\ref{N}(c)], then the Hamiltonian can be written as
\begin{eqnarray}\label{HN3}
	H^{[N,3]} &=& \sum_{j^{\prime }=(N+3)/2}^{N}E_{j^{\prime }}\vert
	j^{\prime }\rangle \langle j^{\prime }\vert +\sum_{j=1}^{(N+1)/2}E_{j}\left\vert j\right\rangle \left\langle
	j\right\vert \notag \\
	&&+\sum_{j^{\prime }=(N+3)/2}^{N}[\Omega _{[ j^{\prime }-(N+1)/2] j^{\prime }}  \vert j^{\prime	}\rangle \langle j^{\prime }-(N+1)/2\vert  \notag \\
	&&\times e^{-i\omega _{[j^{\prime }-(N+1)/2] j^{\prime }}t}+\Omega
	_{[ j^{\prime }-(N-1)/2] j^{\prime }}  \vert j^{\prime
	}\rangle \langle j^{\prime }-(N-1)/2\vert \notag \\ 
	&&\times e^{-i\omega _{[j^{\prime }-(N-1)/2] j^{\prime }}t}+\text{H.c.}].
\end{eqnarray}%
In a rotating frame with respect
to $H_{0}=\sum_{j^{\prime
	}=(N+3)/2}^{N}E_{j^{\prime }}|j^{\prime }\rangle \langle j^{\prime }\vert$ $+(E_{(N+1)/2}+\Delta _{[(N+1)/2]N})\vert (N+1)/2\rangle \langle
(N+1)/2\vert  + \sum_{i=2}^{(N-1)/2}(E_{i}+\Delta _{i[i+(N+1)/2]})\vert
i\rangle \langle i\vert +(E_{1}+\Delta _{1[(N+3)/2]})\vert 1\rangle \langle
1\vert$, the transformed time-independent Hamiltonian reads
\begin{eqnarray}\label{HN3_t}
	\tilde{H}^{[N,3]} &=&-\Delta _{[(N+1)/2]N}\left\vert (N+1)/2\right\rangle \left\langle(N+1)/2	\right\vert    \notag 	 \\
	&&-\sum\limits_{i=2}^{(N-1)/2}\Delta _{i[
		i+(N+1)/2] }\left\vert i\right\rangle \left\langle i\right\vert - \Delta _{1[ (N+3)/2] }\left\vert 1\right\rangle
		\left\langle 1\right\vert \notag 	 \\
	&&+\sum_{j^{\prime }=(N+3)/2}^{N}[\Omega _{[j^{\prime }-(N+1)/2]j^{\prime }}|j^{\prime }\rangle \langle j^{\prime }-(N+1)/2|\notag
	\\
	&&+\Omega _{[j^{\prime }-(N-1)/2]j^{\prime
	}}|j^{\prime }\rangle \langle j^{\prime }-(N-1)/2|+\text{H.c.}],
\end{eqnarray}%
with the resonance conditions $\Delta _{i[ i+(N-1)/2] }=\Delta _{i[i+(N+1)/2] }$ for $i=2,3,\ldots,(N-1)/2$. By defining the basis vectors $\vert U_{1}\rangle =\vert u_{1}\rangle =\vert N\rangle =(
1,0,\ldots,0,0,0,\ldots,0) ^{T},$ $\vert U_{2}\rangle =\vert u_{2}\rangle =\vert N-1\rangle =(
0,1,\ldots,0,0,0,\ldots,0) ^{T},$ $\ldots,$ $\vert U_{(N-1)/2}\rangle =\vert u_{(N-1)/2}\rangle =\vert (N+3)/2\rangle =(
0,0,\ldots,1_{({N-1})/{2}},0,0,\ldots,0) ^{T},$ $\vert L_{1}\rangle =\vert l_{1}\rangle =\vert
(N+1)/2 \rangle =( 0,0,\ldots,0,1_{({N+1})/{2}},0,\ldots,0) ^{T},$ $\vert L_{2}\rangle =\vert l_{2}\rangle =\vert
(N-1)/2 \rangle =( 0,0,\ldots,0,0,1_{({N+3})/{2}},\ldots,0) ^{T},$ $\ldots,$ and $\vert
L_{(N+1)/2}\rangle =\vert
l_{(N+1)/2}\rangle =\vert 1\rangle =( 0,0,\ldots,0,0,0,\ldots,1_{N}) ^{T}$,
the Hamiltonian $\tilde{{H}}_{D}^{[N,3]}$ can be expressed as
\begin{equation}\label{HDN3}
	\tilde{{H}}_{D}^{[N,3]}=\left( 
	\begin{array}{c|c}
		\mathbf{H}_{U} & \mathbf{C} \\ \hline
		\mathbf{C}^{\dag } & \mathbf{H}_{L}%
	\end{array}%
	\right),
\end{equation}%
where the submatrices are given by
\begin{subequations}\label{HeM}
	\begin{align}
		\mathbf{H}_{U} =&\mathbf{0}_{[(N-1)/2]\times [(N-1)/2]}, \\
		\mathbf{H}_{L}=&\text{diag}(-\Delta _{[(N+1)/2]N},-\Delta _{[(N-1)/2]
			N},\ldots,-\Delta _{i[i+(N+1)/2 ]}, \notag \\
			&\ldots,-\Delta _{2[ (N+5)/2 ]},-\Delta _{1[ (N+3)/2] }) , \\
			\mathbf{C} =& \scalebox{0.83}{$\displaystyle
				\begin{pmatrix}
					\Omega_{[(N+1)/2]N} & \Omega_{[(N-1)/2]N} & \ldots & 0 & 0 & 0 \\ 
					0 & \Omega_{[(N-1)/2](N-1)} & \ldots & 0 & 0 & 0 \\ 
					\ldots & \ldots & \ldots & \ldots & \ldots & \ldots \\ 
					0 & 0 & \ldots & \Omega_{3[(N+5)/2]} & \Omega_{2[(N+5)/2]} & 0 \\ 
					0 & 0 & \ldots & 0 & \Omega_{2[(N+3)/2]} & \Omega_{1[(N+3)/2]}
				\end{pmatrix}$}\label{CN3}. 
	\end{align}%
\end{subequations}

When these dressed lower states are degenerate $\Delta _{[(N+1)/2]N}=\Delta _{i[ 	i+(N+1)/2] }=\Delta _{1[(N+3)/2] }=\Delta $ with $i=2,3,$ $\ldots,(N-1)/2$, the dark state can be
analyzed with the above coupling matrix. Theoretically, we can obtain the orthogonal bright and dark states by calculating the SVD of the coupling matrix $\mathbf{C}$. However, it is difficult to obtain the analytical result of the SVD for such an $[(N-1)/2]\times[(N+1)/2]$ matrix. Therefore, here we solve the null space of the coupling matrix $\mathbf{C}$ to obtain the dark states. Since the coupling matrix in Eq.~(\ref{CN3}) is a row-echelon matrix with rank $r=(N-1)/2$, the number of the dark state, which is equal to the dimension of the null space of the coupling matrix, is one. We assume that the dark state composed of all the dressed lower states can be expressed as
\begin{equation}\label{D_def}
\left\vert D\right\rangle =\sum_{i=0}^{ (N-1)/2}x_{i}\left\vert
L_{i+1}\right\rangle.
\end{equation}%
By solving the null space equation $ \mathbf{C}\vert D\rangle=\mathbf{0}$, we obtain the relations of the undetermined coefficients 
\begin{equation}
	\Omega _{[(N+1)/2-i](N-i)}x_{i}+\Omega _{[(N-1)/2-i](N-i)}x_{i+1}=0,
\end{equation}
with $i=0,1,\ldots,(N-3)/2$. Based on these relations, the unique dark state in this system can be obtained as%
\begin{eqnarray}\label{N-M}
\left\vert D^{[N,3]}_{1}\right\rangle  &=&\sum_{i=0}^{(N-1)/2}\frac{(-1)^{(N-1)/2-i}}{\mathcal{N}%
	_{1}^{[N,3]}}\notag \\
	&&\times \prod_{j=i}^{(N-3)/2}\frac{\Omega _{[(N-1)/2-j](N-j)}}{\Omega
	_{[(N+1)/2-j](N-j)}}\left\vert L_{i+1}\right\rangle,
	\end{eqnarray}%
where $\mathcal{N}^{[N,3]}_{1}$ is the normalization constant.

\subsection{Configuration 4: $\boldsymbol{N_{l}=N_{u}}$}

We consider the shared-edge $\mathrm{N}$-chain system (namely a chain of the letter $\mathrm{N}$-type configuration with shared edges) with a zigzag coupling and satisfying $N_{l}=N_{u}=N/2$ (i.e., $\vert u_{1}\rangle =\vert N\rangle,$ $\vert u_{2}\rangle =\vert N-1\rangle,$ $\ldots,$ $\vert u_{N/2}\rangle =\vert N/2+1\rangle $, $\vert l_{1}\rangle =\vert N/2 \rangle $, $\vert l_{2}\rangle =\vert N/2-1 \rangle $, $\ldots$, and $\vert
l_{N/2}\rangle =\vert 1\rangle $), note that $N$ is an even number in this case~[see Fig.~\ref{N}(d)]. Then the Hamiltonian of this system can be written as 
\begin{eqnarray}
	H^{[N,4]} &=&\sum_{j^{\prime }=N/2+1}^{N}E_{j^{\prime }}|j^{\prime }\rangle
	\langle j^{\prime }|+\sum_{j=1}^{N/2}E_{j}\left\vert j\right\rangle
	\left\langle j\right\vert     \notag \\
	&&+\sum\limits_{i^{\prime}=2}^{N/2}[\Omega _{i^{\prime}( N/2+i^{\prime}-1) }e^{-i\omega _{i^{\prime}(
			N/2+i^{\prime}-1) }t} \vert
	 N/2+i^{\prime}-1 \rangle \langle i^{\prime}\vert  \notag \\
	&&+\Omega _{i^{\prime}( N/2+i^{\prime} ) }e^{-i\omega _{i^{\prime}( N/2+i^{\prime} ) }t}\vert N/2+i^{\prime} \rangle \langle
	i^{\prime}\vert +\text{H.c.}]\notag \\
	&&+[\Omega _{1( N/2+1) }e^{-i\omega _{1( N/2+1) }t}\left\vert N/2+1\right\rangle
	\left\langle 1\right\vert +\text{H.c.}]  .
\end{eqnarray}%
In a rotating frame with respect to $%
H_{0}=\sum_{j^{\prime }=N/2+1}^{N}E_{j^{\prime
}}|j^{\prime }\rangle \langle j^{\prime }|+\sum_{j=1}^{N/2}(E_{j}+\Delta _{j(N/2+j )})\left\vert j\right\rangle
\left\langle j\right\vert ,$ the transformed
time-independent Hamiltonian reads 
\begin{eqnarray}
	\tilde{H}^{[N,4]} &=&\sum\limits_{j=1}^{N/2}-\Delta _{j( N/2+ j)
	}\left\vert j\right\rangle \left\langle j\right\vert +\sum\limits_{i^{\prime}=2}^{N/2}[\Omega _{i^{\prime}(N/2+ i^{\prime} ) }\vert N/2+i^{\prime} \rangle \langle
	i^{\prime}\vert    \notag \\
	&& + \Omega _{i^{\prime}(  N/2 +i^{\prime}-1) }\vert
	N/2+i^{\prime}-1 \rangle \langle i^{\prime}\vert +\text{H.c.}] \notag \\
	&&
	+[ \Omega
	_{1( N/2+1) }\left\vert N/2+1\right\rangle \left\langle 1\right\vert +%
	\text{H.c.}],
\end{eqnarray}%
with the resonance condition $\Delta _{i^{\prime}( N/2 +i^{\prime}-1) }=\Delta
_{i^{\prime}( N/2+i^{\prime} ) }$ for $i^{\prime}=2,3,\ldots,N/2$.
By defining the following basis vectors: $\vert U_{1}\rangle = \vert u_{1}\rangle =\vert N\rangle =(
1,0,\ldots,0,0,0,\ldots,0) ^{T},$ $\vert U_{2}\rangle = \vert u_{2}\rangle =\vert N-1\rangle =(
0,1,\ldots,0,0,0,\ldots,0) ^{T},$ $\ldots,$ $\vert U_{N/2}\rangle = \vert u_{N/2}\rangle =\vert N/2+1\rangle =( 0,0,\ldots,1_{N/2},0,0,\ldots,0) ^{T},$ $\vert L_{1}\rangle = \vert l_{1}\rangle =\vert N/2 \rangle =( 0,0,\ldots,0,1_{N/2+1},0,\ldots,0) ^{T},$ $\vert L_{2}\rangle = \vert l_{2}\rangle =\vert N/2-1 \rangle =( 0,0,\ldots,0,0,1_{N/2+2},\ldots,0) ^{T},$ $\ldots,$ and $\vert L_{N/2}\rangle = \vert l_{N/2}\rangle =\vert 1\rangle =( 0,0\ldots,0,0,0,\ldots,1_{N}) ^{T}$,
the Hamiltonian $\tilde{{H}}_{D}^{[N,4]}$ can be expressed as
\begin{equation}
	\tilde{{H}}_{D}^{[N,4]}=\left( 
	\begin{array}{c|c}
		\mathbf{H}_{U} & \mathbf{C} \\ \hline
		\mathbf{C}^{\dag } & \mathbf{H}_{L}%
	\end{array}%
	\right),
\end{equation}%
where these submatrices are given by%
\begin{subequations}\label{HN4D}
	\begin{align}
\mathbf{H}_{U} =&\mathbf{0}_{( N/2) \times ( N/2) }, \\
\mathbf{H}_{L}=& \text{diag}(-\Delta _{ ( N/2)N},\ldots,-\Delta _{j( N/2+ j)
},\ldots,-\Delta _{1( N/2 +1) }), \\
\mathbf{C}=& \small{\left( 
\begin{array}{cccccc}
	\Omega _{(N/2)N} & 0 &  \ldots & 0 & 0 \\ 
	\Omega _{(N/2)( N-1) } & \Omega _{( N/2-1) ( N-1)
	} & \ldots & 0 & 0 \\ 
	\ldots & \ldots & \ldots &  \ldots & \ldots \\ 
	0 & 0 &  \ldots & \Omega _{2( N/2+2) } & 0 \\ 
	0 & 0 &  \ldots & \Omega _{2( N/2+1) } & \Omega _{1(
		N/2+1) }%
\end{array}%
\right)} .
	\end{align}%
\end{subequations}
When the dressed lower states are degenerate $\Delta _{j( N/2+j)
}=\Delta $ for $j=1,2,\ldots,N/2$, the dark state can be analyzed with the 
coupling matrix in Eq.~(\ref{HN4D}).
In this case, the coupling matrix $\mathbf{C}$ is an ${(N/2)\times( N/2)}$ lower triangular matrix, therefore, when all the matrix elements on the main diagonal are nonzero, it is full rank. As a result, for nonzero $\Omega_{(N/2+1-j)(N+1-j)}$, there are $N/2$ bright states coupled with the dressed upper states, and there is no dark state in the shared-edge $\mathrm{N}$-chain system.

\subsection{Configuration 5: $\boldsymbol{N_{l}=N_{u}-1}$}

We consider the $\mathrm{V}$-chain system with a zigzag coupling and satisfying $N_{l}=N_{u}-1$ (i.e., $\vert u_{1}\rangle =\vert N\rangle,$ $\vert u_{2}\rangle =\vert N-1\rangle,$  $\ldots,$ $\vert u_{(N+1)/2}\rangle =\vert (N+1)/2\rangle $, $\vert l_{1}\rangle =\vert (N-1)/2 \rangle $, $\ldots$, and $\vert	l_{(N-1)/2}\rangle =\vert 1\rangle $)~[see Fig.~\ref{N}(e)], then the Hamiltonian of this system can be written as 
\begin{eqnarray}
H^{[N,5]} &=&\sum\limits_{j^{\prime }=(N+1)/2}^{N}E_{j^{\prime }}\vert
j^{\prime }\rangle \langle j^{\prime }\vert
+\sum\limits_{j=1}^{(N-1)/2}E_{j}\vert j\rangle \langle
j\vert   \notag \\
&&+\sum\limits_{j=1}^{(N-1)/2}[\Omega _{j( j+(N-1)/2) } e^{-i\omega _{j(	j+(N-1)/2) }t}\vert j+(N-1)/2\rangle \langle j\vert  \notag \\
&&+\Omega _{j( j+(N+1)/2) }e^{-i\omega _{j( j+(N+1)/2) }t}\vert j+(N+1)/2\rangle \langle
j\vert +\text{H.c.}].\notag \\
\end{eqnarray}%
In a rotating frame with respect
to $H_{0}=\sum_{j^{\prime }=(N+1)/2}^{N}E_{j^{\prime
}}|j^{\prime }\rangle \langle j^{\prime }|$ $+\sum_{j=1}^{(N-1)/2}(E_{j}+\Delta _{j[j+(N-1)/2]})\left\vert j\right\rangle
\left\langle j\right\vert $, the transformed
time-independent Hamiltonian reads
\begin{eqnarray}
\tilde{H}^{[N,5]} &=&\sum\limits_{j=1}^{(N-1)/2}-\Delta _{j[ j+(N-1)/2]
}\left\vert j\right\rangle \left\langle j\right\vert \notag \\
&& +\sum\limits_{j=1}^{(N-1)/2}[\Omega _{j[j+(N-1)/2] }\left\vert j+(N-1)/2\right\rangle \left\langle j\right\vert   \notag \\
&&+\Omega _{j[ j+(N+1)/2] }\left\vert j+(N+1)/2\right\rangle \left\langle
j\right\vert +\text{H.c.}],
\end{eqnarray}%
with the resonance condition $\Delta _{j[ j+(N-1)/2] }=\Delta _{j[
j+(N+1)/2] }$ for $j=1,2,\ldots,(N-1)/2.$ Similarly, by defining the following basis vectors: $%
\vert U_{1}\rangle = \vert u_{1}\rangle =\vert N\rangle
=(1,0,\ldots,0,0,0,\ldots,0)^{T},$ $\vert U_{2}\rangle = \vert u_{2}\rangle =\vert
N-1\rangle =(0,1,\ldots,0,0,0,\ldots,0)^{T},$ $\ldots,$ $\vert U_{(N+1)/2}\rangle = \vert u_{(N+1)/2}\rangle =\vert (N+1)/2\rangle =(
0,0,\ldots,1_{(N+1)/2},0,0,\ldots,0) ^{T},$ $\vert L_{1}\rangle = \vert l_{1}\rangle=\vert
(N-1)/2 \rangle =( 0,0,\ldots,0,1_{(N+3)/2},0,\ldots,0) ^{T},$ $\vert L_{2}\rangle =\vert l_{2}\rangle =\vert
(N-3)/2 \rangle =( 0,0,\ldots,0,0,1_{(N+5)/2},\ldots,0) ^{T},$ $\ldots,$ and $\vert
L_{(N-1)/2}\rangle =\vert
l_{(N-1)/2}\rangle =\vert 1\rangle =( 0,0,\ldots,0,0,0,\ldots,1_{N}) ^{T}$, the
Hamiltonian $\tilde{{H}}_{D}^{[N,5]}$ can be expressed as
\begin{equation}
\tilde{{H}}_{D}^{[N,5]}=\left( 
\begin{array}{c|c}
	\mathbf{H}_{U} & \mathbf{C} \\ \hline
	\mathbf{C}^{\dag } & \mathbf{H}_{L}%
\end{array}%
\right) ,
\end{equation}%
where these submatrices are given by 
\begin{subequations}
\begin{align}
	\mathbf{H}_{U}=& \mathbf{0}_{[(N+1)/2]\times [(N+1)/2]}, \\
	\mathbf{H}_{L}=& \text{diag}(-\Delta _{[(N-1)/2]( N-1) },\ldots,-\Delta
	_{j[ j+(N-1)/2] },\ldots,-\Delta _{1[ (N+1)/2] }), \\
	\mathbf{C}=& \small{\left( 
	\begin{array}{cccccc}
		\Omega _{[(N-1)/2]N} & 0 &  \ldots & 0 & 0 \\ 
		\Omega _{[(N-1)/2]( N-1) } & \Omega _{[ (N-3)/2 ]( N-1)
		} &  \ldots & 0 & 0 \\ 
		0 & \Omega _{[(N-3)/2] ( N-2) } &  \ldots & 0 & 0 \\ 
		\ldots & \ldots & \ldots &  \ldots & \ldots \\ 
		0 & 0 &  \ldots & \Omega _{2[(N+3)/2 ]} & \Omega _{1[
			(N+3)/2 ]} \\ 
		0 & 0 &  \ldots & 0 & \Omega _{1[(N+1)/2] }%
	\end{array}%
	\right) .}
\end{align}%
\end{subequations}
When $\Delta _{j( j+(N-1)/2) }=\Delta $ for $j=1,2,\ldots,(N-1)/2,$ these
dressed lower states are degenerate. Since the numbers of the dressed upper
and lower states satisfy $N_{l}=N_{u}-1$, the coupling matrix $\mathbf{C}$
is an $[(N+1)/2]\times [(N-1)/2]$ matrix, whose transpose has the same form as that in
Eq.~(\ref{HeM}). Therefore, the matrix $\mathbf{C}
$ is full rank $r=(N-1)/2$, and then there is no dark state in the $\mathrm{V}$-chain system.

Note that if we define the states $\{\vert
1\rangle,\vert 2\rangle,\ldots,\vert (N-1)/2\rangle\}$ as the upper states, and other states $\{\vert (N+1)/2\rangle,\vert (N+3)/2\rangle,$ $\ldots,\vert N\rangle\}$ as the lower states, then configuration $5$ ($\mathrm{V}$-chain system) is reduced to configuration $3$ ($\Lambda$-chain system) in Sec.~\ref{expanded M-type}.
	
\begin{figure*}[t!]
	\centering\includegraphics[width=0.85\textwidth]{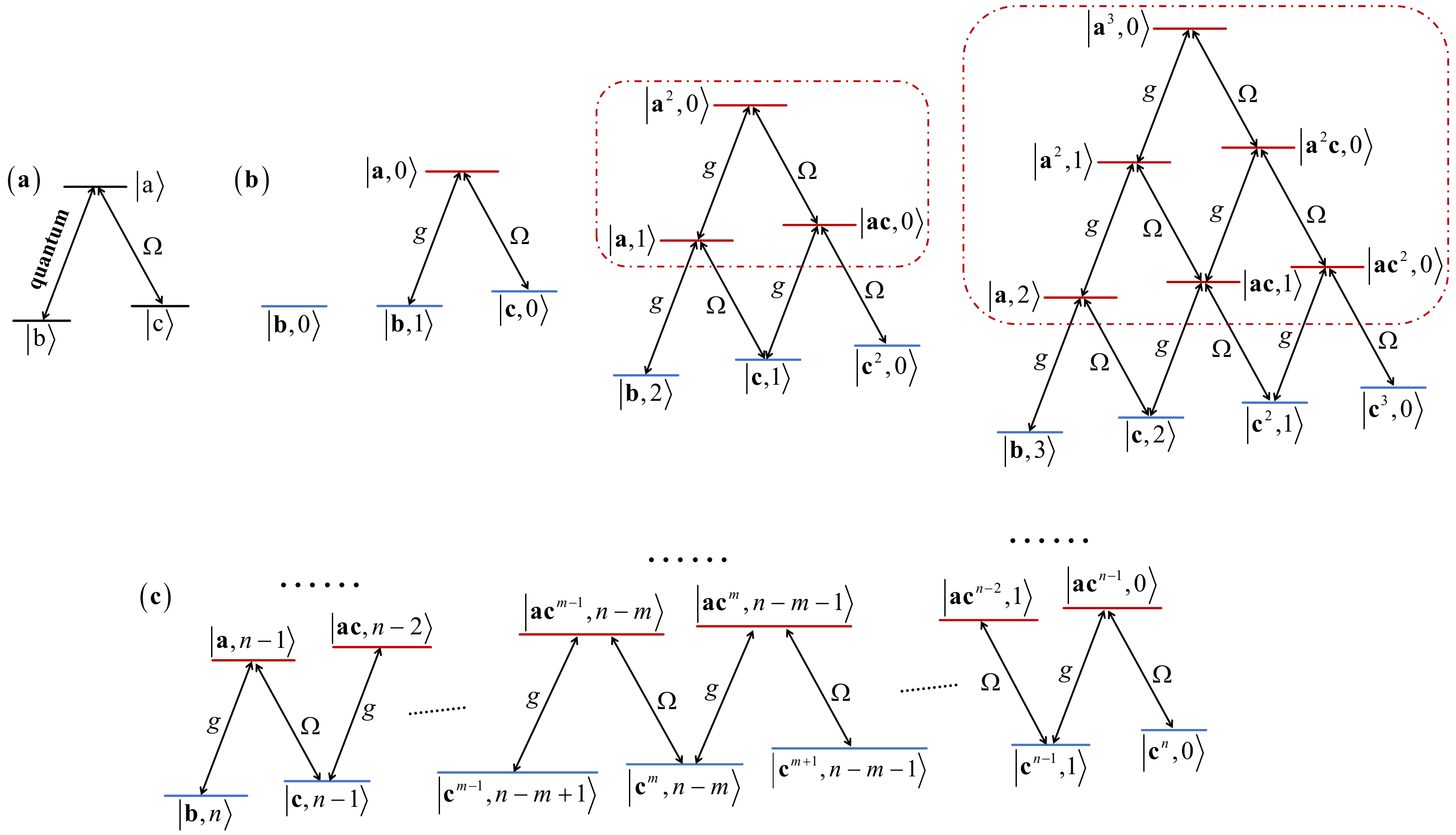}
	\caption{(a) Schematic of a $\Lambda$-type three-level atom coupled to a single-mode quantum field with coupling strength $g$ and a classical control field with the Rabi frequency $\Omega ( t)$. (b) The transition diagram of the bare states with excitation number at most three. The red (blue) lines denote the upper (lower) states. (c) The reduced transition diagram of the bare states with excitation number $n$, which only contain the directly couplings between the upper states and lower states.}\label{DSP}
\end{figure*}
\section{Rederivation of the dark-state polaritons in driven three-level systems with the arrowhead-matrix method}
\label{DSp}

In previous sections, we have discussed the dark states in some typical multilevel quantum systems in detail. In this section, we recover the results of the dark-state polaritons in driven three-level systems~\cite{DSP:PRA2002,DPiEIT2000} with the arrowhead-matrix method.

We adopt the same model in Refs.~\cite{DSP:PRA2002,DPiEIT2000}, which describes an ensemble of $N$ $\Lambda$-type three-level atoms with an excited state $\vert a\rangle$ and two metastable lower states $\vert b\rangle$ and $\vert c\rangle$~[see Fig.~\ref{DSP}(a)]. In particular, the system is resonantly driven by two single-mode fields: the transition $\vert a\rangle \leftrightarrow	 \vert b\rangle $ is driven by a single-mode quantum field with coupling strength $g$, while the transition $\vert a\rangle \leftrightarrow	 \vert
c\rangle $ is coupled with a classical control field associated with the Rabi
frequency $\Omega ( t)$ and frequency $v$. Here, all the frequencies and coupling strengths are assumed to be equal for all atoms for simplicity. The Hamiltonian of this system can be expressed by%
\begin{eqnarray}
	H^{\mathrm{DSP}} &=&\omega _{a}\hat{a}^{\dag }\hat{a}+ \sum_{i=1}^{N}(E_{a_{i}}\left\vert
	a_{i}\right\rangle \left\langle a_{i}\right\vert +E_{b_{i}}\left\vert
	b_{i}\right\rangle \left\langle b_{i}\right\vert +E_{c_{i}}\left\vert
	c_{i}\right\rangle \left\langle c_{i}\right\vert) \notag  \\
	&&+\sum_{i=1}^{N}( g\hat{a}\left\vert a_{i}\right\rangle
	\left\langle b_{i}\right\vert  +\Omega ( t)
	e^{-ivt}\left\vert a_{i}\right\rangle \left\langle c_{i}\right\vert 
	+\text{H.c.}),
\end{eqnarray}%
where $\omega _{a}$ is the resonance frequency of the single-mode quantum field described by the creation (annihilation) operator $\hat{a}^{\dag }$ $(\hat{a})$, $E_{\mu_i}$ is the energy of the state $\vert \mu \rangle$ ($\mu=a,b, $ and $c$) in the $i$th atom, and $\vert \mu_i \rangle \langle \mu^{\prime}_{i}\vert$ is the atomic transition operator  between the states $\vert \mu \rangle $ and $\vert\mu^{\prime}\rangle $ of the $i$th atom.
The Hamiltonian in the interaction picture can be expressed as
\begin{equation}
\tilde{H}^{{\mathrm{DSP}}}=\sum_{i=1}^{N} (g \hat{a}\left\vert a_{i}\right\rangle
\left\langle b_{i}\right\vert  +\Omega ( t)\left\vert a_{i}\right\rangle \left\langle c_{i}\right\vert
+\text{H.c.}).
\end{equation}%
The total excitation number operator $N^{\mathrm{DSP}}=\sum_{i=1}^{N}(\vert
a_{i}\rangle \langle a_{i}\vert+ \vert
c_{i}\rangle \langle c_{i}\vert)+\hat{a}^{\dag }\hat{a}$ of the system is a conserved quantity due to $[H^{{\mathrm{DSP}}},N^{\mathrm{DSP}}]=0$. Therefore, we can obtain the transition diagram in different excitation-number subspaces according to the Hamiltonian.

Since we focus on the dark state composed of the metastable lower states $\vert b\rangle$ and $\vert c\rangle$ relative to the excited states $\vert a\rangle$, here we define the states containing the excited state as the upper states [as marked by the red lines in Fig.~\ref{DSP}(b)], and the states only containing the atomic lower states are defined as the lower states of the system [as marked by the blue lines in Fig.~\ref{DSP}(b)]. In particular, we find that only part of the upper states are directly coupled to the lower states, therefore, only the coupling submatrix corresponding to this part is valid for analyzing the dark-state effect while other rows are zero vectors. As a result, the transition diagram can be reduced to $\Lambda$-chain configuration with a zigzag coupling in Sec.~\ref{expanded M-type}~[see Fig.~\ref{DSP}(c)]. 
The basis states in the $n$-excitation subspace of the reduced model can be introduced as%
\begin{widetext}
\begin{subequations}\label{OrderSum}
	\begin{align}
\left\vert \mathbf{b},n\right\rangle  =&\left\vert
		b_{1}, b_{2},\ldots,b_{N},n\right\rangle , \\
\left\vert \mathbf{ac}^{m-1},n-m\right\rangle  =&\frac{1%
}{\sqrt{ (m-1) !A_{N}^{m}}}%
\sum_{k_{1}\neq k_{2}\neq\ldots \neq k_{m}=1}^{N}\left\vert
b_{1},\ldots,a_{k _{1}},\ldots,c_{k _{2}},\ldots,c_{k
_{3}},\ldots,c_{k _{m}},\ldots,b_{N},n-m\right\rangle
,   \\ 
\left\vert \mathbf{c}^{m},n-m\right\rangle  =&\frac{1}{\sqrt{m!A_{N}^{m}}}%
\sum_{k_{1}\neq k_{2}\neq\ldots \neq k_{m}=1}^{N}\left\vert
b_{1},\ldots,c_{k _{1}},\ldots,c_{k _{2}},\ldots,c_{k
_{3}},\ldots,c_{k _{m}},\ldots,b_{N},n-m\right\rangle ,
	\end{align}%
\end{subequations}
\end{widetext}
where $m=1,2,\ldots,n$, $A_{N}^{m}=N!/(N-m)!$ is the permutation number, and the boldfaced states (i.e., $ \vert \mathbf{b}\rangle,\vert \mathbf{ac}^{m-1}\rangle, $ and $\vert \mathbf{c}^{m}\rangle$) represent collective states of the ensemble of $N$ three-level atoms. In particular, the state $\vert \mathbf{b}, n\rangle$ is identical to the state $\vert \mathbf{c}^{m},n-m\rangle$ in the case of $m=0$, namely, $ \vert \mathbf{c}^{0},n\rangle=\vert \mathbf{b}, n\rangle$.
Note that we adopted the ordered summation in Eqs.~(\ref{OrderSum}) and there is an additional factorial term related to the repetitive cases.

Similarly, we arrange these basis states and define the basis vectors in order: $\vert U_{m}\rangle= \vert u_{m}\rangle= \vert \mathbf{ac}^{m-1},n-m\rangle=(0,0,\ldots,1_{m},\ldots,0,0,0,\ldots,0)^{T} $ for $m=1,2,\ldots,n$, and $\vert L_{m^{\prime}+1}\rangle= \vert l_{m^{\prime}+1}\rangle= \vert \mathbf{c}^{m^{\prime}},n-m^{\prime}\rangle =(0,0,\ldots,0,0,0,\ldots,1_{m+m^{\prime}+1},\ldots,0)^{T}$ for $m^{\prime}=0,1,\ldots,n$, then the Hamiltonian can be expressed as a thick arrowhead matrix. The upper- and lower-state submatrices are zero matrices, which means the dressed lower states are degenerate, and the coupling matrix has the same form as the row-echelon matrix in Eq.~(\ref{CN3}), which can be obtained as
\begin{subequations}
	\begin{align}
		&\left\langle \mathbf{ac}^{m-1},n-m\right\vert \tilde{H}^{{\mathrm{DSP}}}\left\vert \mathbf{c}%
		^{m-1},n-m+1\right\rangle  \notag \\ &={g\sqrt{N-m+1}\sqrt{n-m+1}},\label{gN}\\
		&\left\langle \mathbf{ac}^{m-1},n-m\right\vert \tilde{H}^{{\mathrm{DSP}}}\left\vert \mathbf{c}%
		^{m},n-m\right\rangle  =\sqrt{m}\Omega ( t) ,
	\end{align}%
\end{subequations}
with $m=1,2,\ldots,n$. We consider the case where the excitation number $n$ is  much less than the atom number $N$, therefore the result in Eq.~(\ref{gN}) can then be approximately rewritten as $g\sqrt{N}{\sqrt{n-m+1}}$. By introducing the mixing angle $\theta ( t)$ by 
\begin{equation}
	\tan \theta ( t) ={  g\sqrt{N}}/{\Omega ( t)  },
\end{equation}
the coupling matrix can be expressed as
\begin{equation}
	{\frac{\mathbf{C}}{{\tilde{\Omega }( t)}}}
	=\scalebox{0.84}{$
	\left( 
	\begin{array}{ccccccc}
		{\sqrt{n}\sin \theta}  & {\cos \theta}  & 0 & 0 & 0 & 0 \\ 
		0 & \ldots & \ldots & 0 & 0 & 0 \\ 
		0 & 0 & {\sqrt{n+1-m}\sin \theta}   & {\sqrt{m}\cos \theta}  & 0 & 0
		\\ 
		0 & 0 & 0  & \ldots & \ldots & 0 \\ 
		0 & 0 & 0 & 0 & {\sin \theta}  & {\sqrt{n}\cos \theta }
	\end{array}%
	\right)$},
\end{equation}%
with $\tilde{\Omega }( t)=\sqrt{g^{2}{N}+\Omega ^{2}( t)}$.
The coupling matrix here is an $n\times (n+1)$ row-echelon matrix and full rank, which takes the same form as that in Eq.~(\ref{CN3}). Therefore, based on the analyses in Sec.~\ref{expanded M-type}, there are $n$ bright states and one dark state.
Based on Eq.~(\ref{D_def}), we define this unique dark state as 
\begin{equation}
	\left\vert D\right\rangle =\sum_{i=0}^{n}x_{i}\left\vert L_{i+1}\right\rangle=\sum_{i=0}^{n}x_{i}\vert \mathbf{c}^{i},n-i\rangle,
\end{equation}
where $x_{i}$ are the superposition coefficients of these lower states $\vert \mathbf{c}^{i},n-i\rangle$.
With the relation 
\begin{equation}
(\sqrt{{n-i}}\sin \theta) x_{i}+(\sqrt{{i+1}}\cos \theta) x_{i+1}=0,
\end{equation}
for $i=0,1,\ldots,n-1$, the dark state can be obtained from Eq.~(\ref{N-M}) 
\begin{eqnarray}
		\left\vert D^{{\mathrm{DSP}}}_{n}\right\rangle &=&\sum_{i=0}^{n}{( -1) ^{n-i}}\prod_{j=i}^{n-1}\frac{\sqrt{{j+1}}\cos \theta}{\sqrt{{n-j}}\sin \theta} \vert \mathbf{c}^{i},n-i\rangle \notag \\
		&=&\sum_{i=0}^{n}\sqrt{\frac{n!}{i!( n-i) !}}( -\cos
		\theta ) ^{n-i}\sin ^{i}\theta \vert \mathbf{c}%
		^{i},n-i\rangle ,
\end{eqnarray}%
which is the same as Eq.~(8) in Ref.~\cite{DSP:PRA2002}.
Therefore, we recover the same results of the dark-state polaritons in a $\Lambda$-type three-level atom ensemble using the arrowhead-matrix method.

\section{Conclusion}\label{conclusion}

In conclusion, we have presented a general theory for studying the dark states in arbitrary multilevel systems with the arrowhead-matrix method. We have also generalized the concept of dark state in the sense of decoupling. Concretely, we have divided the basis states into the upper- and lower-state subspaces, diagonalized the Hamiltonian within the two subspaces to obtain the dressed upper and lower states, and introduced the bipartite-graph description of the quantum system. In this way, we have transformed the Hamiltonian matrix into an arrowhead matrix, then the number of the dark states can be determined by analyzing the ranks of the coupling submatrices associated with the degenerate dressed-lower-state subspaces, and the form of the dark states can be obtained by solving the null space of these coupling submatrices. Note that our theory was proposed to determine the dark-state effect for a closed multilevel system.

Based on the arrowhead-matrix method, we have calculated the dark states in three-, four-, and five-level quantum systems in detail. By classifying the systems based on the numbers of the upper and lower states, they can be divided into different configurations and the dark states in these systems can be analyzed systematically.
We have also extended the situation to the $N$-level quantum systems and given some typical examples in which there is no intracoupling within the upper- and lower-state subspaces, such as multipod quantum system, shared-lower-state multiple-$\Lambda$ system, $\Lambda$-chain system, shared-edge $\mathrm{N}$-chain system, and $\mathrm{V}$-chain system.
Concretely, we found that for the multipod quantum system with one upper state and $N-1$ lower states, when $j$ ($j=2,3,\ldots,N-1$) lower states are degenerate, there is a $(j-1)$-dimensional dark-state subspace.
For the shared-lower-state multiple-$\Lambda$ system with $N-2$ upper states and two lower states and the $\Lambda$-chain system with a zigzag coupling satisfying  $N_{l}=N_{u}+1$, only when all lower states are degenerate, there exists a unique dark state composed of all lower states. 
For the shared-edge $\mathrm{N}$-chain system with a zigzag coupling satisfying $N_{l}=N_{u}$ and the $\mathrm{V}$-chain system with a zigzag coupling satisfying $N_{l}=N_{u}-1$, there is no dark state.
Finally, we have recovered the results of the dark-state polaritons in driven three-level systems using the arrowhead-matrix method. 
Our method is general and it can be used to study the dark-state effect in any multilevel quantum system. It will also motivate the future research concerning dark-state preparation, manipulation, and application.

\begin{acknowledgments}
J.-Q.L. was supported in part by National Natural Science Foundation of China (Grants No.~12175061, No.~12575015, No.~12247105, No.~11935006, and No.~12421005), National Key Research and Development Program of China (Grant No.~2024YFE0102400), and Hunan Provincial Major Sci-Tech Program (Grant No.~2023ZJ1010).  
L.M.K. is supported by the NSFC (Grant Nos.~12247105 and~12421005), Quantum Science and Technology-National Science and Technology Major Project (Grant No.~2024ZD0301000), the Hunan Provincial Major Sci-Tech Program (Grant No.~2023ZJ1010),  and the XJ-Lab key project (Grant No.~23XJ02001).

\end{acknowledgments}

\appendix

\section{Derivation of the time-independent Hamiltonian in Eq.~(\ref{HN_tilde})}\label{A1}

In this Appendix, we present the detailed derivation of the time-independent Hamiltonian for the general $N$-level quantum system described by the Hamiltonian in Eq.~(\ref{HN}). To this end, we assume
\begin{equation}
 H_{0}=\sum_{{j}=1}^{N}x_{j}\left\vert j\right\rangle \left\langle
j\right\vert,
\end{equation}
where $x_{j=1-N}$ are the coefficients to be determined by eliminating the time-oscillating factors. To derive the transformed Hamiltonian, we
introduce the transformation $\vert \Psi(t)\rangle=U\vert \Phi(t)\rangle $ with $U=\text{exp}(-iH_{0}t)=\text{exp}\left(-{i}t\sum_{{j}=1}^{N}x_{j}\vert j\rangle
\langle j\vert \right)$, where $\vert \Psi(t)\rangle$ and $\vert \Phi(t)\rangle$ are the states of the system in the Schr\"odinger picture and the rotating frame, respectively. In the rotating frame defined by $U$, the Hamiltonian in Eq.~(\ref{HN}) becomes 
\begin{eqnarray}
	\tilde{H}^{[N]} &=&U^{\dagger } {H^{[N]}}U -i{ U^{\dagger } }\dot{U}\notag\\
	&=&\sum_{{j}=1}^{{N}}( E_{j}-x_{j}) \left\vert j\right\rangle
	\left\langle j\right\vert 
	+ \sum_{j,j^{\prime }=1,j<j^{\prime
	}}^{N}(\Omega _{jj^{\prime }}\vert j^{\prime }\rangle \left\langle
	j\right\vert\notag \\
	&&\times e^{i( x_{j^{\prime }}-x_{j}-\omega _{jj^{\prime }})
		t}+\text{H.c.}),
\end{eqnarray}%
where the time-oscillating factors can be eliminated under the condition
\begin{eqnarray}
	x_{j^{\prime }}-x_{j} -\omega _{jj^{\prime }} =0.
\end{eqnarray}%
By introducing the detuning $\Delta _{jj^{\prime }} =E_{j^{\prime }}-E_{j}- \omega_{jj^{\prime }}$, which is the energy separation of the transition between the states $\left\vert j\right\rangle$ and $\vert j^{\prime }\rangle$ with respect to the driving frequency $\omega_{jj^{\prime}}$ of the field, we have the relation $x_{j^{\prime }}-x_{j} =\omega _{jj^{\prime }} =E_{j^{\prime }}-E_{j}- \Delta_{jj^{\prime }}$.

For simplicity, we take $x_{N}=E_{N}$, then the remaining coefficients can be obtained as
\begin{eqnarray}
	x_{r}=E_{r}+\Delta _{rN}, \hspace{0.3cm} r=1,2,\ldots,N-1.
\end{eqnarray}%
For different $r,r^{\prime }=1,2,\ldots,N-1$ with $r<r^{\prime
}$, there is
\begin{eqnarray}
	x_{r^{\prime }}-x_{r}  &=&E_{r^{\prime }}-E_{r}-\Delta _{rr^{\prime }}\notag \\
 &=&E_{r^{\prime }}+\Delta _{r^{\prime }N}-(E_{r}+\Delta _{rN})	,
\end{eqnarray}%
hence the detunings satisfy the conditions $\Delta _{rN}-\Delta _{r^{\prime }N} =\Delta _{rr^{\prime }}$. Here, we should point out that, to obtain the time-independent Hamiltonian for this multilevel system with complex transitions, the above conditions should be satisfied. Physically, these conditions imply the resonance in the multi-photon processes of loop transitions. For the time-independent Hamiltonian in the rotating frame, the effective energy associated with the level is determined by its free energy term, and the effective frequency of the light is zero. Therefore, the detuning related to a certain transition is completely determined by the free energies of the two involved levels.

Based on the above analyses, we see that, in a rotating frame with respect to $H_{0}=  E_{N}\left\vert
N\right\rangle \left\langle N\right\vert+ \sum_{{r}=1}^{N-1}( E_{r}+\Delta _{rN}) \left\vert
r\right\rangle \left\langle r\right\vert$, the time-independent Hamiltonian in Eq.~(\ref{HN_tilde}) can be obtained.

\section{Proof of the assertions for determining the bright and dark states with the arrowhead-matrix method}\label{A2}

In this Appendix, we present the detailed proof of the assertions for determining the bright and dark states with the arrowhead-matrix method introduced in Sec.~\ref{assertion}. Below we give the proof of assertions 2(ii) and 4.

\subsection{Proof of assertion 2(ii)}

In the $l$-dimensional degenerate dressed-lower-state subspace, the Hamiltonian can be written as 
\begin{eqnarray}
	\tilde{H}_{D}^{[N_{(l)}]} &=&\sum_{{n_{u}}=1}^{N_{u}}\Delta
	_{n_{u}}|U_{n_{u}}\rangle \langle U_{n_{u}}|+\sum_{{n_{l}}=1}^{l}\Omega |L_{{%
					n_{l}}}\rangle \langle L_{{n_{l}}}|  \notag \\
	&&+\sum_{{n_{u}}=1}^{N_{u}}\sum_{{n_{l}}=1}^{l}(G_{n_{u}n_{l}}|U_{n_{u}}%
	\rangle \langle L_{n_{l}}|+\text{H.c.}),
\end{eqnarray}
where the superscript \textquotedblleft${[N_{(l)}]}$\textquotedblright\  of $\tilde{H}_{D}^{[N_{(l)}]}$ denotes that there is a degenerate dressed-lower-state subspace with dimension $l$ in the $N$-level system. Then we adopt the method of mathematical induction to prove the results in Eqs.~(\ref{Bj}) and~(\ref{Dj}).

(1) Step 1: When $l=2$ and the coupling column vectors $%
\mathbf{C}_{2}$ and $\mathbf{C}_{1}$ are linearly dependent, i.e., 
$	\mathbf{C}_{2}=\lambda _{2}\mathbf{C}_{1}$, namely, $(G_{12},G_{22},\ldots,G_{N_{u}2})^{T}=(\lambda _{2}G_{11},\lambda _{2}G_{21},\ldots,\lambda _{2}G_{N_{u}1})^{T}$, then the Hamiltonian is reduced to%
\begin{eqnarray}
	\tilde{H}_{D}^{[N_{(2)}]}&=&\sum_{{n_{u}}=1}^{N_{u}}\Delta _{n_{u}}|U_{n_{u}}\rangle \langle
	U_{n_{u}}|+\Omega( |L_{1}\rangle \langle L_{1}|+ |L_{2}\rangle \langle
	L_{2}|)
	\notag \\
	&&+\sum_{{n_{u}}=1}^{N_{u}}[G_{n_{u}1}|U_{n_{u}}\rangle (\langle
	L_{1}|+\lambda _{2}\langle L_{2}|) +\text{H.c.}].
\end{eqnarray}%
We further introduce 
\begin{subequations}\label{B1D1}
	\begin{align}
		|B_{1}\rangle  =&\frac{1}{\mathcal{N}_{1}}( |L_{1}\rangle +\lambda
		_{2}^{\ast }|L_{2}\rangle ) , \\
		|D_{1}\rangle  =&\frac{1}{\mathcal{N}_{1}}( \lambda _{2}|L_{1}\rangle
		-|L_{2}\rangle ) ,
		\end{align}%
\end{subequations}
with $\mathcal{N}_{1}=\sqrt{1+\left\vert \lambda _{2}\right\vert ^{2}}>0$.
Then we can obtain $	|L_{1}\rangle \langle L_{1}|+|L_{2}\rangle \langle L_{2}|=|B_{1}\rangle
	\langle B_{1}|+|D_{1}\rangle \langle D_{1}|$,
and the Hamiltonian $\tilde{H}_{D}^{[N_{(2)}]}$ can be expressed as 
\begin{eqnarray}
	\tilde{H}_{D}^{[N_{(2)}]} &=&\sum_{{n_{u}}=1}^{N_{u}}\Delta
	_{n_{u}}|U_{n_{u}}\rangle \langle U_{n_{u}}|+\Omega (|B_{1}\rangle \langle
	B_{1}|+|D_{1}\rangle \langle D_{1}|) \notag \\
	&&+\sum_{{n_{u}}=1}^{N_{u}}(\tilde{G}_{n_{u}1}|U_{n_{u}}\rangle \langle
	B_{1}|+\text{H.c.}),
\end{eqnarray}%
with $\tilde{G}_{n_{u}1}=\mathcal{N}_{1}G_{n_{u}1}$. Here we can see that
there is one bright state $|B_{1}\rangle $ coupled with the dressed upper states and
one dark state $|D_{1}\rangle $ decoupled from all the dressed upper states.

(2) Step 2: When $l=3$ and the coupling column vectors $\mathbf{C}_{3}$, $\mathbf{C}_{2}$, and $\mathbf{C}_{1}$ are linearly dependent, i.e., 
$	\mathbf{C}_{3}=\lambda _{3}\mathbf{C}_{1}$ and $	\mathbf{C}_{2}=\lambda _{2}\mathbf{C}_{1}$, namely, $(G_{13},G_{23},\ldots,G_{N_{u}3})^{T}=(\lambda _{3}G_{11},\lambda _{3}G_{21},\ldots,\lambda _{3}G_{N_{u}1})^{T}$ and $(G_{12},G_{22},\ldots,G_{N_{u}2})^{T}=(\lambda _{2}G_{11},\lambda _{2}G_{21},\ldots,\lambda _{2}G_{N_{u}1})^{T}$, then the Hamiltonian reads%
\begin{eqnarray}
	\tilde{H}_{D}^{[N_{(3)}]} &=&	\tilde{H}_{D}^{[N_{(2)}]}+\Omega |L_{3}\rangle \langle
	L_{3}| +\sum_{{n_{u}}=1}^{N_{u}}(G_{n_{u}3}|U_{n_{u}}\rangle \langle
	L_{3}|+\text{H.c.})  \notag \\
	&=&\sum_{{n_{u}}=1}^{N_{u}}\Delta _{n_{u}}|U_{n_{u}}\rangle \langle
	U_{n_{u}}|+\Omega|D_{1}\rangle \langle D_{1}|
	\notag \\
	&&+\Omega( |B_{1}\rangle \langle B_{1}|+ |L_{3}\rangle \langle
		L_{3}|)  \notag \\
		&&+\sum_{{n_{u}}=1}^{N_{u}}[{G}_{n_{u}1}|U_{n_{u}}\rangle(\mathcal{N}_{1}\langle
	B_{1}|+\lambda _{3}\langle
	L_{3}|)  +\text{H.c.}].
\end{eqnarray}%
Similarly, we introduce 
\begin{subequations}
	\begin{align}
			|B_{2}\rangle  =&\frac{1}{\mathcal{N}_{2}}( \mathcal{N}%
		_{1}|B_{1}\rangle +\lambda _{3}^{\ast }|L_{3}\rangle ) ,\\
		|D_{2}\rangle  =&\frac{1}{\mathcal{N}_{2}}( \lambda
		_{3}|B_{1}\rangle -\mathcal{N}_{1}|L_{3}\rangle ) ,
	\end{align}%
\end{subequations}
with $\mathcal{N}_{2}=\sqrt{\mathcal{N}_{1}^{2} +\vert \lambda _{3}\vert ^{2}} =\sqrt{1+\vert \lambda _{2}\vert ^{2}+\vert \lambda _{3}\vert ^{2}}>0$.
Then we can obtain $|B_{1}\rangle
\langle B_{1}|+	|L_{3}\rangle \langle L_{3}|=|B_{2}\rangle
\langle B_{2}|+|D_{2}\rangle \langle D_{2}|$,
and the Hamiltonian $\tilde{H}_{D}^{[N_{(3)}]}$ can be simplified to
\begin{eqnarray}
		\tilde{H}_{D}^{[N_{(3)}]}	&=&\sum_{{n_{u}}=1}^{N_{u}}\Delta _{n_{u}}|U_{n_{u}}\rangle \langle
		U_{n_{u}}|+\Omega( |D_{1}\rangle \langle D_{1}|+|D_{2}\rangle \langle D_{2}| )
	\notag \\
	&&+\Omega |B_{2}\rangle \langle B_{2}|+\sum_{{n_{u}}=1}^{N_{u}}(\tilde{G}_{n_{u}2}|U_{n_{u}}\rangle \langle
	B_{2}|+\text{H.c.}),
\end{eqnarray}%
with $\tilde{G}_{n_{u}2}=\mathcal{N}_{2}G_{n_{u}1}$. Here we can see that
there is one bright state $|B_{2}\rangle $ and
two dark states $|D_{1}\rangle $ and $|D_{2}\rangle $.

(3) Step 3: We assume that the statement is valid for $l=j$, and all the coupling column vectors $%
\mathbf{C}_{j^{\prime}=2,3,\ldots,j}$ and $\mathbf{C}_{1}$ are linearly dependent $	\mathbf{C}_{j^{\prime}}=\lambda _{j^{\prime}}\mathbf{C}_{1}$, namely, $(G_{1j^{\prime}},G_{2j^{\prime}},\ldots,G_{N_{u}j^{\prime}})^{T}=(\lambda _{j^{\prime}}G_{11},\lambda _{j^{\prime}}G_{21},\ldots,\lambda _{j^{\prime}}G_{N_{u}1})^{T}$. Then the
Hamiltonian can be expressed as 
\begin{eqnarray}
	\tilde{H}_{D}^{[N_{(j)}]} &=&\tilde{H}_{D}^{[N_{(j-1)}]}+\Omega |L_{j}\rangle
	\langle L_{j}|  +\sum_{{n_{u}}=1}^{N_{u}}(G_{n_{u}j}|U_{n_{u}}\rangle \langle
	L_{j}|+\text{H.c.})  \notag \\
	&=&\sum_{{n_{u}}=1}^{N_{u}}\Delta _{n_{u}}|U_{n_{u}}\rangle \langle
	U_{n_{u}}|+\Omega |B_{j-1}\rangle \langle
	B_{j-1}|+\sum_{s=1}^{j-1}\Omega|D_{s}\rangle \langle D_{s}| \notag \\
	&&+\sum_{{n_{u}}=1}^{N_{u}}[\tilde{G}_{n_{u}(j-1)}|U_{n_{u}}\rangle \langle
	B_{j-1}|+\text{H.c.}],
\end{eqnarray}%
where $\tilde{G}_{n_{u}(j-1)}=\mathcal{N}_{j-1}G_{n_{u}1}$ with $\mathcal{N}%
_{j-1}=\sqrt{\mathcal{N}_{j-2}^{2}+|\lambda _{j}|^{2}}=\sqrt{1+\sum_{j^{\prime }=2}^{j}|\lambda _{j^{\prime }}|^{2}}>0$,
and these states are defined by%
\begin{subequations}
	\begin{align}
		|B_{j-1}\rangle  =&\frac{1}{\mathcal{N}_{j-1}}( \mathcal{N}%
		_{j-2}|B_{j-2}\rangle +\lambda _{j}^{\ast }|L_{j}\rangle ) ,\\
			|D_{j-1}\rangle  =&\frac{1}{\mathcal{N}_{j-1}}( \lambda
		_{j}|B_{j-2}\rangle -\mathcal{N}_{j-2}|L_{j}\rangle ) .
		\end{align}%
\end{subequations}
Here, $|B_{j-1}\rangle $ is the bright state, and these $j-1$ states $%
|D_{1}\rangle $, $|D_{2}\rangle $, \ldots, and $|D_{j-1}\rangle $ are dark states.

(4) Step 4: We show that when all the coupling column vectors are linearly dependent $	\mathbf{C}_{j^{\prime}}=\lambda _{j^{\prime}}\mathbf{C}_{1}$, namely, $(G_{1j^{\prime}},G_{2j^{\prime}},\ldots,G_{N_{u}j^{\prime}})^{T}=(\lambda _{j^{\prime}}G_{11},\lambda _{j^{\prime}}G_{21},\ldots,\lambda _{j^{\prime}}G_{N_{u}1})^{T}$ for $j^{\prime}=2,3,\ldots,j,j+1$, then the statement is valid for $l=j+1$. For the case of $l=j+1$, the Hamiltonian can be expressed as 
\begin{eqnarray}
	\tilde{H}_{D}^{[N_{(j+1)}]} &=&\tilde{H}_{D}^{[N_{(j)}]}+\Omega |L_{j+1}\rangle
	\langle L_{j+1}|  \notag \\
	&&+\sum_{{n_{u}}=1}^{N_{u}}[G_{n_{u}(j+1)}|U_{n_{u}}\rangle \langle
	L_{j+1}|+\text{H.c.}]  \notag 	\\
	&=&\sum_{{n_{u}}=1}^{N_{u}}\Delta _{n_{u}}|U_{n_{u}}\rangle \langle
	U_{n_{u}}|
	+\Omega (|L_{j+1}\rangle \langle L_{j+1}|+|B_{j-1}\rangle \langle
	B_{j-1}| )
	 \notag \\
	&&+\sum_{s=1}^{j-1}\Omega|D_{s}\rangle \langle D_{s}|+\sum_{{n_{u}}=1}^{N_{u}}[{G}_{n_{u}1}|U_{n_{u}}\rangle( \mathcal{N}_{j-1}\langle 
	B_{j-1}|\notag \\
	&&
	+ \lambda _{j+1}\langle L_{j+1}|)+\text{H.c.}].
\end{eqnarray}%
Similarly, we introduce the states 
\begin{subequations}
	\begin{align}
		|B_{j}\rangle  =&\frac{1}{\mathcal{N}_{j}}( \mathcal{N}%
		_{j-1}|B_{j-1}\rangle +\lambda _{j+1}^{\ast }|L_{j+1}\rangle ) ,\\
		|D_{j}\rangle  =&\frac{1}{\mathcal{N}_{j}}( \lambda
		_{j+1}|B_{j-1}\rangle -\mathcal{N}_{j-1}|L_{j+1}\rangle ) ,
		\end{align}%
\end{subequations}
with $\mathcal{N}_{j}=\sqrt{\mathcal{N}_{j-1}^{2}+|\lambda _{j+1}|^{2}}=%
\sqrt{1+\sum_{j^{\prime }=2}^{j+1}|\lambda _{j^{\prime }}|^{2}}>0$.
It can be shown that $
	|B_{j-1}\rangle \langle B_{j-1}|+|L_{j+1}\rangle \langle
	L_{j+1}|=|B_{j}\rangle \langle B_{j}|+|D_{j}\rangle \langle D_{j}|$, then the Hamiltonian can be rewritten as%
\begin{eqnarray}\label{HD_j+1}
	\tilde{H}_{D}^{[N_{(j+1)}]} &=&\sum_{{n_{u}}=1}^{N_{u}}\Delta
	_{n_{u}}|U_{n_{u}}\rangle \langle U_{n_{u}}|+\Omega |B_{j}\rangle \langle
	B_{j}|+\sum_{s=1}^{j}\Omega|D_{s}\rangle \langle D_{s}|  \notag \\
	&&+\sum_{{n_{u}}=1}^{N_{u}}(\tilde{G}_{n_{u}j}|U_{n_{u}}\rangle \langle
	B_{j}|+\text{H.c.}),
\end{eqnarray}%
where $\tilde{G}_{n_{u}j}=\mathcal{N}_{j}G_{n_{u}1}$. We see form Eq.~(\ref{HD_j+1}) that there is one
bright state $|B_{j}\rangle $ and $j$ dark states $\{|D_{1}\rangle
,|D_{2}\rangle , \ldots, |D_{j-1}\rangle ,|D_{j}\rangle \}$. Therefore, the
statement is valid for $l=j+1.$ 

Based on the above analyses, we can conclude
that assertion 2(ii) is valid for an arbitrary positive integer $l$.

\subsection{Proof of assertion 4}

Based on the assertions in Sec.~\ref{assertion}, we can obtain that 

(1) When $\mathbf{C}_{k=1-N_{l}}\neq \mathbf{0}$, the state $\left\vert
L_{k}\right\rangle $ is always coupled to some (or all) of the dressed upper states and it will not be a dark state.

(2) When there is no degeneracy in the dressed lower states, there is no dark state in the system. To show this assertion, we consider the $\Lambda$-type three-level system as an example to show that the dark state for the two degenerate-lower-state case will no longer be a dark state in the nondegenerate-lower-state case. For the three-level system with an upper state $\vert U_{1}\rangle$ and two nondegenerate lower states $\vert	L_{1}\rangle$ and $\vert L_{2}\rangle$, the Hamiltonian reads
\begin{eqnarray}\label{H-e}
	\tilde{H}_{D} &=&\Delta _{1}|U_{1}\rangle
	\langle U_{1}|+\Omega _{1}|L_{1}\rangle \langle L_{1}|+\Omega
	_{2}|L_{2}\rangle \langle L_{2}|  \notag \\
	&&+[G_{11}|U_{1}\rangle (\langle
	L_{1}|+\lambda _{2}\langle L_{2}|)+\text{H.c.}],
\end{eqnarray}%
where the column vectors $	\mathbf{C}_{2}$ and $\mathbf{C}_{1}$ are linearly dependent $	\mathbf{C}_{2}=\lambda _{2}\mathbf{C}_{1}$, namely, $G_{12}=\lambda _{2}G_{11}$. Similarly, we introduce the bright state $|B_{1}\rangle $ coupled with the upper state
and its orthogonal state $|D_{1}\rangle $ in Eq.~(\ref{B1D1}). Then the term $\Omega _{1}\vert L_{1}\rangle \langle L_{1}\vert
+\Omega _{2}\vert L_{2}\rangle \langle L_{2}\vert$ can be rewritten based on the states $|B_{1}\rangle $ and $|D_{1}\rangle $ as
\begin{eqnarray}\label{non-deg}
	&&\Omega _{1}\left\vert L_{1}\right\rangle \left\langle L_{1}\right\vert
	+\Omega _{2}\left\vert L_{2}\right\rangle \left\langle L_{2}\right\vert \notag \\
	&=&\frac{1}{\mathcal{N}^{2}_{1}}[( \Omega _{1}+\Omega _{2}\vert \lambda _{2}\vert
	^{2}) |B_{1}\rangle \left\langle B_{1}\right\vert +( \Omega
	_{1}\vert \lambda _{2}\vert ^{2}+\Omega _{2}) |D_{1}\rangle
	\left\langle D_{1}\right\vert \notag \\
	&&+( \Omega _{1}-\Omega _{2}) \lambda _{2}|B_{1}\rangle
	\left\langle D_{1}\right\vert +( \Omega _{1}-\Omega _{2}) \lambda
	_{2}^{\ast }|D_{1}\rangle \left\langle B_{1}\right\vert ].
\end{eqnarray}%
We can see from Eq.~(\ref{non-deg}) that, in the nondegenerate-lower-state case $ \Omega _{1} \neq \Omega _{2}$, the state $|D_{1}\rangle $ is coupled
to the state $|B_{1}\rangle $, and further coupled to the upper state. Only when the two lower states $\vert L_{1}\rangle $ and $\vert
L_{2}\rangle $ are degenerate, namely, $\Omega _{1}=\Omega _{2}$, the
state $|D_{1}\rangle $ becomes a dark state decoupled from the upper state. Therefore, the dark states will only exist in the degenerate-state subspace.

The degenerate condition of the lower states can also be proved based on the energy eigenvalue equation $H\vert\psi\rangle=E\vert\psi\rangle$. Since the dark states are the eigenstates of the Hamiltonian $ H$ and are merely superposed by the lower states, then for the Hamiltonian $ \tilde{H}_{D}$ in Eq.~(\ref{H-e}), we have
	\begin{eqnarray}\label{e-eq}
	\tilde{H}_{D}\vert D\rangle &=&\left( 
	\begin{array}{c|cc}
		\Delta _{1} & G_{11 }& \lambda_{2} G_{11 } \\ \hline
		G_{11 }^{\ast } & \Omega _{1}& 0 \\ 
		\lambda_{2}^{\ast } G_{11 }^{\ast } & 0 & \Omega _{2}%
	\end{array}%
	\right) \left( 
	\begin{array}{c}
		 0\\ \hline
		x_{1 } \\ 
		x_{2 }%
	\end{array}%
	\right)\notag \\
	&=&\left( 
	\begin{array}{c}
		G_{11}(x_1+\lambda_2 x_2) \\ \hline
		\Omega _{1}x_{1 } \\ 
		\Omega _{2}x_{2 }%
	\end{array}%
	\right)=E\left( 
	\begin{array}{c}
		0 \\ \hline
		x_{1 } \\ 
		x_{2 }%
	\end{array}%
	\right), 
\end{eqnarray}
where the dark state defined as $\vert D\rangle=x_{1 }\vert	L_{1}\rangle+x_{2 }\vert L_{2}\rangle $. By solving Eq.~(\ref{e-eq}), we can obtain the relations $G_{11}(x_1+\lambda_2 x_2)=0$ and $\Omega _{1}=\Omega _{2}=E$.
Consequently, to satisfy the energy eigenvalue equation, the energies of these lower states must be equal and therefore the dark states should exist in the degenerate lower-state subspace.

In a word, when $\mathbf{C}_{k=1-N_{l}}\neq \mathbf{0}$ and there is no
degeneracy in these dressed lower states, there is no dark state in the system.


\begin{thebibliography}{99}
		

\bibitem{DS1976} G. Alzetta, A. Gozzini, M. Moi, and G. Orriols, An experimental method for the observation of r.f. transitions and laser beat resonances in oriented Na vapour, Nuovo Cimento B \textbf{36}, 5 (1976).

\bibitem{ScullyQO1997}
M. O. Scully and M. S. Zubairy, \textit{Quantum optics} (Cambridge University Press, Cambridge, England, 1997).

\bibitem{CAbook2011} C. Cohen-Tannoudji and D. Gu\'{e}ry-Odelin, \textit{Advances in Atomic Physics: An Overview} (World Scientific, Singapore, 2011).



\bibitem{CPT1978} H. R. Gray, R. M. Whitley, and C. R. Stroud, Coherent trapping of atomic populations, Opt. Lett. \textbf{3}, 218 (1978).

\bibitem{CPT1982} P. M. Radmore and P. L. Knight, Population trapping and dispersion in a three-level system, J. Phys. B: At. Mol. Phys. \textbf{15}, 561 (1982).

\bibitem{CPT1988} F. T. Hioe and C. E. Carroll, Coherent population trapping in \textit{N}-level quantum systems, Phys. Rev. A \textbf{37}, 3000 (1988).

\bibitem{CPT1996} E. Arimondo, Coherent population trapping in laser spectroscopy, Prog. Opt. \textbf{35}, 257 (1996).


\bibitem{EITPRL1991} K.-J. Boller, A. Imamo\v{g}lu, and S. E. Harris, Observation of Electromagnetically Induced Transparency, Phys. Rev. Lett. \textbf{66}, 2593 (1991).

\bibitem{Arimondo1996} E. Cerboneschi and E. Arimondo, Matched pulses and electromagnetically induced transparency for the interaction of laser pulse pairs with a double-vee system, Opt. Commun. \textbf{127}, 55 (1996).

\bibitem{EIT:H1997} S. E. Harris, Electromagnetically Induced Transparency, Phys. Today \textbf{50}, 36 (1997).

\bibitem{DPiEIT2000} M. Fleischhauer and M. D. Lukin, Dark-State Polaritons in Electromagnetically Induced Transparency, Phys. Rev. Lett. \textbf{84}, 5094 (2000).

\bibitem{EIT2003} L.-M. Kuang and L. Zhou, Generation of atom-photon entangled states in atomic Bose-Einstein condensate via electromagnetically induced transparency, Phys. Rev. A \textbf{68}, 043606 (2003).

\bibitem{EIT2005} M. Fleischhauer, A. Imamoglu, and J. P. Marangos, Electromagnetically induced transparency: Optics in coherent media, Rev. Mod. Phys. \textbf{77}, 633 (2005).

\bibitem{EIT2007} L. M. Kuang, Z. B. Chen, and J. W. Pan, Generation of entangled coherent states for distant Bose-Einstein condensates via electromagnetically induced transparency, Phys. Rev. A \textbf{76}, 052324 (2007).


\bibitem{STIRAP:A1989} J. R. Kuklinski, U. Gaubatz, F. T. Hioe, and K. Bergmann, Adiabatic population transfer in a three-level system driven by delayed laser pulses, Phys. Rev. A \textbf{40}, 6741 (1989).

\bibitem{STIRAP1998} K. Bergmann, H. Theuer, and B. W. Shore, Coherent population transfer among quantum states of atoms and molecules, Rev. Mod. Phys. \textbf{70}, 1003 (1998).

\bibitem{STIRAP2015} K. Bergmann, N. V. Vitanov, and B. W. Shore, Perspective: Stimulated Raman adiabatic passage: The status after 25 years, J. Chem. Phys. \textbf{142}, 170901 (2015).

\bibitem{STRIP2017} N. V. Vitanov, A. A. Rangelov, B. W. Shore, and K. Bergmann, Stimulated Raman adiabatic passage in physics, chemistry, and beyond, Rev. Mod. Phys. \textbf{89}, 015006 (2017).

\bibitem{DS:jpa2017} J. Peng, C. X. Zheng, G. J. Guo, X. Y. Guo, X. Zhang, C. S. Deng, G. X. Ju, Z. Z. Ren, L. Lamata, and E. Solano, Dark-like states for the multi-qubit and multi-photon Rabi models, J. Phys. A \textbf{50}, 174003 (2017).

\bibitem{DS:jpl2021} J. Peng, J. Zheng, J. Yu, P. Tang, G. A. Barrios, J. Zhong, E. Solano, F. A. Arriagada, and L. Lamata, One-Photon Solutions to the Multiqubit Multimode Quantum Rabi Model for Fast \textit{W}-State Generation, Phys. Rev. Lett. \textbf{127}, 043604 (2021).


\bibitem{CuEIT2000} G. Morigi, J. Eschner, and C. H. Keitel, Ground State Laser Cooling Using Electromagnetically Induced Transparency, Phys. Rev. Lett. \textbf{85}, 4458 (2000).

\bibitem{CEwEIT2000} C. F. Roos, D. Leibfried, A. Mundt, F. Schmidt-Kaler, J. Eschner, and R. Blatt, Experimental Demonstration of Ground State Laser Cooling with Electromagnetically Induced Transparency, Phys. Rev. Lett. \textbf{85}, 5547 (2000).

\bibitem{LCiTS2010} J. Cerrillo, A. Retzker, and M. B. Plenio, Fast and Robust Laser Cooling of Trapped Systems, Phys. Rev. Lett. \textbf{104}, 043003 (2010).

\bibitem{LCiTi2012} S. Zhang, C.-W. Wu, and P.-X. Chen, Dark-state laser cooling of a trapped ion using standing waves, Phys. Rev. A \textbf{85}, 053420 (2012).



\bibitem{DS:CIarX} C. J. Villas-Boas, C. E. M\'{a}ximo, P. J. Paulino, R. P. Bachelard, and G. Rempe, Bright and Dark States of Light: The Quantum Origin of Classical Interference, Phys. Rev. Lett. \textbf{134}, 133603 (2025). 


\bibitem{DS_AF2007} G. Hernandez, J. Zhang, and Y. Zhu, Vacuum Rabi splitting and intracavity dark state in a cavity-atom system, Phys. Rev. A \textbf{76}, 053814 (2007).

\bibitem{DS_AF2014} T. Kampschulte, W. Alt, S. Manz, M. Martinez-Dorantes, R. Reimann, S. Yoon, D. Meschede, M. Bienert, and G. Morigi, Electromagnetically-induced-transparency control of single-atom motion in an optical cavity, Phys. Rev. A \textbf{89}, 033404 (2014).

\bibitem{DS_AF2020}  P. Samutpraphoot, T. \DH or\dj evi\'{c}, P. L. Ocola, H. Bernien, C. Senko, V. Vuleti\'{c}, and M. D. Lukin, Strong Coupling of Two Individually Controlled Atoms via a Nanophotonic Cavity, Phys. Rev. Lett. \textbf{124}, 063602 (2020).

\bibitem{DS_AF2021} Y. Zhang, C. Shan, and K. M\o lmer, Ultranarrow Superradiant Lasing by Dark Atom-Photon Dressed States, Phys. Rev. Lett. \textbf{126}, 123602 (2021).
		
\bibitem{DS:prx2022} A. Pi\~neiro Orioli, J. K. Thompson, and A. M. Rey, Emergent Dark States from Superradiant Dynamics in Multilevel Atoms in a Cavity, Phys. Rev. X \textbf{12}, 011054 (2022). 

\bibitem{DS_AF2023} J. Skulte, P. Kongkhambut, S. Rao, L. Mathey, H. Ke\ss ler, A. Hemmerich, and J. G. Cosme, Condensate Formation in a Dark State of a Driven Atom-Cavity System, Phys. Rev. Lett. \textbf{130}, 163603 (2023).


\bibitem{DS_CQEDe2024}  B. Sundar, D. Barberena, A. M. Rey, and A. P. Orioli, Squeezing Multilevel Atoms in Dark States via Cavity Superradiance, Phys. Rev. Lett. \textbf{132}, 033601 (2024).



\bibitem{DS_TiCe2009} J. D. Jost, J. P. Home, J. M. Amini, D. Hanneke, R. Ozeri, C. Langer, J. J. Bollinger, D. Leibfried, and D. J. Wineland, Entangled mechanical oscillators, Nature (London) \textbf{459}, 683 (2009).

\bibitem{DS_TiCe2013} Y. Lin, J. P. Gaebler, T. R. Tan, R. Bowler, J. D. Jost, D. Leibfried, and D. J. Wineland, Sympathetic Electromagnetically-Induced-Transparency Laser Cooling of Motional Modes in an Ion Chain, Phys. Rev. Lett. \textbf{110}, 153002 (2013).

\bibitem{DS_Tie2015} J. Ro\ss nagel, K. N. Tolazzi, F. Schmidt-Kaler, and K. Singer, Fast thermometry for trapped ions using dark resonances, New J. Phys. \textbf{17}, 045004 (2015).

\bibitem{DS_TiCe2020} L. Feng, W. L. Tan, A. De, A. Menon, A. Chu, G. Pagano, and C. Monroe, Efficient Ground-State Cooling of Large Trapped-Ion Chains with an Electromagnetically-Induced-Transparency Tripod Scheme, Phys. Rev. Lett. \textbf{125}, 053001 (2020).

\bibitem{DS_Tie2024} C. Huang, C. Wang, H. Zhang, H. Hu, Z. Wang, Z. Mao, S. Li, P. Hou, Y. Wu, Z. Zhou, and L. Duan, Electromagnetically Induced Transparency Cooling of High-Nuclear-Spin Ions, Phys. Rev. Lett. \textbf{133}, 113204 (2024).


\bibitem{DS_QCe2014} X. Zhu, Y. Matsuzaki, R. Ams\"{u}ss, K. Kakuyanagi, T. Shimo-Oka, N. Mizuochi, K. Nemoto, K. Semba, W. J. Munro, and S. Saito, Observation of dark states in a superconductor diamond quantum hybrid system, Nat. Commun. \textbf{5}, 3424 (2014).

\bibitem{DS_QCe2015} Y. H. Liu, D. Lan, X. Tan, J. Zhao, P. Zhao, M. Li, K. Zhang, K. Dai, Z. Li, Q. Liu, S. Huang, G. Xue, P. Xu, H. Yu, S.-L. Zhu, and  Y. Yu, Realization of dark state in a three-dimensional transmon superconducting qutrit, Appl. Phys. Lett. \textbf{107}, 202601 (2015).

\bibitem{DS_QCe2018} A. Poto\v{c}nik, A. Bargerbos, F. A. Y. N. Schr\"{o}der, S. A. Khan, M. C. Collodo, S. Gasparinetti, Y. Salath\'{e}, C. Creatore, C. Eichler, H. E. T\"{u}reci, A. W. Chin, and A. Wallraff, Studying light-harvesting models with superconducting circuits, Nat. Commun. \textbf{9}, 904 (2018). 

\bibitem{DS_QCe2022} R. Holzinger, R. Guti\'{e}rrez-J\'{a}uregui, T. H\"{o}nigl-Decrinis, G.  Kirchmair, A. Asenjo-Garcia, and H. Ritsch, Control of Localized Single- and Many-Body Dark States in Waveguide QED, Phys. Rev. Lett. \textbf{129}, 253601 (2022).

\bibitem{DS_QCe2025} C. Wang, F.-M. Liu, H. Chen, Y.-F. Du, C. Ying, J.-W. Wang, Y.-H. Huo, C.-Z. Peng, X. Zhu, M.-C. Chen, C.-Y. Lu, and J.-W. Pan, Longitudinal and Nonlinear Coupling for High-Fidelity Readout of a Superconducting Qubit, Phys. Rev. Lett. \textbf{135}, 060803 (2025).


\bibitem{MSt:PRA2006} A. A. Rangelov, N. V. Vitanov, and B. W. Shore, Extension of the Morris-Shore transformation to multilevel ladders, Phys. Rev. A \textbf{74}, 053402 (2006).

\bibitem{CoDS:PRA2019} D. Finkelstein-Shapiro, S. Felicetti, T. Hansen, T. Pullerits, and A. Keller, Classification of dark states in multilevel dissipative systems, Phys. Rev. A \textbf{99}, 053829 (2019).

\bibitem{DS_ML2022} X. Yuan, Y. Li, M. Zhang, C. Liu, M. Zhu, X. Qin, N. V. Vitanov, Y. Lin, and J. Du, Preserving multilevel quantum coherence by dynamical decoupling, Phys. Rev. A \textbf{106}, 022412 (2022).

\bibitem{DDSiMS2025} K. Zhou, J. Wu, J. Shi, and T. Byrnes, Simple determination of dark states in a general multi-level system, J. Phys. A: Math. Theor. \textbf{58}, 095303 (2025).


\bibitem{ELtMS1998} N. V. Vitanov, Adiabatic population transfer by delayed laser pulses in multistate systems, Phys. Rev. A \textbf{58}, 2295 (1998).

\bibitem{DSoMM2013} G.-F. Xu and C. K. Law, Dark states of a moving mirror in the single-photon strong-coupling regime, Phys. Rev. A \textbf{87}, 053849 (2013).

\bibitem{ELtMS2020} N. V. Vitanov, High-fidelity multistate stimulated Raman adiabatic passage assisted by shortcut fields, Phys. Rev. A \textbf{102}, 023515 (2020).

\bibitem{ELtMS:PRA1991} P. Marte, P. Zoller, and J. L. Hall, Coherent atomic mirrors and beam splitters by adiabatic passage in multilevel systems, Phys. Rev. A \textbf{44}, R4118 (1991). %

\bibitem{ELtMS1992} A. V. Smith, Numerical studies of adiabatic population inversion in multilevel systems, J. Opt. Soc. Am. B \textbf{9}, 1543 (1992).%

\bibitem{ELtMS:JCP1991} Y. B. Band and P. S. Julienne, Population transfer by multiple stimulated Raman scattering, J. Chem. Phys. \textbf{95}, 5681 (1991).%



\bibitem{MAPT1991} B. W. Shore, K. Bergmann, J. Oreg, and S. Rosenwaks, Multilevel adiabatic population transfer, Phys. Rev. A \textbf{44}, 7442 (1991).

\bibitem{APT1993} P. Pillet, C. Valentin, R.-L. Yuan, and J. Yu, Adiabatic population transfer in a multilevel system, Phys. Rev. A \textbf{48}, 845 (1993). 

\bibitem{EoSTIRAP1997} V. S. Malinovsky and D. J. Tannor, Simple and robust extension of the stimulated Raman adiabatic passage technique to \textit{N}-level systems, Phys. Rev. A \textbf{56}, 4929 (1997).

\bibitem{OSTIRAP2005} W. Chalupczak and K. Szymaniec, Adiabatic passage in an open multilevel system, Phys. Rev. A \textbf{71}, 053410 (2005). 

\bibitem{M_STIRAP2008} E. Kuznetsova, P. Pellegrini, R. C\^{o}t\'{e}, M. D. Lukin, and S. F. Yelin, Formation of deeply bound molecules via chainwise adiabatic passage, Phys. Rev. A \textbf{78}, 021402(R) (2008).


\bibitem{QCaQI2000} M. A. Nielsen and I. L. Chuang, \textit{Quantum Computation and Quantum Information} (Cambridge University Press, Cambridge, 2000).


\bibitem{AO1994} M. Weitz, B. C. Young, and S. Chu, Atomic Interferometer Based on Adiabatic Population Transfer, Phys. Rev. Lett. \textbf{73}, 2563 (1994).
 
\bibitem{AO1998} P. D. Featonby, G. S. Summy, C. L. Webb, R. M. Godun, M. K. Oberthaler, A. C. Wilson, C. J. Foot, and K. Burnett, Separated-Path Ramsey Atom Interferometer, Phys. Rev. Lett. \textbf{81}, 495 (1998).
	
	
\bibitem{DC1981} F. T. Hioe and J. H. Eberly, \textit{N}-Level Coherence Vector and Higher Conservation Laws in Quantum Optics and Quantum Mechanics, Phys. Rev. Lett. \textbf{47}, 838 (1981).
		
		
\bibitem{MS:PRA1983} J. R. Morris and B. W. Shore, Reduction of degenerate two-level excitation to independent two-state systems, Phys. Rev. A \textbf{27}, 906 (1983).
		
\bibitem{MSN:JMO2014}B. W. Shore, Two-state behavior in \textit{N}-state quantum systems: The Morris-Shore transformation reviewed, J. Mod. Opt. \textbf{61}, 787 (2014).
		
\bibitem{MSfNS:PRA2020} K. N. Zlatanov, G. S. Vasilev, and N. V. Vitanov, Morris-Shore transformation for nondegenerate systems, Phys. Rev. A \textbf{102} 063113 (2020).
		
\bibitem{SVDiCQED2013} A. Wickenbrock, M. Hemmerling, G. R. M. Robb, C. Emary, and F. Renzoni, Collective strong coupling in multimode cavity QED, Phys. Rev. A \textbf{87}, 043817 (2013).		
	
\bibitem{SVDiJC2013} C. Emary, Dark-states in multi-mode multi-atom Jaynes-Cummings systems, J. Phys. B: At., Mol. Opt. Phys. \textbf{46}, 224008 (2013).
		
\bibitem{SVDoA2024} X. Li, Y. Zhou, and H. Zhang, Tunable atom-cavity interactions with configurable atomic chains, Phys. Rev. Appl. \textbf{21}, 044028 (2024).
		
			
\bibitem{huang2023dark} J. Huang, C. Liu, X.-W. Xu, and J.-Q. Liao, Dark-Mode Theorems for Quantum Networks, arXiv:2312.06274.


\bibitem{DM_OIT2020} D.-G. Lai, X. Wang, W. Qin, B.-P. Hou, F. Nori, and J.-Q. Liao, Tunable optomechanically induced transparency by controlling the dark-mode effect, Phys. Rev. A \textbf{102}, 023707 (2020).

\bibitem{DM_GSC2020} D.-G. Lai, J.-F. Huang, X.-L. Yin, B.-P. Hou, W. Li, D. Vitali, F. Nori, and J.-Q. Liao, Nonreciprocal ground-state cooling of multiple mechanical resonators, Phys. Rev. A \textbf{102}, 011502(R) (2020).

\bibitem{DM_OC2022} J. Huang, D.-G. Lai, C. Liu, J.-F. Huang, F. Nori, and J.-Q. Liao, Multimode optomechanical cooling via general dark-mode control, Phys. Rev. A \textbf{106}, 013526 (2022).

\bibitem{DM_OE2022} D.-G. Lai, J.-Q. Liao, A. Miranowicz, and F. Nori, Noise Tolerant Optomechanical Entanglement via Synthetic Magnetism, Phys. Rev. Lett. \textbf{129}, 063602 (2022).

\bibitem{DM_QS2022} J. Huang, D.-G. Lai, and J.-Q. Liao, Thermal-noise-resistant optomechanical entanglement via general dark-mode control, Phys. Rev. A \textbf{106}, 063506 (2022).

\bibitem{DM_QS2023} J. Huang, D. G. Lai, and J. Q. Liao, Controllable generation of mechanical quadrature squeezing via dark-mode engineering in cavity optomechanics, Phys. Rev. A \textbf{108}, 013516 (2023).	

		
\bibitem{DSiFSL} X. Zhao, Y. Xu, L.-M. Kuang, and J.-Q. Liao, Dark-state engineering in Fock-state lattices, Phys. Rev. Res. \textbf{7}, 033070 (2025).

		
\bibitem{Leon2014Applications}
S. J. Leon, \textit{Linear Algebra with Applications} (Pearson, Upper Saddle River, NJ, 2014).

		
\bibitem{DSP:PRA2002} M. Fleischhauer and M. D. Lukin, Quantum memory for photons: Dark-state polaritons, Phys. Rev. A \textbf{65}, 022314 (2002).
		
		
\end{thebibliography}
\end{document}